\documentclass{ws-ijmpb}
\usepackage{bm}

\begin{document}

\markboth{V.P. GUSYNIN, S.G.~SHARAPOV, J.P.~CARBOTTE} {GRAPHENE:
FROM TIGHT-BINDING MODEL TO QED$_{2+1}$}

%
\catchline{}{}{}{}{}
%

\title{AC CONDUCTIVITY OF GRAPHENE: \\FROM TIGHT-BINDING MODEL TO \\
$2+1$-DIMENSIONAL QUANTUM ELECTRODYNAMICS}

\author{V.P. GUSYNIN}

\address{Bogolyubov Institute for Theoretical Physics, \\14-b
        Metrologicheskaya Str., Kiev, 03680, Ukraine\\
vgusynin@bitp.kiev.ua}

\author{S.G.~SHARAPOV}

\address{Department of Physics and Astronomy, McMaster University,\\
        Hamilton, Ontario, L8S 4M1, Canada\\
Department of Physics, Western Illinois University, \\
Macomb, IL 61455, USA \\
sharapov@bitp.kiev.ua}

\author{J.P.~CARBOTTE}

\address{Department of Physics and Astronomy, McMaster University,\\
        Hamilton, Ontario, L8S 4M1, Canada\\
carbotte@mcmaster.ca}

\maketitle

\begin{history}
\received{Day Month Year}
\revised{Day Month Year}
\end{history}

\begin{abstract}
We consider the relationship between the tight-binding Hamiltonian
of the two-dimensional honeycomb lattice of carbon atoms with
nearest neighbor hopping only and the  $2+1$ dimensional Hamiltonian
of quantum electrodynamics which follows in the continuum limit. We
pay particular attention to the symmetries of the free Dirac
fermions including spatial inversion, time reversal, charge
conjugation and chirality. We illustrate the power of such a mapping
by considering the effect of the possible symmetry breaking which
corresponds to the creation of a finite Dirac mass, on various
optical properties. In particular, we consider the diagonal AC
conductivity with emphasis on how the finite Dirac mass might
manifest itself in experiment.  The optical sum rules for the
diagonal and Hall conductivities are discussed.
\end{abstract}


\vspace{2cm}

{\large Preprint: NSF-KITP-07-126}

\section{Introduction}

A single sheet of carbon atoms tightly packed into a two-dimensional
(2D) honeycomb lattice is called graphene. The band structure of
graphene consists approximately of a valence (full) and conduction
(empty) band both conical in shape with vertex meeting at a point
called a Dirac point. There are two inequivalent pairs of such
cones. This model, which is based on the continuum limit of
tight-binding bands can adequately describe low energy phenomena.
Deviations from simple cones can become important however near the
ends of the bands where additional details of the tight-binding
bands need to be considered. An example of the need to go beyond the
continuum approximation, which will be considered in this review, is a
discussion of optical sum rules which involve the integration of the
optical conductivity or closely related quantity over all energies
within the band. An important aspect of the graphene problem is that
it can be mapped, in the continuum approximation, into the
Hamiltonian of $2+1$ dimensional quantum electrodynamics
(QED$_{2+1}$) with Dirac fermions. It is remarkable that
historically this  field theoretical view of graphene started 20
years before\cite{Semenoff1984PRL} its actual
discovery\cite{Novoselov2004Science} and has motivated  theoretical
work on a condensed-matter realization of the parity
anomaly,\cite{Fradkin1986PRL,Haldane1988PRL,Schakel1991PRD} the
renormalization group approach in this
system,\cite{Gonzales1993NP,Gonzales1996PRL} and nonperturbative
dynamics of the generation of an excitonic-like gap in
graphene.\cite{Khveshchenko2001PRL,Gorbar2002PRB} The interest in
Dirac fermions in condensed matter theory was also related to
studies of the integer quantum Hall effect
(IQHE).\cite{Ludwig1994PRB} Nowadays graphene inspires work on the
fractionalization of
fermions\cite{Hou2007PRL,Jackiw2007,Herbut2007frac} which still
waits for the discovery.

The experimental proof of the existence of  Dirac fermions in
graphene\cite{Geim2005Nature,Kim2005Nature} that came from the
observation of the theoretically expected unconventional
QHE\cite{Zheng2002PRB,Gusynin2005PRL,Peres2006PRB} with the
quantized filling factor
\begin{equation}
\label{Hall-Dirac} \sigma_{xy} = \frac{e^2}{h}\nu, \qquad \nu = \pm
2 (2n+1), \qquad n=0,1,\ldots
\end{equation}
has promoted Dirac fermions from a beautiful theoretical toy to a
real object that one day may perform in a "graphenium inside"
processor\cite{review1} or even sooner become a resistance standard
operational above liquid-nitrogen
temperature.\cite{GeimKim2007Science} This fortunate situation has
created much excitement and it is now possible to consider doing
bench top experiment to explore the properties QED$_{2+1}$. This
mapping into QED also means that insight obtained in the study of
relativistic fermions can be brought to bare on the study of a
solid state system. Examples of relativistic effects
that could be studied include the Klein paradox which deals with the
perfect transmission of electrons through high potential
barrier\cite{Katsnelson2006NatPhys,Calogeracos2006NatPhys,review2}
and Zitterbewegung or "trembling"  of the center of a free wave
packet.\cite{Cserti2006PRB} The experimental realization of graphene
calls for an overview of the theoretical link between the condensed
matter lattice description of graphene and its continuum QED$_{2+1}$
formulation. This should help one judge better which theoretical
concepts developed during previous theoretical studies are important
for a deeper understanding of the analogies between the behavior of
electrons in graphene and in QED$_{2+1}$.

In this short review we consider in detail this correspondence
between a tight-binding lattice model and QED$_{2+1}$.  We
concentrate on symmetries of free Dirac fermion problem and the
consequences of possible lacking of such symmetries. In particular,
we consider spatial inversion, time reversal, charge conjugation and
a continuum $U(4)$ symmetry. We explain the difference between 2D
Dirac fermions in graphene and the 3D Dirac fermions that are
studied in the more familiar QED$_{3+1}$. In particular, there is a
difference in the physical meaning of  such an important quantum
number as {\em chirality} in QED$_{2+1}$ and QED$_{3+1}$. Because of
this and due to the fact that the notion of chirality is now widely
used in the literature on graphene (see
e.g.~Ref.~\refcite{Wu2007PRL}), we found that it is useful to
explain the similarities and differences of this notion in field
theory and graphene.

Another important issue is the possible effect of correlations on
various properties of graphene. For example, the new IQHE
states\cite{Zhang2006PRL,Abanin2007,Kim2007APS} with the filling
factor $\nu =0, \pm 1$ seen in the strong magnetic field $B$ greater than or equal to
$20\,\mbox{T}$ are likely to originate from many-body interactions. In
particular, the zero filling factor ($\nu=0$) state is related to a
spin polarized state\cite{Abanin2007,Kim2007APS}, while $\nu=\pm 1$
states are probably related to the lifting of sublattice degeneracy.
There also exists evidence\cite{Jiang2007} that Coulomb interactions
may be partially responsible for an observed increase in the energy
difference between second and first Landau levels in fields $\sim 18
\, \mbox{T}$. For the conventional case a theorem by
Kohn\cite{Kohn1961PRev} based on nearly parabolic bands shows that
there should be no such shifts due to correlations, but the theorem
may not apply here. An interesting feature of the QED description of
graphene is that there is a phenomenon of spontaneous symmetry
breaking when the interactions generate a {\em Dirac mass} or gap.
Historically this was exactly the interest in the condensed matter
realization of the anomalous properties of {\em massive} QED$_{2+1}$,
which brought the attention of
theoreticians\cite{Semenoff1984PRL,Fradkin1986PRL,Haldane1988PRL,Schakel1991PRD}
to graphene before its actual discovery. Now it is an {\em open
question\/} (see e.g. Ref.~\refcite{Yang2007} for a brief overview
and more references are provided below in
Sec.~\ref{sec:Dirac-mass}) whether the spontaneous mass generation
and symmetry breaking indeed occur in graphene or the observed $\nu
= \pm 1$ IQHE states have a different origin. Because the experiment
does not yet provide a definitive answer, in this review we discuss
the theoretical possibility of breaking the $U(4)$ symmetry of the
Lagrangian of graphene by eight Dirac masses. We illustrate how
the creation of a finite Dirac mass affects the discrete and continuum
symmetries and trace in one case its consequence in the diagonal AC
conductivity.  It is left to future experiments to verify this
theoretical possibility.

Although graphene is only 3 years old, there already exists several
popular articles at the introductory\cite{Wilson2006PT,Neto2006PW}
and more advanced\cite{Chakraborty2006PC,Katsnelson2007MT,Yang2007}
level. Also there is an excellent review article by A.K.~Geim and
K.~Novoselov\cite{review1} and a more theoretical review by
M.I.~Katsnelson and K.S.~Novoselov.\cite{review2} Because graphene
is a basic element of all graphitic forms, all existing literature
from late 40's onwards considers graphene as a starting point of the
analysis. Particularly useful are the results for carbon nanotubes
(see e.g. Refs.~\refcite{Saito:book,Ando2005JPSJ}).

Despite this variety of works on graphene we found that a more
formal view which can form the bridge between condensed matter and
quantum field theory is still missing, and we decided to
present it here. The review is organized as follows. In
Sec.~\ref{sec:lattice-model} we introduce the tight-binding model
with an external vector potential and obtain expressions for
electric current and diamagnetic term. In Sec.~\ref{sec:QED} we
derive the continuum QED$_{2+1}$ Lagrangian from the tight-binding
model of graphene. This allows one to trace the link between the
underlying lattice structure of graphene and clarify the physical
meaning of the spinor components in the Dirac theory. The physical
meaning of chirality in graphene is discussed in
Sec.~\ref{sec:chirality}. In Sec.~\ref{sec:discrete-QED}  the
discrete symmetry operations of the effective Dirac theory are
defined in {\em accordance\/} with the discrete symmetries of the
lattice model considered in Sec.~\ref{sec:discrete-lattice}. Basing
on the discrete symmetries  we discuss in
Sec.~\ref{sec:discrete-difference} the difference between massless
neutrinos in QED$_{3+1}$ and quasiparticles in graphene. Eight
possible Dirac masses are introduced in Sec.~\ref{sec:Dirac-mass}
and their transformation properties under discrete symmetries are
classified. The AC conductivity in zero and finite magnetic field
and optical sum rules are considered in Sec.~\ref{sec:conductivity}.
In Sec.~\ref{sec:concl} we present our conclusions and discuss open
questions.

\section{Tight-binding description of graphene}
\label{sec:lattice-model}

\subsection{Tight-binding model}

\begin{figure}[bt]
\centerline{\psfig{file=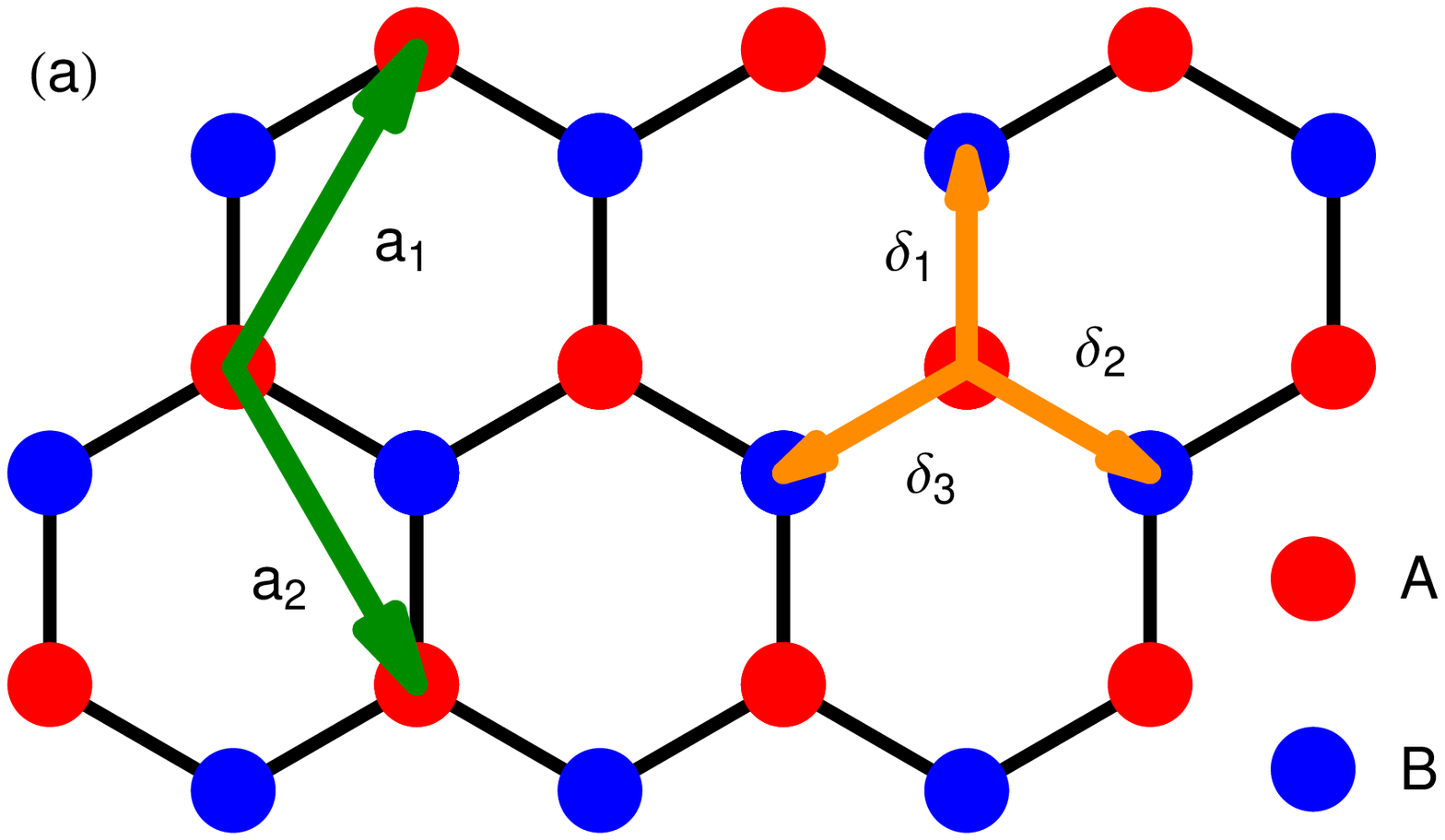,width=2.2in}\psfig{file=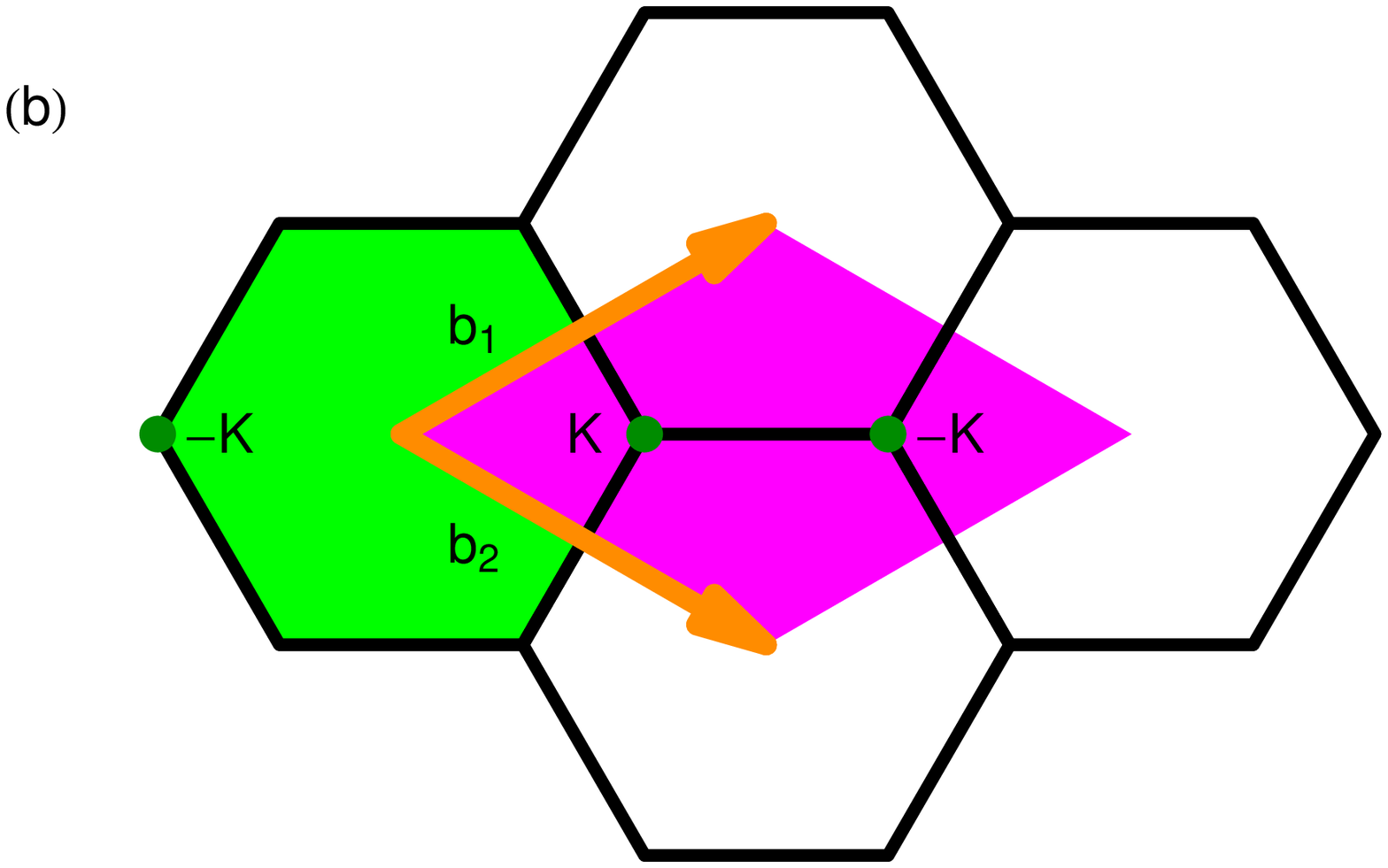,width=2.2in}}
\vspace*{8pt} \caption{(Colour online) (a) Graphene hexagonal
lattice constructed as a superposition of two triangular lattices
$\mathrm{A}$ and $\mathrm{B}$, with basis vectors $\mathbf{a}_{1,2}$
for lattice $\mathrm{A}$ and vectors $\bm{\delta}_i$ with $i=1,2,3$
connecting $\mathrm{A}$ to $\mathrm{B}$.  (b) The green hexagon is a
Brillouin zone (BZ) and pink diamond is the extended BZ for the
honeycomb lattice. The reciprocal lattice vectors are $\mathbf{b}_1$
and $\mathbf{b}_2$.} \label{fig:1}
\end{figure}

The honeycomb lattice can be described in terms of two triangular
sublattices, $\mathrm{A}$ and $\mathrm{B}$ (see Fig.~\ref{fig:1}~a).
A unit cell contains two atoms, one of type $A$ and one of type $B$.
The vectors
\begin{equation}
\label{basis} \mathbf{a}_1= a(\frac{1}{2}, \frac{\sqrt{3}}{2}),
\quad \mathbf{a}_2=a(\frac{1}{2},-\frac{\sqrt{3}}{2}),
\end{equation}
shown there are primitive translations,
where the lattice constant $a = |\mathbf{a}_1| = |\mathbf{a}_2| =
\sqrt{3} a_{CC}$ and $a_{CC}$ is the distance between two nearest
carbon atoms.
The corresponding reciprocal lattice whose vectors are $\mathbf{b}_1= \frac{2\pi}{a}(1,
1/\sqrt{3})$ and $\mathbf{b}_2= \frac{2\pi}{a}(1,-1/\sqrt{3})$ is shown in
Fig.~\ref{fig:1}~b together with the reduced (symmetrical and extended) Brillouin
zone. The reciprocal vectors satisfy the relation $\mathbf{a}_i \cdot \mathbf{b}_j =
2 \pi \delta_{ij}$.

Any $\mathrm{A}$ atom at the position $\mathbf{n}= \mathbf{a}_1 n_1
+ \mathbf{a}_2 n_2$, where $n_1,n_2$ are integers, is connected to
its nearest neighbors on $B$ sites by the three vectors $\bm{
\delta}_i $:
\begin{equation}
\label{delta_i} \bm{\delta}_1 =(\mathbf{a}_1-\mathbf{a}_2)/3, \quad
\bm{\delta}_2 =\mathbf{a}_1/3 + 2\mathbf{a}_2/3,\quad \bm{\delta}_3
= - \bm{\delta}_1-\bm{\delta}_2=-2\mathbf{a}_1/3 -\mathbf{a}_2/3.
\end{equation}
Besides translations, the symmetry group of honeycomb lattice
includes rotations ($R$ and $R^{-1}$) on $\pm 2\pi/3$ and mirror
reflections ($Y_{1},Y_{2},Y_{3}$) about planes passing through the
center and three hexagon vertices. Together the rotations and
reflections form a nonabelian group $C_{3v}$  with $6$ elements.
Their explicit matrix representation may be defined by matrices
giving the transformations of the basis vectors $\mathbf{a}_1$ and
$\mathbf{a}_2$,
\begin{eqnarray}
\label{group} &&E=\left(\begin{array}{cc}1 &   0 \\0 & 1 \\
\end{array}\right),\, R=\left(\begin{array}{cc} -1 &   -1 \\1 & 0 \\
\end{array}\right),\,
R^{-1}=\left(\begin{array}{cc}0 &   1 \\-1 & -1\\ \end{array}\right),\nonumber\\
&&Y_{1}=\left(\begin{array}{cc}-1& 0\\ 1&1 \\ \end{array}\right),\,
Y_{2}=\left(\begin{array}{cc}1&1 \\ 0&-1 \\ \end{array}\right),\,
Y_{3}=\left(\begin{array}{cc}0&-1\\-1&0 \\ \end{array}\right).
\end{eqnarray}
In addition there is a reflection Z in the plane of the sheet
(changes $\mathbf{a}_i\rightarrow - \mathbf{a}_i$, followed by the
translation by one of the vectors $3\bm{\delta}_i$) and,
accordingly, combinations of Z with any of the above operations. All
these operations do not interchange $\mathrm{A}$- and
$\mathrm{B}$-type atoms. The one-electron eigenfunctions  can be
classified according to the subgroup ($E,Z$) whether the states are
even ($\sigma$ states) or odd ($\pi$ states) under reflection $Z$.
There are also operations which interchange $\mathrm{A}$ and
$\mathrm{B}$ atoms, for example, reflections $X_{1},X_{2},X_{3}$ in
mirror planes perpendicular to the corresponding $Y_{i}$ planes.
They are symmetry operations when accompanied by some fractional
translations.\cite{Lomer1955PRS}
The rotation and rotation-reflection operations (leaving aside $Z$
reflection) form the point-group of graphene which contains 12
elements. The effect of group operations $G$ on a function of the
coordinates is defined as
$\psi^\prime(\mathbf{r})=\psi(G^{-1}\mathbf{r})=T(G)\psi(\mathbf{r})$,
and it allows one to determine the irreducible representations for
graphene. In particular, one should note that the subgroup
(\ref{group}) has a two-dimensional spinor representation which we
use below. Full consideration of the symmetry group of graphene was
made in Refs.~\refcite{Lomer1955PRS,Slonczewski1958PRev}.

The carbon atoms in graphene plane are connected by strong covalent
$\sigma$-bonds due to the $sp^2$ hybridization of the atomic
$2s,2p_x,2p_y$ orbitals. The $2p_z$ ($\pi$) orbitals are
perpendicular to the plane and have a weak overlap. Therefore we
start with the simplest tight-binding description for $\pi$ orbitals
of carbon in terms of the Hamiltonian
\begin{equation}
\label{Hamilton-lattice} H =-
t\sum_{\mathbf{n},\bm{\delta}_i,\sigma}\left[a_{\mathbf{n},\sigma}^\dagger
 \exp \left( \frac{ie}{\hbar c}\bm{\delta}_i\mathbf{A} \right)
b_{\mathbf{n}+\bm{\delta},\sigma} + \mbox{c.c.}\right],
\end{equation}
where $t$ is the nearest neighbor hopping parameter,
$a_{\mathbf{n},\sigma}$ and $b_{\mathbf{n}+ \bm{\delta},\sigma}$
are the Fermi operators of electrons with spin $\sigma =
\uparrow,\downarrow$ on $A$ and $B$ sublattices, respectively. Since
we are interested in the current response, the vector potential
$\mathbf{A}$ is introduced in the Hamiltonian
(\ref{Hamilton-lattice}) by means of the Peierls substitution
$a_{\mathbf{n},\sigma}^\dagger b_{\mathbf{m},\sigma}\rightarrow
a_{\mathbf{n},\sigma}^\dagger\exp\left(- \frac{i e}{\hbar
c}\int_{\mathbf{m}}^{\mathbf{n}}{\bf A}d
\mathbf{r}\right)b_{\mathbf{m},\sigma}$ that introduces the phase
factor $\exp (\frac{ie}{\hbar c}\bm{\delta}_i\mathbf{A})$ in the
hopping term (see Ref.~\refcite{Millis:book} for a review). We keep
Planck constant $\hbar$ and the velocity of light $c$, but set
$k_B=1$. The charge of electron is $-e<0$. Note that we consider a
2D model. The experiment shows that graphene, in order to exist
without a substrate, spreads itself out of the 2D
plane.\cite{Meyer2007Nature}


Expanding the Hamiltonian (\ref{Hamilton-lattice}) to the second
order in the vector potential, one has
\begin{equation}
\label{Hamitonian-expand}
H=H_0-\sum_{\mathbf{n}}\left[\frac{1}{c}\mathbf{A}(\mathbf{n})
\mathbf{j}(\mathbf{n})-\frac{1}{2c^2}A_\alpha(\mathbf{n})\tau_{\alpha
\beta}(\mathbf{n})A_\beta(\mathbf{n})\right], \qquad \alpha,
\beta=1,2.
\end{equation}
The total current density operator is obtained by differentiating
Eq.~(\ref{Hamitonian-expand}) with respect to
$A_\alpha(\mathbf{n})$,
\begin{equation}
j_\alpha(\mathbf{n})=-\frac{\partial
H}{\partial\left(A_\alpha/c\right)}=j_\alpha^P(\mathbf{n})-
\tau_{\alpha \beta}(\mathbf{n})A_\beta(\mathbf{n})/c,
\end{equation}
and consists of the usual paramagnetic  part,
\begin{equation}
\label{current-param} j_\alpha^P(\mathbf{n})=\frac{ite}{\hbar}
\sum_{\bm{\delta}_i,\sigma}(\delta_{i})_{\alpha}\left[a_{\mathbf{n},\sigma}^\dagger
b_{\mathbf{n}+\bm{\delta},\sigma} -
b_{\mathbf{n}+\bm{\delta},\sigma}^\dagger
a_{\mathbf{n},\sigma}\right],
\end{equation}
and diamagnetic part,
\begin{equation}
\label{diamagnetic-term} \tau_{\alpha
\beta}(\mathbf{n})=\frac{\partial^2
H}{\partial\left(A_\alpha/c\right)\partial\left(A_\beta/c\right)}
=\frac{te^2}{\hbar^2}
\sum_{\bm{\delta}_i,\sigma}(\delta_{i})_{\alpha}(\delta_{i})_{\beta}\left[a_{\mathbf{n},\sigma}^\dagger
b_{\mathbf{n}+\bm{\delta},\sigma}
+b_{\mathbf{n}+\bm{\delta},\sigma}^\dagger
a_{\mathbf{n},\sigma}\right].
\end{equation}
We stress that to obtain the correct form of the current
(\ref{current-param}) and the diamagnetic part
(\ref{diamagnetic-term}), the Peierls substitution has to be made in
the initial Hamiltonian (\ref{Hamilton-lattice}) or its momentum
representation (see Eq.~(\ref{H_0}) below) rather than after the
diagonalization of these Hamiltonians is made (see
Ref.~\refcite{Fukuyama2007JPSJ} and references therein). We will
return to the paramagnetic and diamagnetic terms of the interacting
Hamiltonian (\ref{Hamitonian-expand}) later when we discuss coupling
to the vector potential in Sec.~\ref{sec:Lagrangian} and the sum
rules in Sec.~\ref{sec:sum}. For now we consider the noninteracting
Hamiltonian $H_0$.

\subsection{Spinor representation of the noninteracting Hamiltonian}

In the momentum representation the Hamiltonian $H_0$ reads
\begin{equation}
\label{H_0} H_0 =  \sum_{\sigma} \int_{BZ} \frac{d^2 k}{(2 \pi)^2}
\Upsilon_{\sigma}^\dagger (\mathbf{k}) \mathcal{H}_0
\Upsilon_{\sigma} (\mathbf{k}), \qquad \mathcal{H}_0 = \left(
  \begin{array}{cc}
    0 &   \phi (\mathbf{k}) \\
    \phi^\ast (\mathbf{k}) & 0 \\
  \end{array}
\right)
\end{equation}
with $ \phi(\mathbf{k}) = -t \sum_{\bm{\delta}_i} e^{i \mathbf{k}
\bm{\delta}_i} \equiv- \epsilon(\mathbf{k}) e^{i \varphi
(\mathbf{k})}$. In the basis (\ref{basis}) we get
\begin{eqnarray}
\label{phi} \phi(\mathbf{k}) &=&
 -te^{i\mathbf{k}(\mathbf{a}_{1}-\mathbf{a}_{2})/3}
\left[1+e^{i\mathbf{k}\mathbf{a}_{2}}+e^{-i\mathbf{k}\mathbf{a}_{1}}\right]\nonumber \\
&=& -t \left[\exp \left(i \frac{k_y a}{\sqrt{3}}\right) + \exp
\left(-i \frac{k_y a}{2\sqrt{3}}\right) 2 \cos \frac{k_x a}{2}
\right]
\end{eqnarray}
and
\begin{equation}
\label{dispersion} \epsilon(\mathbf{k})=  t \sqrt{1+ 4 \cos^2
\frac{k_x a}{2} +4 \cos \frac{k_x a}{2} \cos \frac{\sqrt{3} k_y
a}{2}}.
\end{equation}
\begin{figure}[bt]
\centerline{\psfig{file=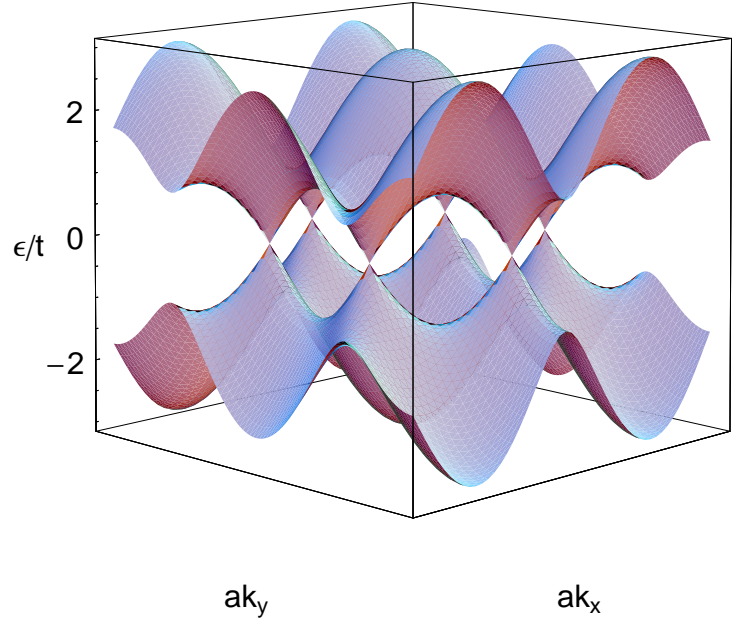,width=4in}} \vspace*{8pt}
\caption{(Colour online) The energy band structure of graphene.
Valence and conduction bands meet at six $\mathbf{K}$ points.}
\label{fig:2}
\end{figure}
Accordingly, because the graphene structure contains two atoms per
unit cell (two sublattices), the spectrum of quasiparticles
excitations has two branches (bands) with the
dispersion\cite{Wallace1947PRev} $E_{\pm} = \pm
\epsilon(\mathbf{k})$ shown in Fig.~\ref{fig:2}. In Eq.~(\ref{H_0})
we introduced the spinors
\begin{equation}
\label{spinor} \Upsilon_{\sigma} (\mathbf{k})= \left(
                                                 \begin{array}{c}
                                                   a_{\sigma} (\mathbf{k}) \\
                                                   b_{\sigma} (\mathbf{k})\\
                                                 \end{array}
                                               \right)
\end{equation}
with the operator $\Upsilon_{\sigma} (\mathbf{k})$ being the Fourier
transform of the spinor $\Upsilon_{\sigma}(\mathbf{n}) = \left(
                                          \begin{array}{c}
                                            a_{\mathbf{n},\sigma} \\
                                            b_{\mathbf{n},\sigma}
                                          \end{array}
                                        \right)$:
\begin{equation}
\label{Fourier} \Upsilon_{\sigma}(\mathbf{n}) =\sqrt{S}
\int_{BZ}\frac{d^2{\mathbf{k}}}{(2\pi)^2}\, e^{i \mathbf{k}
\mathbf{n}}  \Upsilon_{\sigma} (\mathbf{k}).
\end{equation}
Here  $S = \sqrt{3} a^2/2$ is the area of a unit cell and the
integration in Eqs.~(\ref{H_0}) and (\ref{Fourier}) goes over the
extended rhombic Brillouin zone (BZ). We also add to $H_0$ the
Zeeman term and the chemical potential
\begin{equation}
\label{H-mu} H_Z= -\sum_{\sigma} \mu_\sigma \int_{BZ} \frac{d^2
k}{(2 \pi)^2} \Upsilon_{\sigma}^\dagger (\mathbf{k})
\Upsilon_{\sigma}(\mathbf{k})
\end{equation}
with $\mu_{\sigma} = \mu - \sigma g/2 \mu_B B$, where $\mu_B = e
\hbar/(2m_ec)$ is the Bohr magneton, and $g$ is the Lande factor. We
count the Zeeman energy from the chemical potential $\mu$, so that
our subsequent consideration is based on the grand canonical
ensemble.

The corresponding imaginary time Green's function (GF) is defined as
a thermal average
\begin{equation}
\label{GF-def} G_\sigma(\tau_1-\tau_2,\mathbf{n}_1 - \mathbf{n}_2) =
- \langle T_\tau \Upsilon_\sigma(\tau_1,\mathbf{n}_1)
\Upsilon_\sigma^{\dagger}(\tau_2,\mathbf{n}_2) \rangle
\end{equation}
and  its Fourier transform is
\begin{eqnarray}
\label{GF-Fourier} && G_\sigma(\tau_1-\tau_2,\mathbf{n}_1 -
\mathbf{n}_2) = \\&&S T \sum_n \int_{BZ}
\frac{d^2{\mathbf{k}}}{(2\pi)^2} G_\sigma(i \omega_n, \mathbf{k})
\exp[- i \omega_n (\tau_1 - \tau_2)+ i \mathbf{k}(\mathbf{n}_1 -
\mathbf{n}_2)] \nonumber
\end{eqnarray}
with
\begin{equation}
\label{GF-basic} G_\sigma(i \omega_n,\mathbf{k}) = \frac{(i \omega_n
+ \mu_\sigma) \hat{I} + \tau_+ \phi(\mathbf{k}) + \tau_-
\phi^\ast(\mathbf{k})}{(i \omega_n + \mu_\sigma)^2 -
\epsilon^2(\mathbf{k})},
 \qquad \omega_n = \pi (2n+1)T,
\end{equation}
where the matrices $\tau_{\pm} = (\tau_1 \pm i \tau_2)/2$ made from
Pauli matrices, operate in the sublattice space. The GF
(\ref{GF-basic}) describes the electron- and hole-like excitations
with the energies $E_{\pm}(\mathbf{k}) = \pm
\epsilon(\mathbf{k})-\mu_\sigma$, respectively.

The dispersion $\epsilon(\mathbf{k})$ near the six $\mathbf{K}$
points $\pm 2\pi/a (1/3, 1/\sqrt{3})$, $\pm 2\pi/a(2/3, 0)$, $\pm
2\pi/a(1/3, -1/\sqrt{3})$  at the corners of the hexagonal BZ (see
Fig.~\ref{fig:2}) is linear,\cite{Wallace1947PRev} $E_{\pm}({\bf
p})=\pm \hbar v_F|\mathbf{p}| -\mu_\sigma$ (see Fig.~\ref{fig:3}~(a)
in Sec.~\ref{sec:Dirac-mass}), where the wave vector ${\bf
p}=(p_1,p_2)$ is now measured from the $\mathbf{K}$ points and the
Fermi velocity is $v_F = \sqrt{3} ta/(2\hbar)$. Its experimental
value\cite{Geim2005Nature,Kim2005Nature} is $v_F \approx 10^6
\mbox{m/s}$ (see also Ref.~\refcite{Deacon2007} for more recent
values of $v_F$ and $t$). Since only two $\mathbf{K}$ points are
inequivalent, in what follows, we select them to be $\mathbf{K}_{\pm}
= \pm 2\pi/a(2/3,0)$ (see Fig.~\ref{fig:1}~b), so that they are
inside the extended BZ. The degeneracy of the two $\mathbf{K}_\pm$
points is protected by the point-group symmetry of the hexagonal
lattice.

\subsection{Discrete symmetries of the tight-binding Hamiltonian}
\label{sec:discrete-lattice}

The $\mathbf{K}_{\pm}$ points in single layer graphene are stable
against perturbations which preserve the discrete spacetime
inversion symmetry and which do not mix the $\mathbf{K}_{\pm}$
points.\cite{Manes2007PRB} We consider these symmetries for the
tight-binding model and define below the corresponding discrete
symmetry operations for the effective QED$_{2+1}$ theory of
graphene. A topological stability for the appearance of massless
Dirac fermions was studied in Ref.~\refcite{Hatsugai2007}, and
in Sec.~\ref{sec:e-h-lattice} we discuss it.

\subsubsection{The spatial inversion $\mathcal{P}$}
\label{sec:inversion-lattice}

Choosing the center of symmetry to be the center of the
hexagon\cite{Slonczewski1958PRev} in Fig.~\ref{fig:1}~a, we observe
that the spatial inversion $\mathcal{P}$: $(x,y) \to (-x,-y)$ would
be the symmetry of the system if one also exchanges $\mathrm{A}$ and
$\mathrm{B}$ atoms, i.e.
\begin{equation}
\label{inversion-lattice-space} a_{\mathbf{n},\sigma} \rightarrow
\mathcal{P} a_{\mathbf{n},\sigma} \mathcal{P}^{-1} =
b_{-\mathbf{n},\sigma} \qquad b_{\mathbf{n},\sigma} \rightarrow
\mathcal{P} b_{\mathbf{n},\sigma} \mathcal{P}^{-1} =
a_{-\mathbf{n},\sigma}.
\end{equation}
As we will discuss in Sec.~\ref{sec:discrete-QED}, the definition of
$\mathcal{P}$ is different from that commonly used in
QED$_{2+1}$.\cite{Jackiw2007}  In momentum space $\mathcal{P}$ acts
as follows
\begin{equation}
\label{inversion-lattice} a_\sigma(\mathbf{k}) \rightarrow
\mathcal{P} a_\sigma(\mathbf{k}) \mathcal{P}^{-1} =
b_\sigma(-\mathbf{k}), \qquad b_\sigma(\mathbf{k}) \rightarrow
\mathcal{P} b_\sigma(\mathbf{k}) \mathcal{P}^{-1} =
a_\sigma(-\mathbf{k}),
\end{equation}
i.e. it reverses the sign of the momentum $\mathbf{k} \to
-\mathbf{k}$ and exchanges $\mathbf{K}$ points. For the spinors
$\Upsilon_{\sigma}(\mathbf{k})$, the action of $\mathcal{P}$ is
defined as
\begin{equation}
\label{inversion-lattice-spinor} \Upsilon_{\sigma}(\mathbf{k})
\longrightarrow  \mathcal{P} \Upsilon_{\sigma}(\mathbf{k})
\mathcal{P}^{-1} =\tau_1 \Upsilon_{\sigma}(-\mathbf{k}).
\end{equation}
One can easily check that the noninteracting Hamiltonian $H_0$ is
invariant under $\mathcal{P}$
\begin{equation}
\label{H0-inversion-invariant} H_0 \longrightarrow \mathcal{P} H_0
\mathcal{P}^{-1} =H_0
\end{equation}
because the Hamiltonian density $\mathcal{H}_0$ satisfies the
condition
\begin{equation}
\label{inversion-H0-lattice} \tau_1 \mathcal{H}_0(\mathbf{k}) \tau_1
= \mathcal{H}_0(-\mathbf{k}).
\end{equation}
However, if one assumes that the densities of particles on
the $\mathrm{A}$ and $\mathrm{B}$ sublattices are different, i.e.
that
\begin{eqnarray}
\label{H_1-def} H_1  & = & \sum_{\sigma} \int_{BZ}
\frac{d^2{\mathbf{k}}}{(2\pi)^2} [m_a a_\sigma^\dagger(\mathbf{k})
a_\sigma(\mathbf{k}) + m_b
b_\sigma^\dagger(\mathbf{k}) b_\sigma(\mathbf{k})]\\
& = & \sum_{\sigma} \int_{BZ} \frac{d^2{\mathbf{k}}}{(2\pi)^2}
\Upsilon_\sigma^\dagger(\mathbf{k}) [m_{+} \, \tau_0 + m_{-}
\,\tau_3]\Upsilon_\sigma(\mathbf{k})  \nonumber
\end{eqnarray}
with $m_{+} = (m_a + m_b)/2$ [$\tau_0$ is $2\times 2$ unit matrix]
and $m_{-} =(m_a - m_b)/2$ such term would break the inversion
symmetry by swapping now inequivalent sublattices:
\begin{equation}
H_1 \longrightarrow \mathcal{P} H_1 \mathcal{P}^{-1} = \sum_\sigma
\int_{BZ} \frac{d^2 k}{(2 \pi)^2} [m_b a_\sigma^\dagger(\mathbf{k})
a_\sigma(\mathbf{k}) + m_a b_\sigma^\dagger(\mathbf{k})
b_\sigma(\mathbf{k}) ]
\end{equation}
or in spinor notation
\begin{equation}
H_1 \longrightarrow \mathcal{P} H_1 \mathcal{P}^{-1} = \sum_\sigma
\int_{BZ} \frac{d^2 k}{(2 \pi)^2} \Upsilon_{\sigma}^\dagger
(\mathbf{k}) (m_{+} \tau_0 - m_{-} \tau_3)\Upsilon_{\sigma}
(\mathbf{k}).
\end{equation}
One can see that the term with $m_+$ that corresponds to the same
carrier density on $\mathrm{A}$ and $\mathrm{B}$ sublattices can be
absorbed in the chemical potential $\mu$, and the carrier imbalance
term with $m_-$ is parity breaking. In practice, it has been
suggested that one can introduce a Dirac mass term by placing the
graphene sheet on a substrate made of hexagonal boron nitride. In
this circumstance the two carbon sublattices become inequivalent
because of its interaction with the substrate. A recent band
structure calculation\cite{Giovannetti2007} for such a configuration
has given a gap of about $53\,\mbox{meV}$.

\subsubsection{Time reversal $\mathcal{T}$}
\label{sec:time-lattice}

The time-reversal operation, $t \to -t$ changes the signs of  the
momentum and spin leaving the sign of coordinates
unchanged:\cite{Davydov.book}
\begin{equation}
\left(
\begin{array}{c}
a_+(\mathbf{k}) \\
a_-(\mathbf{k}) \\
\end{array}
\right) \rightarrow \mathcal{T} \left(
\begin{array}{c}
a_+(\mathbf{k}) \\
a_-(\mathbf{k}) \\
\end{array}
\right) \mathcal{T}^{-1} = \left(
\begin{array}{c}
a_-(-\mathbf{k}) \\
-a_+(-\mathbf{k}) \\
\end{array}
\right) = i \sigma_2 \left(
\begin{array}{c}
a_+(-\mathbf{k}) \\
a_-(-\mathbf{k}) \\
\end{array}
\right),
\end{equation}
The operator $b_\pm(\mathbf{k})$ obeys the same rule, and
$a_\pm^\dagger(\mathbf{k})$, $b_\pm^\dagger(\mathbf{k})$ transform
as follows
\begin{equation}
\left(
\begin{array}{c c}
a_+^\dagger(\mathbf{k}) &  a_-^\dagger(\mathbf{k})
\end{array}
\right) \rightarrow \mathcal{T} \left(
\begin{array}{c c}
a_+^\dagger(\mathbf{k}) & a_-^\dagger(\mathbf{k})
\end{array}
\right) \mathcal{T}^{-1} = \left(
\begin{array}{c c}
a_+^\dagger(\mathbf{k}) & a_-^\dagger(\mathbf{k})
\end{array}
\right) (- i \sigma_2).
\end{equation}
To invert the direction of time, the  operator $\mathcal{T}$ has to
be antiunitary. The action of $\mathcal{T}$ on the sublattice
spinors is given by
\begin{equation}
\label{time-lattice-spinor} \Upsilon(\mathbf{k}) \longrightarrow
\mathcal{T} \Upsilon(\mathbf{k}) \mathcal{T}^{-1} = i \sigma_2
\Upsilon(-\mathbf{k}),
\end{equation}
where $\sigma_2$ acts on the spin indices of the spinor
$\Upsilon_\sigma(\mathbf{k})$.

One can check that the noninteracting Hamiltonian $H_0$ is invariant
under $\mathcal{T}$
\begin{equation}
\label{H0-time-invariant} H_0 \longrightarrow \mathcal{T} H_0
\mathcal{T}^{-1} =H_0
\end{equation}
because $\mathcal{H}_0$ satisfies the condition
\begin{equation}
\label{time-H0-lattice} \tau_0 \mathcal{H}_0^\ast(\mathbf{k})\tau_0
= \mathcal{H}_0(-\mathbf{k}).
\end{equation}
Since a fixed external magnetic field breaks  time reversal
symmetry,  both the Hamiltonian (\ref{Hamilton-lattice}) which
includes this field and the Zeeman term (\ref{H-mu}) break it. On
the other hand, the Hamiltonian $H_1$ given by (\ref{H_1-def}) is
invariant under time reversal, but if we consider spin dependent
masses $m_{a,b,\sigma}$, it will also break this symmetry.

The presence of combined  $\mathcal{P}\mathcal{T}$ symmetry enforces
a condition on $\mathcal{H}_0$:
\begin{equation}
\label{PT} \tau_1\mathcal{H}^\ast_0(\mathbf{k})\tau_1 = {\cal
H}_0(\mathbf{k})
\end{equation}
or in matrix components: $\mathcal{H}_{0}^{11} =
\mathcal{H}_{0}^{22}$ and $\mathcal{H}_{0}^{12} =
\mathcal{H}_{0}^{21 \ast}$. One can say that the first equality
forbids $m_- \tau_3$ terms in the Hamiltonian protecting
$\mathbf{K}_{\pm}$ points when they are not
mixed.\cite{Manes2007PRB} The energy spectrum is $E =
\mathcal{H}_{0}^{11} \pm |\mathcal{H}_{0}^{12}|$. For
$\mathcal{H}_{0}^{11} = \mbox{const}$, a constant term can be
absorbed in the chemical potential, so that $\mathcal{P}\mathcal{T}$
symmetry results in the symmetry of the spectrum similarly to the
relation (\ref{tau_3}) considered below.

\subsubsection{Particle-hole symmetry of the spectrum and stability of Fermi point}
\label{sec:e-h-lattice}

The Hamiltonian $\mathcal{H}_0(\mathbf{k})$  also satisfies two
other relations\cite{Ryu2002PRL,Hou2007PRL,Hatsugai2007,Ziegler2007}
(see also the end of Sec.~\ref{sec:chirality})
\begin{equation}
\label{tau_3} \tau_3 \mathcal{H}_0(\mathbf{k}) \tau_3 = -
\mathcal{H}_0(\mathbf{k})
\end{equation}
and
\begin{equation}
\label{tau_2} \tau_2 \mathcal{H}_0^\ast(\mathbf{k}) \tau_2 = -
\mathcal{H}_0(\mathbf{k}).
\end{equation}
It is easy to check that they guarantee that if there is a state $|
\psi\rangle$ with energy $E$, then the states $\tau_3 | \psi\rangle$
and $\tau_2 | \psi\rangle^\ast$ correspond to the energy $-E$ which
implies that the energy bands are symmetric about $E=0$. We note
that while the parity breaking term $m_- \Upsilon^\dagger \tau_3
\Upsilon$ obviously violates the condition (\ref{tau_3}), the more
general condition (\ref{tau_2}) is still satisfied. Indeed, even for
$m_- \neq 0$ the spectrum is symmetric about $E=0$:
$E(\mathbf{k})=\pm\sqrt{m^{2}_{-}+|\phi(\mathbf{k})|^{2}}$. In order
for the Fermi point $E(\mathbf{k}=0)=0$ to exist, we must require
the relation (\ref{tau_3}) which is a necessary but not sufficient
condition for its existence. The existence and stability of the
Fermi point is dictated by topology in momentum space\cite{Volovik1}
that is by nonzero topological invariant (winding number) which is
expressed analytically as
\begin{equation}
N= \oint_{C}\frac{d \mathbf{k}}{4\pi i}{\rm
tr}[\sigma_{3}H^{-1}(\mathbf{k})\partial_{\mathbf{k}}H(\mathbf{k})].
\end{equation}
Here the integral is taken over an arbitrary contour $C$ around one
of the $\mathbf{K}_{\pm}$ points and ${\rm tr}$ is the trace over
sublattice indices. For the Hamiltonian ${\cal H}_{0}$ given by
(\ref{H_0}) the topological invariant can be rewritten in the form
\begin{equation}
N=\frac{1}{4\pi i}\oint_{C}\frac{\phi(\mathbf{k}) d\phi^{*}(\mathbf{k})-\phi^{*}(\mathbf{k})
d\phi(\mathbf{k})}{|\phi(\mathbf{k})|^{2}}.
\end{equation}
For $\mathbf{K}_{\pm}$ points this gives $N(\mathbf{K}_{\pm})=\pm1$.
On the other hand, when the spectrum becomes gapped, the topological
invariant $N=0$ and the Fermi surfaces disappear. This can happen
because the total topological charge at two $\mathbf{K}_{\pm}$
points $N=0$. The trivial total topological charge of the Fermi
surfaces allows for their annihilation, which occurs  when the
energy spectrum becomes fully gapped.

Now we come to the  question of topological stability of the Dirac
points with respect to the presence of the other hopping
terms.\cite{Hatsugai2007}

The next neighbor (intrasublattice) hopping considered, for example,
in Ref.~\refcite{Gaididei2006FNT}, introduces only the diagonal term
in the Hamiltonian $\mathcal{H}_0$ in Eq.~(\ref{H_0}) which
eliminates the particle-hole symmetry of original energy bands and for
which small deviations from $\mathbf{K}_\pm$ can be absorbed in
the chemical potential.\cite{Peres2006PRB} Thus, we address here the
role of the diagonal transfer (third neighbor hopping) $t^\prime$
which is described by
\begin{equation}
\label{Hamilton-lattice-diagonal} H_{t^\prime}
=-t^{\prime}\sum_{\mathbf{n},\sigma}a_{\mathbf{n},\sigma}^\dagger
b_{\mathbf{n}-2\bm{\delta}_{1}} + \mbox{c.c.}.
\end{equation}
Accordingly, in the presence of $H_{t^\prime}$, Eq.~(\ref{phi}) for
$\phi(\mathbf{k})$ takes the form
\begin{eqnarray}
\phi(\mathbf{k}) &=& -t \sum_{\bm{\delta}_i} e^{i \mathbf{k}
\bm{\delta}_i}-t^{\prime}e^{-2i\mathbf{k}\bm{\delta}_1} \\
&=&-t e^{i \mathbf{k} \bm{\delta}_1}\left[1+ e^{i
\mathbf{k}(\bm{\delta}_2-\bm{\delta}_1)}+ e^{i
\mathbf{k}(\bm{\delta}_3-\bm{\delta}_1)}+\frac{t^{\prime}}{t}e^{-3i\mathbf{k}\bm{\delta}_1}\right].
\nonumber
\end{eqnarray}
Using the relations (\ref{delta_i}) we write
\begin{equation}
\phi(\mathbf{k}) = -te^{i \mathbf{k}
\bm{\delta}_1}\left[1+e^{ik_{2}}+e^{-ik_{1}}\left(1+\frac{t^{\prime}}{t}e^{ik_{2}}\right)\right]
\equiv -te^{i \mathbf{k} \bm{\delta}_1}\bar \phi(\mathbf{k}),
\end{equation}
where we introduced the notations
$k_{1}=\mathbf{k}\mathbf{a}_1, k_{2}=\mathbf{k}\mathbf{a}_2 $. Note
that these variables change from $-\pi$ to $\pi$. In the complex
plane $\bar \phi(\mathbf{k})$ delineates a circle centered at
$C_{0}=1+e^{ik_{2}}$ with a radius $r=\sqrt{1+2(t^{\prime}/t)\cos
k_{2}+ (t^{\prime}/t)^{2}}$ when $k_{1}$ is changed from $-\pi$ to
$\pi$ for a fixed value $k_{2}$. In order to have a Dirac cone, the
circle must cross the origin of coordinates. This is guaranteed for
the values $-3\leq t^{\prime}/t <1 $. Thus, the Dirac cone is stable
against additional interactions, diagonal hopping, when the last
condition is satisfied. The appearance of gapless Dirac fermions is
not accidental to the honeycomb lattice, but is rather generic for a
class of two-dimensional lattices that interpolate between square
($t^{\prime}=t$) and $\pi$-flux ($t^{\prime}=-t$) lattices. This
indicates a certain robustness of the topological quantum number.

\section{QED$_{2+1}$ description of graphene}
\label{sec:QED}

\subsection{Noninteracting Dirac Hamiltonian}
\label{Dirac-Hamiltonian}

Around two inequivalent $\mathbf{K}_\pm$ points, where
$\epsilon(\mathbf{K}_\pm) =0$, the function (\ref{phi}) can be
expanded as $\phi(\mathbf{K}_\pm + \mathbf{p}) = \pm \hbar v_F(p_1 \mp i
p_2)$, so that the Hamiltonian (\ref{H_0}) is
linearized\cite{Semenoff1984PRL,DiVincenzo1984PRB}
\begin{eqnarray}
\label{H-linear} H_0   =  \sum_{\sigma} \int_{DC} \frac{d^2 p}{(2
\pi)^2}&& \left[   \Upsilon_{\sigma}^\dagger (\mathbf{K}_+
+\mathbf{p}) \mathcal{H}_{\mathbf{K}_+} (\mathbf{p})
\Upsilon_{\sigma}
(\mathbf{K}_+ + \mathbf{p}) \right.\\
+ && \left. \Upsilon_{\sigma}^\dagger (\mathbf{K}_- +\mathbf{p})
\mathcal{H}_{\mathbf{K}_-} (\mathbf{p}) \Upsilon_{\sigma}
(\mathbf{K}_- +\mathbf{p})\right], \nonumber
\end{eqnarray}
with the Hamiltonian densities for $\mathbf{K}_\pm$ points
\begin{equation}
\mathcal{H}_{\mathbf{K}_+} = \hbar v_F (\tau_1 p_1 + \tau_2 p_2),
\qquad \mathcal{H}_{\mathbf{K}_-} = \hbar v_F (-\tau_1 p_1 + \tau_2
p_2).
\end{equation}
The integration in Eq.~(\ref{H-linear}) is done over the Dirac cone
(DC) and the energy cutoff which preserves the number of states is
\begin{equation}
\label{cutoff-energy} W = \hbar v_F \sqrt{\frac{\Omega_B}{2\pi}} =
\frac{\hbar v_F}{a}\sqrt{\frac{4\pi}{\sqrt{3}}} = \sqrt{\pi
\sqrt{3}} t \approx 2.33 t,
\end{equation}
where $\Omega_B= (2\pi)^2/S$ is  BZ area.

It is convenient to use a 4-component spinor
$\Psi_\sigma(\mathbf{p})$ made from the two spinors for
$\mathbf{K}_\pm$ points. When the spinors
$\Upsilon_\sigma(\mathbf{K}_{\pm} + \mathbf{p})$ are combined into
one, we exchange the sublattices in the spinor for $\mathbf{K}_{-}$
point,\footnote{This allows to use the same Hamiltonian for both
$\mathbf{K}$ points, i.e. $\mathcal{H}_{\mathbf{K}_{\pm}} = \pm
\hbar v_F (\tau_1 p_1 + \tau_2 p_2)$ and obtain the standard
representation of gamma matrices.} so that
\begin{equation}
\label{4-spinor} \Psi_\sigma(\mathbf{p})= \left(
\begin{array}{c}
\psi_{\mathbf{K}_+,\sigma}(\mathbf{p}) \\
\psi_{\mathbf{K}_-,\sigma}(\mathbf{p}) \\
\end{array}
\right)= \left(
\begin{array}{c}
a_{\sigma} (\mathbf{K}_{+} + \mathbf{p}) \\
b_{\sigma} (\mathbf{K}_{+} + \mathbf{p})\\
b_{\sigma} (\mathbf{K}_{-} + \mathbf{p}) \\
a_{\sigma} (\mathbf{K}_{-} + \mathbf{p})
\end{array}
\right).
\end{equation}
Accordingly, the Hamiltonian (\ref{H-linear}) acquires the form
\begin{eqnarray}
\label{Hamilton-Dirac} &  & \qquad \qquad  \qquad \qquad \qquad
\qquad \quad
K_{+}A \quad  K_{+}B \qquad \quad K_{-}B \quad \quad K_{-}A \nonumber \\
H_0 &= &\hbar v_F \sum_{\sigma} \int_{DC} \frac{d^2 p}{(2 \pi)^2}
\Psi_\sigma^\dagger(\mathbf{p}) \left(
\begin{array}{cccc}
    0 & p_x-i p_y & 0 & 0 \\
    p_x+i p_y & 0 & 0 & 0 \\
    0 & 0 & 0 & -p_x+i p_y \\
    0 & 0 & -p_x-i p_y & 0 \\
\end{array}
\right)
\Psi_\sigma(\mathbf{p}) \nonumber \\
&=& \hbar v_F \sum_{\sigma} \int_{DC} \frac{d^2 p}{(2 \pi)^2}
\Psi_\sigma^\dagger(\mathbf{p})\mathcal{H}_0(\mathbf{p})
\Psi_\sigma(\mathbf{p}), \qquad \mathcal{H}_0(\mathbf{p}) = \alpha^1
p_1 + \alpha^2 p_2.
\end{eqnarray}
Here the $4 \times 4$ matrices $\alpha^{1,2}$ are the first two
$\alpha$ matrices out of the three $\alpha^i$ ($i=1,2,3$) matrices
\begin{equation}
\bm{\alpha} = (\alpha^1, \alpha^2, \alpha^3) = {\tilde \tau}_3
\otimes (\tau_1,\tau_2,\tau_3) = \left(
                                 \begin{array}{cc}
                                   \bm{\tau} & 0 \\
                                   0 & -\bm{\tau} \\
                                 \end{array}
                               \right)
\end{equation}
that anticommute among themselves and with the matrix $\beta$:
\begin{equation}
\beta = {\tilde \tau}_1 \otimes I_2 = \left(
          \begin{array}{cc}
            0 & I_{2} \\
            I_2 & 0 \\
          \end{array}
        \right).
\end{equation}
The irreducible representation of the Dirac algebra in $(2+1)$
dimensions is given by $2 \times 2$ matrices and there are two such
irreducible representations differing in sign. They are both used in
Eq.~(\ref{Hamilton-Dirac}), reflecting the fact that in addition to
two degrees of freedom associated with the $\mathrm{A}$ and
$\mathrm{B}$ sublattices, one should also take into account fermions
at two distinct $\mathbf{K}_{\pm}$ points. These degrees of freedom
are described by $\tilde{\tau}$ matrices that act in the valley
($\mathbf{K}_{\pm}$) subspace and as a result, we use a $4\times 4$
reducible (in $2+1$ dimensions) representation of the Dirac
matrices.

\subsection{Lagrangian of graphene}
\label{sec:Lagrangian}

To finish the mapping of the tight-binding model for graphene into
QED$_{2+1}$, we introduce $\gamma$ matrices and include an external
magnetic field. We choose
\begin{equation}
\label{Weyl} \gamma^0 = \beta, \qquad \bm{\gamma} = \beta
\bm{\alpha} = -i {\tilde \tau}_2 \otimes (\tau_1,\tau_2,\tau_3)=
\left(
          \begin{array}{cc}
            0 & - \bm{\tau} \\
            \bm{\tau} & 0 \\
          \end{array}
        \right)
\end{equation}
and introduce the Dirac conjugated spinor ${\bar
\Psi}_\sigma(\mathbf{p}) = \Psi_\sigma^\dagger(\mathbf{p})\gamma^0$.
The matrices $\gamma^{\nu}$ with $\nu=0,1,2,3$ satisfy the usual
anticommutation relations
\begin{equation}
\left\{\gamma^\mu,\gamma^\nu\right\}=2g^{\mu\nu} I_4, \quad
g^{\mu\nu}=(1,-1,-1,-1)\,, \quad \mu,\nu=0,1,2,3,
\end{equation}
but we remind that in QED$_{2+1}$, we do not use $\gamma^3$ in the
Dirac representation of the Hamiltonian
\begin{eqnarray}
\label{Dirac-Hamiltonian-density} && H_0= \sum_{\sigma} \int_{DC}
\frac{d^2 p}{(2 \pi)^2} {\bar\Psi}_\sigma(\mathbf{p})\mathcal{H}_0^D
(\mathbf{p}) \Psi_\sigma(\mathbf{p}),\nonumber \\ && \mathcal{H}_0^D
(\mathbf{p}) =\gamma^0\mathcal{H}_0 (\mathbf{p})=\hbar v_F (\gamma^1
p_1 + \gamma^2 p_2).
\end{eqnarray}
The representation (\ref{Weyl}) is  called the Weyl or chiral
representation of $\gamma$ matrices and because these matrices are
$4 \times 4$, one can construct the chiral $\gamma^5$ matrix as
\begin{equation}
\gamma^5 = i \gamma^0 \gamma^1 \gamma^2 \gamma^3 = \left(
          \begin{array}{cc}
            I_2 & 0 \\
            0 & -I_2 \\
          \end{array}
        \right)
\end{equation}
which commutes with the $\alpha^i$ and anticommutes with the
$\gamma^\nu$. We note that the chiral representation
Eq.~(\ref{Weyl}) by means of the matrix
\begin{equation}
\label{S2our-old}
S=\frac{1}{\sqrt{2}}\left(\begin{array}{cc}I&\tau_{3}\\
I&-\tau_{3}\end{array}\right), \qquad
S^{-1}=\frac{1}{\sqrt{2}}\left(\begin{array}{cc}I&I\\
\tau_{3}&-\tau_{3}\end{array}\right),
\end{equation}
translates into another representation
\begin{eqnarray}
\label{gamma-old} && \gamma^\nu=\tilde{\tau}_3\otimes
(\tau_3,i\tau_2,-i\tau_1) =\left(\begin{array}{cc}
\hat{\gamma}^0 & 0 \\
0 & \check{\gamma}^0
\end{array}\right),
\left(\begin{array}{cc}
\hat{\gamma}^1 &0 \\
0 & \check{\gamma}^1
\end{array}\right),
\left(\begin{array}{cc}
\hat{\gamma}^2 & 0 \\
0 & \check{\gamma}^2
\end{array}\right),\\
&& \gamma^3 =\left(\begin{array}{cc} 0& I \\ -I &
0\end{array}\right), \qquad \gamma^5=\left(\begin{array}{cc} 0 & I
\\ I & 0\end{array}\right), \nonumber
\end{eqnarray}
which was used in
Refs.~\refcite{Khveshchenko2001PRL,Gusynin2005PRL,Khveshchenko2004nb,Gusynin2006catalysis}.
Note that various researchers used different representations of the
Dirac algebra (see e.g.
Refs.~\refcite{Semenoff1984PRL,Gonzales1993NP,Gorbar2002PRB,Herbut2006PRL}),
so that the comparison of the $U(4)$ symmetry breaking terms in
terms of the underlying lattice and band structure of graphene is
rather complicated. Our choice of the chiral representation of
$\gamma$ matrices in Eq.~(\ref{Weyl}) is motivated by the fact that in this
representation, the spinors with a definite chirality are eigenstates
of $\gamma^5$ which makes the physical interpretation of the
chirality quantum number in graphene in Sec.~\ref{sec:chirality}
more transparent. Another advantage of this representation is that
one can make a comparison with some recent works in
Refs.~\refcite{McCann2004JPCM,Kane2005PRL,Cheianov2006PRL,McCann2006PRL,Ostrovsky2006PRB,Khveshchenko2007PRB},
where the discrete symmetries of graphene were discussed.

One can obtain from Eq.~(\ref{current-param}) the total electric
current
\begin{eqnarray}
\label{current-graphene-QED} \sum_{\mathbf{n}}j_{\alpha}(\mathbf{n})
&= & e \int_{BZ}\frac{d^2\mathbf{k}}{(2\pi)^2}
\sum_{\bm{\delta}_i,\sigma}
 \Upsilon_\sigma^\dagger (\mathbf{k})
\frac{\partial}{\partial k_\alpha} \frac{t}{\hbar} \left(
  \begin{array}{cc}
    0 & e^{i \mathbf{k} \bm{\delta}_i} \\
    e^{-i \mathbf{k} \bm{\delta}_i} & 0 \\
  \end{array}
\right) \Upsilon_\sigma (\mathbf{k})\\
  &\simeq& -e v_F \sum_{\sigma} \int_{DC} \frac{d^2 p}{(2 \pi)^2}
{\bar \Psi}_\sigma(\mathbf{p}) \gamma^\alpha
\Psi_\sigma(\mathbf{p}), \nonumber
\end{eqnarray}
where in the second line, we expanded the matrix near the
$\mathbf{K}_{\pm}$ points and introduced 4-component spinors $
\Psi_\sigma(\mathbf{p})$ and ${\bar \Psi}_\sigma(\mathbf{p})$. One
can recognize that the form of the electric current operator and its
coupling to the vector potential $\mathbf{A}$ in
Eq.~(\ref{current-param}) is standard for QED. Another famous
condensed matter system, a $d$-wave superconductor, is also
described (see e.g.
Refs.~\refcite{Volovik2001PR,Vafek2001PRB,Khveshchenko2001aPRL,Herbut2002PRB,Gusynin2004EPJB,Sharapov2006PRB})
in terms of QED$_{2+1}$, but the coupling to the vector potential is
different due to the presence of supercurrents.

Thus putting together the Hamiltonian,
Eq.~(\ref{Dirac-Hamiltonian-density}), and the interaction term
$\mathbf{A}(\mathbf{n}) \mathbf{j}(\mathbf{n})$ (see
Eq.~(\ref{Hamitonian-expand})), where current density
$\mathbf{j}(\mathbf{n})$ corresponds to the current operator in Eq.~(\ref{current-graphene-QED}),
we arrive at the QED$_{2+1}$
Lagrangian
\begin{equation}
\label{Lagrangian} \mathcal{L} = \sum_{\sigma= \pm 1}
\bar{\Psi}_{\sigma}(t, \mathbf{r})  \left[ i \gamma^0 (\hbar
\partial_t - i \mu_\sigma) + i \hbar v_F \gamma^1 D_x + i \hbar v_F \gamma^2 D_y \right]
\Psi_{\sigma}(t, \mathbf{r})
\end{equation}
with $D_\alpha = \partial_\alpha + i e/\hbar c A_\alpha$, $\alpha
=x,y$ written in the coordinate representation. In what follows we
consider the setup when the external constant magnetic field
$\mathbf{B}= \nabla \times \mathbf{A}$ is applied perpendicular to
the graphene plane along the positive z axis. The magnetic field
also enters the Zeeman term included via $\mu_\sigma$. When the
field $\mathbf{B}$ is not perpendicular to the plane, the total
field still  enter the Zeeman term, while the orbital term
would depend on its perpendicular component.

The theoretical explanation of the basic
experiments\cite{Geim2005Nature,Kim2005Nature} which proved that
Dirac quasiparticles exist in graphene is grounded in the Dirac
Lagrangian (\ref{Lagrangian}). Although the low-energy quasiparticle
excitations described by (\ref{Lagrangian}) are noninteracting, this
Lagrangian captures the Dirac nature of the quasiparticles.
Moreover, for magnetic fields below $10\, \mbox{T}$, there is no need
to include  the Zeeman term\cite{Zhang2006PRL} and the orbital
effect of the magnetic field is sufficient. Nevertheless,
dc\cite{Zhang2006PRL,Abanin2007} and ac\cite{Jiang2007} measurements
in strong fields  $B \gtrsim 10\, \mbox{T}$ already indicate that
the interaction between quasiparticles in graphene plays an
important role. From a theoretical point of view, the effect of the
long range Coulomb interaction had become a research topic long
before the discovery of
graphene.\cite{Gonzales1993NP,Gonzales1996PRL,Khveshchenko2001PRL,Gorbar2002PRB}
The Coulomb interaction Hamiltonian is
\begin{equation}
\label{Hint}
 H_{C}=\frac{\hbar v_F}{2}\sum_{\sigma,\sigma^\prime}\int d^2{\bf r}d^2{\bf
r}^\prime\bar{\Psi}_\sigma({\bf r})\gamma^0 \Psi_\sigma({\bf
r})\frac{g}{|{\bf r}-{\bf r}^\prime|} \bar{\Psi}_{\sigma\prime}({\bf
r}^\prime)\gamma^0 \Psi_{\sigma^\prime}({\bf r}^\prime).
\end{equation}
The coupling constant is $g =e^2/\epsilon \hbar v_F = \alpha
c/\epsilon v_F$, where $\alpha \simeq 1/137$ is the fine-structure
constant. With air on one side of the graphene plane and SiO$_2$ on
the other, the unscreened dielectric constant of the medium is
estimated in Ref.~\refcite{Alicea2006PRB} to be $\epsilon \approx
1.6 \epsilon_0$. The corresponding value of $g \approx 1.37$ is well
above a simplifying  assumption for theoretical analysis  $g \ll 1$
and this explains why the effect of Coulomb interaction in graphene
is a hot topic of ongoing research (see e.g.
Refs.~\refcite{Khveshchenko2006PRB,Mishchenko2006,Barlas2007}).

Besides long range Coulomb interaction, one can also consider other
lattice interactions like on-site repulsion and nearest neighbor
repulsion\cite{Alicea2006PRB,Herbut2006PRL}
\begin{eqnarray}
H_{int} &=& \frac{U}{2} \sum_{\mathbf{n},\sigma,\sigma^\prime}
\Upsilon_{\sigma}^\dagger (\mathbf{n})P_+ \Upsilon_{\sigma}
(\mathbf{n})  \Upsilon_{\sigma^\prime}^\dagger P_+(\mathbf{n})
\Upsilon_{\sigma^\prime}
(\mathbf{n})\\
&+& \frac{U}{2} \sum_{\mathbf{n},\sigma,\sigma^\prime}
\Upsilon_{\sigma}^\dagger (\mathbf{n}+\bm{\delta}_1)P_-
\Upsilon_{\sigma} (\mathbf{n}+\bm{\delta}_1)
\Upsilon_{\sigma^\prime}^\dagger (\mathbf{n}+\bm{\delta}_1)P_-
\Upsilon_{\sigma^\prime}
(\mathbf{n}+\bm{\delta}_1)\nonumber \\
&+& \frac{V}{2}\sum_{\mathbf{n},\bm{\delta}_i,\sigma,\sigma^\prime}
\Upsilon_{\sigma}^\dagger (\mathbf{n}) P_+ \Upsilon_{\sigma}
(\mathbf{n}) \Upsilon_{\sigma^\prime}^\dagger
(\mathbf{n}+\bm{\delta}_i) P_- \Upsilon_{\sigma^\prime}
(\mathbf{n}+\bm{\delta}_i)\nonumber
\end{eqnarray}
with $P_{\pm} = (1\pm \tau_3)/2$. In the low-energy continuum limit, this
leads to several local four-fermion interaction
terms,\cite{Alicea2006PRB,Herbut2006PRL} which in general break
initial $U(4)$ invariance and the generation of corresponding gaps
(see, Sec.~\ref{sec:Dirac-mass}):
\begin{equation}
H_{int}=\sum_i U_i \int d^2{\bf r}
(\bar\Psi(\mathbf{r})\Gamma_i\Psi(\mathbf{r}))^2, \qquad \Gamma_i=
(I,\gamma^3,\gamma^5,\gamma^3\gamma^5)\otimes (\sigma_0,\sigma_3).
\end{equation}

In summary, we would like to emphasize that the statement that the
effective low-energy theory of graphene is massless QED$_{2+1}$ is
based on three nontrivial facts:
\begin{romanlist}[(ii)]

\item Low-energy excitations in graphene are massless quasiparticles with
linear dispersion which have positive and negative branches, $\pm
\hbar v_F |\mathbf{p}|$. It is often underemphasized, that even the
massive quasiparticles (see Sec.~\ref{sec:Dirac-mass} and
Fig.~\ref{fig:3}~(b) below) with the energy $\pm\sqrt{\hbar^2 v_F^2
\mathbf{p}^2 +\Delta^2}$ are still governed by the massive
QED$_{2+1}$ rather than Schr\"{o}dinger theory.

\item
A qualitatively new feature of graphene is that the eigenfunctions
of the low energy quasiparticle excitations obey the Dirac equation.
As we have seen, the spinor structure of the wave functions is a general
consequence of the honeycomb lattice structure of graphene with two
carbon atoms per unit cell.\cite{Semenoff1984PRL,DiVincenzo1984PRB}

\item It is crucial that interaction of the quasiparticles in
graphene with an external electromagnetic field be introduced using
the minimal coupling prescription of quantum field theory.

\end{romanlist}

\subsection{Continuum symmetries of QED$_{2+1}$ model}
\label{sec:continuum-QED}

In the absence of Zeeman splitting, $\mu_\sigma = \mu$ the effective
Lagrangian (\ref{Lagrangian}) possesses  global $U(4)$ symmetry
which is discussed for example in Appendices~A and C in
Refs.~\refcite{Gorbar2002PRB,Gusynin2006catalysis}. This symmetry
operates in the valley-sublattice and spin spaces. It is useful to
ignore, for a moment, spin space and begin by considering a global
$U(2)$ symmetry for the 4-component spinors in the valley-sublattice
space.

\subsubsection{$U(2)$ valley-sublattice symmetry}
\label{sec:U(2)}

One can easily check that matrices
\begin{equation}
T_{1}=\frac{1}{2}i\gamma^{3}= \frac{1}{2}{\tilde
\tau}_{2}\otimes\tau_{3}, \quad
T_{2}=\frac{1}{2}\gamma^{5}=\frac{1}{2}{\tilde\tau}_{3}\otimes\tau_{0},
\quad T_{3}=\frac{1}{2}\gamma^{3}\gamma^{5}=\frac{1}{2}{\tilde
\tau}_{1}\otimes\tau_{3}
\end{equation}
commute with the Hamiltonian (\ref{Hamilton-Dirac}) and satisfy the
commutation relations of $SU(2)$ algebra
\begin{equation}
[T_{i},T_{j}]=i\epsilon_{ijk}T_{k},
\end{equation}
where $\epsilon_{ijk}$ is the Levi-Civita symbol. Together with the
identity matrix ($T_{0}=I_4/2$), they lead to the $U(2)$
symmetry\cite{Appelquist1986PRD,Gorbar2002PRB} which acts in the
space of valley and sublattice indices. We will call this effective
symmetry of the low-energy approximation for the lattice graphene
Hamiltonian as chiral symmetry because it contains the $\gamma^{5}$
matrix and resembles chiral symmetries for massless particles in
high energy physics (see a special Sec.~\ref{sec:chirality} on
chirality below). Because of chiral $SU(2)$ symmetry, there is a
conserving quantum number chirality which is characterized by
eigenvalues of a diagonal generator $T_{2}=\gamma^{5}/2$. Since
there are two eigenvalues $+1/2$ and  $-1/2$ of the diagonal
generator $T_{2}$, the four-component spinor $\Psi_\sigma$ is the
reducible representation of the $SU(2)$ group. The conservation of
chirality number plays an important role and leads to the absence of
backscattering\cite{Ando1998JPSJ} in the presence of impurities that
do not violate chiral symmetry.

We note that the matrix
\begin{equation}
S=\left(\begin{array}{cccc}1&0&0&0\\ 0&0&1&0\\ 0&1&0&0\\
0&0&0&-1\end{array}\right)=S^{T}=S^{-1}
\end{equation}
brings the generators $T_{i}$ to block-diagonal form
\begin{equation}
ST_{1}S^{-1}=\frac{1}{2}\left(\begin{array}{cc}\tau_{2}&0\\
0&\tau_{2}
\end{array}\right), \quad ST_{2}S^{-1}=\frac{1}{2}\left(\begin{array}{cc}\tau_{3}&0\\ 0&\tau_{3}
\end{array}\right),\quad ST_{3}S^{-1}=\frac{1}{2}\left(\begin{array}{cc}\tau_{1}&0\\ 0&\tau_{1}
\end{array}\right).
\end{equation}
In this reducible representation, the spinor $\Psi_\sigma$ acquires
the form
\begin{equation}\label{Psi_irreducible}
\Psi^S_\sigma=S\Psi_\sigma= \left(
\begin{array}{c}
a_{\sigma} (\mathbf{K}_{+} + \mathbf{p}) \\
b_{\sigma} (\mathbf{K}_{-} + \mathbf{p})\\
b_{\sigma} (\mathbf{K}_{+} + \mathbf{p}) \\
-a_{\sigma} (\mathbf{K}_{-} + \mathbf{p})
\end{array}
\right),
\end{equation}
where the two upper and two lower components correspond to the two
irreducible representations of $SU(2)$ group.

\subsubsection{$U(4)$ spin-valley-sublattice symmetry and its breaking by the Zeeman term}
\label{sec:U(4)}

We now generalize the previous section and include rotations in spin
space. The 16 generators of the $U(4)$ that operate in the spin and
valley-sublattice space are
\begin{equation}
\frac{\sigma_\kappa}{2}\otimes I_4,\quad
\frac{\sigma_\kappa}{2}\otimes i \gamma^3,\quad
\frac{\sigma_\kappa}{2}\otimes\gamma^5,\quad\mbox{and}\quad
\frac{\sigma_\kappa}{2}\otimes\frac{1}{2}[\gamma^3,\gamma^5],
\end{equation}
where $I_4$ is the $4 \times 4$  unit matrix and $\sigma_\kappa$,
with $\kappa=0, 1, 2, 3$ being four Pauli matrices connected with
spin degrees of freedom [$\sigma_0$ is the $2 \times 2$ unit
matrix]. It is easy to see that when $B=0$ and there is no Zeeman
splitting, the Lagrangian (\ref{Lagrangian}) and the interaction
term (\ref{Hint}) are invariant under global $U(4)$ group generated
by these 16 generators.

The simplest example of symmetry breaking is provided by the Zeeman
term, $\Psi^\dagger_\sigma \sigma_3 \Psi = \bar{\Psi}\gamma^0
\sigma_3 \Psi$, where $\sigma_3$ acts on the spin indices $\sigma$
of the Dirac spinors. It explicitly breaks the $U(4)$ down to the
$U(2)_a \times U(2)_b$ with the generators
\begin{equation}
\label{ab} \frac{\sigma_{\kappa^{\prime}}}{2}\otimes I_4,\quad
\frac{\sigma_{\kappa^{\prime}}}{2}\otimes i \gamma^3,\quad
\frac{\sigma_{\kappa^{\prime}}}{2} \otimes
\gamma^5,\quad\mbox{and}\quad
\frac{\sigma_{\kappa^{\prime}}}{2}\otimes
\frac{1}{2}[\gamma^3,\gamma^5],
\end{equation}
where $\kappa^{\prime} = 0, 3$. We will discuss other possible
symmetry breaking terms in Sec.~\ref{sec:Dirac-mass} below.

\subsection{Chirality in QED$_{2+1}$ theory of graphene and its difference
from the chirality in QED$_{3+1}$ }
\label{sec:chirality}

Because in the massless Dirac theory $\gamma^5$ commutes with the
Hamiltonian (\ref{Hamilton-Dirac}) (anticommutes with $H_0^D$ given
by (\ref{Dirac-Hamiltonian-density})), it introduces the conserving
{\em chirality\/} quantum number which corresponds to the valley
index. Indeed the spinors
\begin{equation}
\label{chiral-spinors} \Psi_{\mathbf{K}_+} = \left(
                        \begin{array}{c}
                          \psi_{\mathbf{K}_+} \\
                          0 \\
                        \end{array}
                      \right), \qquad
                      \Psi_{\mathbf{K}_-} = \left(
                        \begin{array}{c}
                          0 \\
                           \psi_{\mathbf{K}_-} \\
                        \end{array}
                      \right)
\end{equation}
that describe the quasiparticle excitations at $\mathbf{K}_\pm$
points, respectively, are the eigenstates of $\gamma^5$:
\begin{equation}
\gamma^5 \Psi_{\mathbf{K}_+} = \Psi_{\mathbf{K}_+}, \qquad \gamma^5
\Psi_{\mathbf{K}_-} = - \Psi_{\mathbf{K}_-}.
\end{equation}
The labeling of $\mathbf{K}_{\pm}$ points by eigenstates of the
chiral operator $\gamma^5$ is an advantage of the chiral
representation (\ref{Weyl}) of the $\gamma$ matrices. This is not
the case in the often used basis (\ref{gamma-old}) with diagonal
$\gamma^0$.

For massless particles in $3+1$ dimensions, the chirality quantum
number corresponds to the {\em helicity\/} which characterizes the
projection of its spin on the direction of momentum. In the $2+1$
dimensional case, the usual helicity concept for massless particles
is meaningless and one may only talk about pseudohelicity.

Let us now illustrate  this by considering the Dirac equation which
follows from the Lagrangian (\ref{Lagrangian}) in the simplest $B =
\mu= 0$ case, also dropping the spin index. We consider its positive
and negative energy solutions with a definite chirality
\begin{equation}
\label{spinors-e-h} \Psi_{\mathbf{K}_{\pm}}^e(t, \mathbf{r}) = e^{-i
\frac{E t}{\hbar} + i \mathbf{r} \mathbf{p}}
U^e_{\mathbf{K}_{\pm}}(E,\mathbf{p}), \quad
\Psi_{\mathbf{K}_{\pm}}^h(t, \mathbf{r}) = e^{i \frac{E t}{\hbar} +
i \mathbf{r} \mathbf{p}} U^h_{\mathbf{K}_{\pm}}(E,\mathbf{p})
\end{equation}
with $E = \hbar v_F |\mathbf{p}|$ which correspond to the electrons
and holes from $\mathbf{K}_{\pm}$ valleys, respectively.
Substituting (\ref{spinors-e-h}) in the Dirac equation, we obtain
that the spinors
\begin{equation}
U^{e,h}_{\mathbf{K}_{+}}(E,\mathbf{p}) = \left(
                        \begin{array}{c}
                          \psi^{e,h}_{\mathbf{K}_+}(E,\mathbf{p}) \\
                          0 \\
                        \end{array}
                      \right), \qquad
U^{e,h}_{\mathbf{K}_{-}}(E,\mathbf{p}) = \left(
                        \begin{array}{c}
                          0 \\
                           \psi^{e,h}_{\mathbf{K}_-}(E,\mathbf{p}) \\
                        \end{array}
                      \right)
\end{equation}
satisfy Weyl equations, $\mathcal{H}_0(\mathbf{p})
U^{e,h}(\mathbf{p}) = \pm E U^{e,h}(\mathbf{p})$ or for 2-component
spinors,
\begin{eqnarray}
\label{2D-Weyl}&& \hbar v_F (\tau_1 p_1 + \tau_2 p_2)
\psi^{e,h}_{\mathbf{K}_+} = \pm E \psi^{e,h}_{\mathbf{K}_+},
\nonumber\\ -&&\hbar v_F(\tau_1 p_1 + \tau_2 p_2)
\psi^{e,h}_{\mathbf{K}_-} = \pm E \psi^{e,h}_{\mathbf{K}_-},
\end{eqnarray}
where the upper sign corresponds to the electrons and lower to the
holes, respectively. We stress that in our case the vector
$\mathbf{p} = (p_1,p_2)$ is in the graphene plane. Formally,
Eqs.~(\ref{2D-Weyl}) look similar to the Dirac-Weyl equations that
describe massless neutrinos,\cite{Itzykson.book}
\begin{equation}
\label{Dirac-3D} \hbar c \mathbf{p}\bm{\alpha} \Psi =\hbar c \left(
          \begin{array}{cc}
            \mathbf{p}\bm{\sigma} & 0 \\
            0 & -\mathbf{p}\bm{\sigma} \\
          \end{array}
        \right)\Psi=
 p_0 \Psi,
\end{equation}
but in the latter case, the space is 3D. This allows one to introduce,
for a massless particle in $3+1$ dimension, the {\em helicity\/}
operator
\begin{equation}
\label{helicity-3D} \Lambda = \frac{\mathbf{p}\bm{\Sigma}}{|
\mathbf{p}|}, \qquad \bm{\Sigma} = \left(
                                     \begin{array}{cc}
                                       \bm{\sigma} & 0 \\
                                       0 & \bm{\sigma} \\
                                     \end{array}
                                   \right)
\end{equation}
which commutes with the Dirac Hamiltonian $\hbar c \mathbf{p}
\bm{\alpha}$ and characterizes the projection of particle spin on
the direction of its momentum. Multiplying  Eq.~(\ref{Dirac-3D}) by
$\gamma^5$ and taking into account that massless particles
(antiparticles) have the dispersion $p_0 = \pm \hbar c
|\mathbf{p}|$, one obtains that $\gamma^5 \Psi = \pm \mathbf{p}
\bm{\Sigma}/|\mathbf{p}| \Psi$. This illustrates that, for massless
particles, the helicity coincides with the chirality, while for
antiparticles it has the opposite sign to the chirality.

Let us consider the solutions of  Eq.~(\ref{2D-Weyl}) for the
$\mathbf{K}_{+}$ point
\begin{equation}
\label{+Weyl-solutions} \psi^{e}_{\mathbf{K}_+}(E,\mathbf{p}) =
\frac{1}{\sqrt{2}}\left(
                                       \begin{array}{c}
                                         1 \\
                                          \frac{p_x+ip_y}{|\mathbf{p}|}\\
                                       \end{array}
                                     \right),
                                     \qquad
\psi^{h}_{\mathbf{K}_+}(E,\mathbf{p}) = \frac{1}{\sqrt{2}}\left(
                                       \begin{array}{c}
                                          \frac{-p_x+ip_y}{|\mathbf{p}|}\\
                                          1 \\
                                       \end{array}
                                     \right),
\end{equation}
and $\mathbf{K}_{-}$ point
\begin{equation}
\label{-Weyl-solutions} \psi^{e}_{\mathbf{K}_-}(E,\mathbf{p}) =
\frac{1}{\sqrt{2}}\left(
                                       \begin{array}{c}
                                         1 \\
                                          -\frac{p_x+ip_y}{|\mathbf{p}|}\\
                                       \end{array}
                                     \right),
                                     \qquad
\psi^{h}_{\mathbf{K}_-}(E,\mathbf{p}) = \frac{1}{\sqrt{2}}\left(
                                       \begin{array}{c}
                                         \frac{p_x-ip_y}{|\mathbf{p}|}\\
                                          1 \\
                                       \end{array}
                                     \right).
\end{equation}
The 4-component spinors (\ref{chiral-spinors}) made from the
solutions (\ref{+Weyl-solutions}), (\ref{-Weyl-solutions}) are
mutually orthogonal eigenstates of the Hamiltonian
$\mathcal{H}_0(\mathbf{p})$ and the operator $\gamma^5$:
\begin{equation}\label{orthogonaliy-relations}
\Psi^{i*}_{r}(\mathbf{p})\Psi^{j}_{s}(\mathbf{p})=\delta^{ij}\delta_{rs},\qquad
i,j=e,h,\qquad r,s=\mathbf{K}_+, \mathbf{K}_-.
\end{equation}
In the $2+1$ dimensional case  one cannot make rotations around the
direction of the quasiparticle momentum $\mathbf{p}$ lying in the 2D
plane and the operator $\bm{\tau} = (\tau_1,\tau_2)$ does not have
the physical meaning of the usual spin operator. Therefore the
concept of the helicity for massless particles related to the
Lorentz group and the real space rotations  is meaningless in this
case.\cite{Binegar1981JMP} Furthermore, since there is only one
generator of angular momentum $\tau_3$, there is no non-abelian Lie
algebra that can restrict its possible eigenvalues, giving rise to
the possibility of exotic statistics (see
Ref.~\refcite{Boyanovsky1986NPB} and references therein). The
solutions (\ref{+Weyl-solutions}) and (\ref{-Weyl-solutions}) are
double-valued under the rotations $R(\theta) = \exp(i \theta \tau_3
/2 )$ with $0< \theta <4 \pi$, because they describe  spinor
fields.\cite{Binegar1981JMP}

Formally one can consider the pseudohelicity operator
\begin{equation}
\label{helicity-2D} \Lambda^{\mathrm{2D}} =
\frac{\mathbf{p}\bm{\Sigma}}{| \mathbf{p}|}, \qquad \bm{\Sigma} =
\left(
                                     \begin{array}{cc}
                                       \bm{\tau} & 0 \\
                                       0 & \bm{\tau} \\
                                     \end{array}
                                   \right), \qquad \bm{\tau} = (\tau_1,\tau_2)
\end{equation}
which commutes with the Hamiltonian and thus corresponds to a
conserving quantum number. This operator, however, would correspond
to an internal symmetry rather than to spatial symmetry as in the
$3+1$ dimensional case. Still one can relate chirality and
pseudohelicity operators for massless quasiparticles, $\gamma^5 \Psi
= \pm \Lambda^{\mathrm{2D}} \Psi$.

One can also see that the solutions in Eq.~(\ref{+Weyl-solutions}) and
Eq.~(\ref{-Weyl-solutions}) which obey
Eq.~(\ref{orthogonaliy-relations}) differ only by the sign of
momentum. Thus we arrive at the
conclusion\cite{McEuen1999PRL,McCann2006PRL,Cheianov2006PRL} that at
the $\mathbf{K}_{+}$ point, electronic excitations have energy
$\hbar v_F |\mathbf{p}|$ and $\mathbf{p} \bm{\tau}/|\mathbf{p}|
\psi_{\mathbf{K}_+}^e = \psi_{\mathbf{K}_+}^e $, while for holes the
energy is $-\hbar v_F |\mathbf{p}|$, i.e. $\mathbf{p}
\bm{\tau}/|\mathbf{p}|\psi_{\mathbf{K}_+}^h = -\psi_{\mathbf{K}_+}^h
$. For the $\mathbf{K}_{-}$ point, these relations are inverted: for
electrons, $\mathbf{p} \bm{\tau}/|\mathbf{p}|\psi_{\mathbf{K}_-}^e =
-\psi_{\mathbf{K}_-}^e $ and for holes, $\mathbf{p}
\bm{\tau}/|\mathbf{p}|\psi_{\mathbf{K}_-}^h = \psi_{\mathbf{K}_-}^h
$. For 4-component spinors $\Psi_{\mathbf{K}_\pm}^{e,h}$ these
conditions can be rewritten using the pseudohelicity operator:
$\Lambda^{\mathrm{2D}} \Psi^{e}_{\mathbf{K}_\pm} = \pm
\Psi^{e}_{\mathbf{K}_\pm}$ and $\Lambda^{\mathrm{2D}}
\Psi^{h}_{\mathbf{K}_\pm} = \mp \Psi^{h}_{\mathbf{K}_\pm}$. They
imply that at any given $\mathbf{K}_{\pm}$ point, the direction of the
momentum for electrons and holes with the same absolute value of the
energy is opposite. This property of the massless electrons and
holes is the consequence of equations of motion. However, to forbid
backscattering of the quasiparticles\cite{Ando1998JPSJ} one should
also suppress the transfer of quasiparticles from one valley to
another. This restriction is already associated with chirality
conservation. For example, chirality is a good quantum number in a
monolayer with electrostatic potential scattering.


We were able to define the chiral matrix $\gamma^5$ and the
chirality quantum number, because we used the reducible $4\times 4$
representation of the Dirac matrices. In $2+1$ dimensions, there are
two inequivalent irreducible $2 \times2$ representations of the
Dirac algebra.\cite{Jackiw1981PRD}  The corresponding $2\times2$
${\tilde \gamma}$ matrices are characterized by their
signature\cite{Hosotani1993PLB}
\begin{equation}
\label{signature} \eta = \frac{i}{2} \mbox{Tr} [{\tilde \gamma}^0
{\tilde \gamma}^1 {\tilde \gamma}^2].
\end{equation}
One can check that $\hat{\gamma}^\nu$ and $\check{\gamma}^\nu$
matrices that constitute the $4 \times 4$ representation
(\ref{gamma-old}) have the opposite signatures $\eta = +1$ and $\eta
= -1$, respectively. It turns out that taking into account two
inequivalent $\mathbf{K}_{\pm}$ points demands using two unitary
inequivalent representations of $2\times2$ gamma matrices. As we saw
in the chiral representation of $4 \times 4$ $\gamma$ matrices, the
$\mathbf{K}_{\pm}$ points are distinguished by the chirality
(valley) quantum number. The signature $\eta$ of the corresponding
$2\times 2$ $\gamma$ matrices is sometimes known as
``chirality''.\cite{Hosotani1993PLB} We note that by itself, this
sign is not observable and one should introduce either the Dirac
mass or/and the magnetic field and consider their relative signs
(see Sec.~\ref{sec:Dirac-mass} below).

Finally, we note that in accordance with Eq.~(\ref{tau_3}) the matrix
$\alpha^3$ anticommutes with the Hamiltonian (\ref{Hamilton-Dirac}):
$\{\alpha^3,H_0\}=0$. The consequence of this is that the spectrum
of the Hamiltonian is symmetric with respect to the eigenvalue
$E=0$. The identity $\alpha^2 H_0^\ast(\mathbf{k})\alpha^2 =
-H_0(\mathbf{k})$ corresponds to Eq.~(\ref{tau_2}) and guarantees
the symmetry of the spectrum with respect to $E=0$ even in the
presence of a finite Dirac mass $\bar \Psi \gamma^3 \Psi$. Also the
matrix $\gamma^{0}$ anticommutes with the Hamiltonians
(\ref{Hamilton-Dirac}) and (\ref{Dirac-Hamiltonian-density}):
$\{\gamma^{0},H_0\}=\{\gamma^{0},H_0^D\}=0$. In
Refs.~\refcite{Hatsugai2007,Ziegler2007} the property of
anticommutativeness of $H_0$ with $\alpha^3$ matrix was called
chiral symmetry, due to its analogy with anticommutativeness of
$\gamma^{5}$ with the Dirac Lagrangian in $3+1$ dimensional
case.\cite{Altland2002PR,Garcia2006PRB} Also sometimes  the sign of
the energy, $\varepsilon = \lambda v_F |\mathbf{k}|$, where $\lambda
= 1$ corresponds to the conduction band and $\lambda = -1$
corresponds to the valence band is called the ``chirality label''.
We, however, keep the use of the word ``chiral'' for the
characteristics which really involve the $\gamma^{5}$ matrix.

\subsection{Discrete symmetries of QED$_{2+1}$ model of graphene}
\label{sec:discrete-QED}

In the continuum description, the definitions of the discrete
symmetry operations are not unique. In  field theoretical studies of
QED$_{2+1}$, there is a certain convention on (see e.g.
Refs.~\refcite{Jackiw2007,Jackiw1981PRD}) how to define these
operations. It is based on the assertion that the parity
transformation corresponds to inverting only one axis, say $x$ axis:
$\mathcal{P}(x,y) \to (-x,y)$, because inverting both would be a
rotation.  Bearing in mind the condensed matter roots of the
effective QED$_{2+1}$ model,  the discrete symmetry operations
introduced here are in accord with the operations defined in
Sec.~\ref{sec:discrete-lattice} for the tight-binding model.
Knowledge of these symmetries allows us to properly classify
possible symmetry breaking terms, avoiding the trap vividly described
in Ref.~\refcite{Tchernyshyov2000PRB}.

\subsubsection{The spatial inversion $\mathcal{P}$}
\label{sec:inversion-QED}

As discussed in Sec.~\ref{sec:inversion-lattice}  spatial inversion
should invert both axes and exchange both $\mathrm{A}$ and
$\mathrm{B}$ atoms and $\mathbf{K}_{\pm}$ points. Thus applying the
definition (\ref{inversion-lattice}) and
(\ref{inversion-lattice-spinor}) for the 4-component spinor
$\Psi_\sigma(\mathbf{p})$ given by Eq.~(\ref{4-spinor}), we
define\footnote{The $\psi_{\mathbf{K}_{\pm}}$ components of the
spinor (\ref{4-spinor}) are exchanged by the inversion and because
the sublattices are already exchanged by the definition of
$\psi_{\mathbf{K}_{-}}$, we have $\tau_0$ acting in the sublattice
space instead of $\tau_1$ used in
Eq.~(\ref{inversion-lattice-spinor}). This definition of
$\mathcal{P}$ coincides with the corresponding operator in the
second paper in Ref.~\refcite{Cheianov2006PRL}.}
\begin{equation}
\label{inversion-QED-spinor} \Psi_{\sigma}(\mathbf{p})
\longrightarrow \mathcal{P} \Psi_{\sigma}(\mathbf{p})
\mathcal{P}^{-1} =P \Psi_{\sigma}(-\mathbf{p}), \qquad P = {\tilde
\tau}_1\otimes\tau_0=
 \gamma^0, \qquad P^2 =1.
\end{equation}
Note that this operation of inversion of two spatial coordinates is
not equivalent to a rotation through an angle $\pi$ in the plane
which is given by $\Psi(k_x,k_y)\to
\exp(i\pi(i\gamma^1\gamma^2/2))\Psi(-k_x,-k_y)$ with
$i\gamma^1\gamma^2/2$ being the generator of the corresponding
rotation.

The invariance (\ref{H0-inversion-invariant}) of $H_0$ under
$\mathcal{P}$ now follows from the condition
\begin{equation}
\label{inversion-H0-QED} P \mathcal{H}_0^D(\mathbf{p}) P =
\mathcal{H}_0^D(-\mathbf{p})
\end{equation}
on the Dirac Hamiltonian density (\ref{Dirac-Hamiltonian-density}).
Eq.~(\ref{inversion-H0-QED}) generalizes the condition
(\ref{inversion-H0-lattice}) on the $2\times 2$ Hamiltonian density.
It is easy to see that under this $\mathcal{P}$ transformation, the
current $\bar{\Psi}\sigma_0\gamma^\mu\Psi$ from
Eq.~(\ref{current-graphene-QED}) transforms as follows
\begin{equation}
\bar{\Psi} \sigma_0 \gamma^0\Psi \stackrel{\mathcal{P}}{\rightarrow}
\bar{\Psi} \sigma_0\gamma^0\Psi,\quad \bar{\Psi}
\sigma_0\gamma^1\Psi \stackrel{\mathcal{P}}{\rightarrow}
-\bar{\Psi}\sigma_0\gamma^1\Psi,\quad \bar{\Psi}\sigma_0\gamma^2\Psi
\stackrel{\mathcal{P}}{\rightarrow} -\bar{\Psi}\sigma_0\gamma^2\Psi,
\end{equation}
where the unit matrix $\sigma_0$ acts on the spin indices $\sigma$
of the Dirac spinors. The interaction with  fixed external magnetic
field $e A_\alpha \bar{\Psi}\sigma_0\gamma^\alpha\Psi $ is invariant
under spatial inversion.

\subsubsection{Time reversal $\mathcal{T}$}
\label{sec:time-QED}

The time reversal operator, (\ref{time-lattice-spinor}) interchanges
$\mathbf{K}_{\pm}$ points but not sublattices. Accordingly when we
define the action $\mathcal{T}$ on the 4-component spinors
$\Psi_{\sigma}(\mathbf{p})$, the part acting on the sublattice
degree of freedom is\cite{McCann2006PRL,Cheianov2006PRL}
$\tau_1$\footnote{Compare with the definition of $\mathcal{P}$ above
which includes $\tau_0$.}
\begin{equation}
\label{time-QED-spinor} \Psi(\mathbf{p}) \longrightarrow \mathcal{T}
\Psi(\mathbf{p}) \mathcal{T}^{-1}  = i \sigma_2 T \Psi(-\mathbf{p}),
\qquad T ={\tilde \tau}_1\otimes\tau_1 = \gamma^1 \gamma^5, \qquad
T^2=1,
\end{equation}
where $\sigma_2$ acts on the spin indices of the spinor
$\Psi_\sigma(\mathbf{p})$. Double $\mathcal{T}$-transformation gives
$\mathcal{ T}^{2}\Psi(\mathbf{p})\mathcal{T}^{-2} = -
\Psi(\mathbf{p})$. Hence for time-reversal symmetric systems with
odd number of fermions, the degeneracy of levels cannot be less than
$2$ (Kramers' degeneracy theorem). It is easy to see that
$\mathbf{K}_{\pm}$ points are indeed related by time-reversal
symmetry
\begin{equation}
\label{time-QED-illustration} \Psi(\mathbf{p}) = \left(
\begin{array}{c}
a (\mathbf{K}_{+} + \mathbf{p}) \\
b (\mathbf{K}_{+} + \mathbf{p})\\
b (\mathbf{K}_{-} + \mathbf{p}) \\
a (\mathbf{K}_{-} + \mathbf{p})
\end{array}
\right) \stackrel{\mathcal{T}}{\rightarrow} T  \Psi(-\mathbf{p}) =
\left(
\begin{array}{c}
a (\mathbf{K}_{-} - \mathbf{p}) \\
b (\mathbf{K}_{-} - \mathbf{p})\\
b (\mathbf{K}_{+} - \mathbf{p}) \\
a (\mathbf{K}_{+} - \mathbf{p})
\end{array}
\right),
\end{equation}
where for simplicity, we ignored spin degrees of freedom.

In the time-coordinate representation, the transformation
(\ref{time-QED-spinor}) and the corresponding transformations for
the Dirac conjugated spinor are
\begin{eqnarray}
\label{time-QED-conjspinor} \Psi(t,\mathbf{r}) &\rightarrow&
\mathcal{T} \Psi(t,\mathbf{r}) \mathcal{T}^{-1} =i \sigma_2 T
\Psi(-t,\mathbf{r}), \\
{\bar \Psi}(t,\mathbf{r}) &\rightarrow& (\mathcal{T}
\Psi(t,\mathbf{r}) \mathcal{T}^{-1})^\dagger (\gamma^0)^* = - {\bar
\Psi}(-t,\mathbf{r}) i \sigma_2 T. \nonumber
\end{eqnarray}

The invariance (\ref{H0-time-invariant}) of $H_0$  under
$\mathcal{T}$ now follows from the condition
\begin{equation}
\label{time-H0-QED} T \mathcal{H}_0^{D\ast}(\mathbf{p}) T =
\mathcal{H}_0^D(-\mathbf{p}).
\end{equation}
on the Dirac Hamiltonian density (\ref{Dirac-Hamiltonian-density}).
Eq.~(\ref{time-H0-QED}) generalizes the condition
(\ref{time-H0-lattice}) on the $2\times 2$  Hamiltonian density.

It is easy to see that under this $\mathcal{T}$ transformation, the
current $\bar{\Psi}\sigma_0\gamma^\mu\Psi$ from
Eq.~(\ref{current-graphene-QED}) transforms as follows
\begin{equation}
\bar{\Psi} \sigma_0\gamma^0\Psi \stackrel{\mathcal{T}}{\rightarrow}
\bar{\Psi} \sigma_0\gamma^0\Psi,\quad \bar{\Psi}\sigma_0\gamma^1\Psi
\stackrel{\mathcal{T}}{\rightarrow}
-\bar{\Psi}\sigma_0\gamma^1\Psi,\quad \bar{\Psi}\sigma_0\gamma^2\Psi
\stackrel{\mathcal{T}}{\rightarrow}-\bar{\Psi}\sigma_0\gamma^2\Psi.
\end{equation}
When the external magnetic field is held fixed, the corresponding
interaction term $\bar{\Psi}\sigma_0\gamma^\alpha\Psi $ now breaks
the time-reversal symmetry.

\subsubsection{Charge conjugation $\mathcal{C}$}
\label{sec:charge-QED}

The charge conjugation $\mathcal{C}$ acquires nontrivial meaning
when we consider the second quantized Dirac theory and introduce a
transformation that exchanges particles and antiparticles (electrons
and holes), leaving their spin and momentum unchanged.

Accordingly we define the action of $\mathcal{C}$ on 4-component
spinors $\Psi_{\sigma}(\mathbf{p})$:
\begin{equation}
\label{charge-QED-spinor} \Psi_{\sigma}(\mathbf{p}) \longrightarrow
\mathcal{C} \Psi_{\sigma}(\mathbf{p}) \mathcal{C}^{-1}  = C
{\bar\Psi}_{\sigma}^T(\mathbf{p}), \qquad C= \gamma^1,
\end{equation}
where $T$ denotes the transpose. The matrix $C$ satisfies the
identities\footnote{Notice that our definition of $C = \gamma^1$
coincides with Ref.~\refcite{Jackiw2007} and that
$C(\gamma^3)^TC^{-1}=\gamma^3$.}
\begin{equation}
\label{C} C= -C^{-1} = -C^\dagger = -C^T, \qquad
 C(\gamma^\mu)^TC^{-1}=-\gamma^\mu, \qquad \mu=0,1,2.
\end{equation}
It is easy to see that $\mathcal{C}$ exchanges $\mathrm{A}$ and
$\mathrm{B}$ sublattices
\begin{equation}
\Psi_{\sigma}(\mathbf{p}) = \left(
\begin{array}{c}
a_{\sigma} (\mathbf{K}_{+} + \mathbf{p}) \\
b_{\sigma} (\mathbf{K}_{+} + \mathbf{p})\\
b_{\sigma} (\mathbf{K}_{-} + \mathbf{p}) \\
a_{\sigma} (\mathbf{K}_{-} + \mathbf{p})
\end{array}
\right) \stackrel{\mathcal{C}}{\rightarrow} \gamma^1
\gamma^0\Psi_{\sigma}^{\dagger T}(\mathbf{p}) = \left(
\begin{array}{c}
-b_{\sigma}^\dagger (\mathbf{K}_{+} + \mathbf{p}) \\
-a_{\sigma}^\dagger (\mathbf{K}_{+} + \mathbf{p})\\
a_{\sigma}^\dagger (\mathbf{K}_{-} + \mathbf{p}) \\
b_{\sigma}^\dagger (\mathbf{K}_{-} + \mathbf{p})
\end{array}
\right).
\end{equation}
For the Dirac conjugated spinor we find
\begin{equation}
\label{charge-QED-conjspinor} {\bar \Psi}_{\sigma}(\mathbf{p})
\rightarrow \mathcal{C} \Psi_{\sigma}^\dagger(\mathbf{p})
\mathcal{C}^{-1} \gamma^0 = (\gamma^1 \gamma^0
\Psi_{\sigma}(\mathbf{p}))^T \gamma^0 = (-\gamma^1
\Psi_{\sigma}(\mathbf{p}))^T.
\end{equation}

It is easy to see that under  $\mathcal{C}$ transformation, the
current $\bar{\Psi}\sigma_0\gamma^\mu\Psi$ from
Eq.~(\ref{current-graphene-QED}) transforms as follows
\begin{equation}
\bar{\Psi}\sigma_0\gamma^\mu\Psi \stackrel{\mathcal{C}}{\rightarrow}
-\bar{\Psi}\sigma_0\gamma^\mu\Psi, \qquad \mu =0,1,2.
\end{equation}
One can also check that the Lagrangian (\ref{Lagrangian}) is
invariant under $\mathcal{C}$ if one simultaneously replaces $e \to
-e$.

\subsubsection{Difference between quasiparticles in graphene and massless neutrinos}
\label{sec:discrete-difference}

There is a widespread analogy between quasiparticles in graphene and
massless neutrinos which is based on the similarity of the
Dirac-Weyl equations (\ref{Dirac-3D}) and Eqs.~(\ref{2D-Weyl}) for
quasiparticles in graphene.\cite{Ando2005JPSJ} Our consideration of
the discrete $\mathcal{P}$, $\mathcal{T}$ and $\mathcal{C}$
symmetries in Secs.~\ref{sec:inversion-QED}, \ref{sec:time-QED} and
\ref{sec:charge-QED} shows that this analogy is not complete. The
Dirac equation (\ref{Dirac-3D}) corresponds to a pair of Weyl
equations for 2-component left-handed, $\Psi_L = (1/2)(1-\gamma^5)
\Psi$ and right-handed spinors $\Psi_R = (1/2)(1+\gamma^5) \Psi$.
These two equations are related to each other by parity
transformation. It is assumed in particle physics that in nature,
there exist only left-handed neutrinos, so that it is sufficient to
use one Weyl equation. However, this single equation breaks
$\mathcal{C}$ and $\mathcal{P}$ symmetries, but preserves
$\mathcal{C} \mathcal{P}$ and $\mathcal{T}$
symmetries.\cite{Itzykson.book} The Dirac quasiparticles in graphene
with a definite chirality are also described by one of the Weyl
equations (\ref{2D-Weyl}) for the 2-component spinor. The last
equation, however, breaks $\mathcal{P}$ and $\mathcal{T}$
symmetries, but preserves $\mathcal{P} \mathcal{T}$ and
$\mathcal{C}$ symmetries. This conclusion about the difference of
the discrete symmetries in QED$_{3+1}$ and graphene could not be
made by a formal comparison of Eqs.~(\ref{Dirac-3D}) and
(\ref{2D-Weyl}), because one has to project the discrete symmetries
of the lattice model from Sec.~\ref{sec:discrete-lattice} into the
continuum description of graphene. We also note that the fact that
the $\mathbf{K}_{\pm}$ points are related by time-reversal symmetry
(see Eq.~(\ref{time-QED-illustration})) and this has an important
implication for the Josephson effect in mesoscopic junctions
consisting of a graphene layer contacted by two closely spaced
superconducting electrodes.\cite{Heersche2007Nature} As Cooper pairs
are made up of time reversed electron states, the two electrons in
Cooper pairs that are injected from the superconducting electrodes
into graphene go to opposite $\mathbf{K}_{\pm}$
points.\cite{Titov2006PRB} Finally, we note that the consideration
of discrete symmetries allows one to make an analogy between the
$\mathbf{K}_{\pm}$ index and the circular polarization quantum
number for photons. Indeed, photons with a definite circular
polarization also break parity and time-reversal symmetries, but not
charge conjugation.

\subsection{Dirac masses, their transformation under
$\mathcal{P}$, $\mathcal{T}$, $\mathcal{C}$ and physical meaning}
\label{sec:Dirac-mass}

One of the main reason why graphene is attracting the attention of
many
theoreticians\cite{Khveshchenko2001PRL,Gorbar2002PRB,Yang2007,Gusynin2006catalysis,Herbut2006PRL,Abanin2006PRL,Alicea2006PRB,Nomura2006PRL,Yang2006PRB,Goerbig2006PRB,Fuchs2007PRB,Ezawa2006,Herbut2007,Sheng2007}
is that the electron-electron interactions and/or the magnetic field
result in the breakdown of the $U(4)$ symmetry and in new physics
whose understanding demands going beyond the unconventional yet
rather simple physics of noninteracting Dirac quasiparticles.

For example, it is argued that the new zero filling factor state in
the IQHE in graphene, seen in a strong magnetic field $B \gtrsim
20\,\mbox{T}$ is related to a spin polarized
state\cite{Zhang2006PRL,Abanin2007,Kim2007APS}, while $\nu=\pm 1$
states\cite{Zhang2006PRL,Kim2007APS} are related to the lifting of
sublattice or valley degeneracy. The spin polarized state is
described by the order parameter $\langle {\bar\Psi} \gamma^0
\sigma_3 \Psi \rangle$ (or more general $\langle {\bar \Psi}
\gamma^0 {\bm\sigma} \Psi \rangle$) which makes it similar to the
Zeeman term, but in contrast, it originates from many-body
interactions.

Another channel of spontaneous breaking of the $U(4)$ symmetry  down to
$U(2)_c \times U(2)_d$, is related to a possibility of the
generation by the interactions of the Dirac mass $M_D$ which enters
the Lagrangian (\ref{Lagrangian}) in the following way
\begin{eqnarray}
\label{Lagrangian-M} \mathcal{L} = \bar{\Psi}(t, \mathbf{r}) &&
\left[ i \gamma^0 \left(\hbar \sigma_0
\partial_t - i \sigma_0 \mu  + i \sigma_3 \frac{g}{2} \mu_B B
\right) \right.\nonumber \\ && \left.+ i \hbar v_F \sigma_0 \gamma^1
D_x + i \hbar v_F \sigma_0 \gamma^2 D_y -M_D\right] \Psi(t,
\mathbf{r}).
\end{eqnarray}
The Dirac mass $M_D$\footnote{The energy gap $M_D$ is expressed via
the corresponding Dirac mass $m$ as $M_D = mv_F^{2}$. In what
follows we ignore a difference between ``Dirac gap'' and ``Dirac
mass'' and use the term ``Dirac mass'' for both of them.} has a
general form $\Delta \sigma_\kappa \otimes \Gamma \equiv \Delta
\sigma_\kappa \Gamma$, where $\Delta$ is its absolute value,
$\sigma_\kappa$ is one of the Pauli matrices (in what follows, we
consider only $\sigma_0$ and $\sigma_3$,  more general masses with
${\bm \sigma}$ were recently considered in
Ref.~\refcite{Herbut2007}) and $\Gamma$ is one of the four matrices
\begin{equation}
\label{masses-Dirac} I_4, \qquad \gamma^3, \qquad  i \gamma^5 ,
\qquad  \gamma^3\gamma^5 .
\end{equation}
Under $SU(2)$ group (as in Sec.~\ref{sec:U(2)} we ignore for a
moment the spin degree of freedom) the gaps $\bar
\Psi\Psi,\bar\Psi\gamma^3\Psi,\bar \Psi i\gamma^5\Psi$
 transform as a vector, while the gap $\bar \Psi\gamma^3\gamma^5\Psi$ is a scalar.
Three gaps, $\bar \Psi\Psi$,$\bar \Psi\gamma^3\Psi$, and $\bar \Psi
i\gamma^5\Psi$ break the $SU(2)$ group down to $U(1)$ subgroup (with
generators $T_{3}$, $T_{2}$ and $T_{1}$ being unbroken,
respectively).

\subsubsection{Physical meaning of the Dirac masses}
\label{sec:meaning}

To understand the physical meaning of one of the masses, let us
rewrite the Hamiltonian $H_1$ defined by Eq.~(\ref{H_1-def}) in
terms of 4-component spinor (\ref{4-spinor})
\begin{eqnarray}
\label{Dirac-mass1} H_1 &=& \sum_{\sigma} \int_{DC} \frac{d^2 p}{(2
\pi)^2} \Psi^\dagger_\sigma(\mathbf{p})[m_+ {\tilde \tau}_0 \otimes
\tau_0 + m_-
{\tilde \tau}_3 \otimes \tau_3]  \Psi_\sigma(\mathbf{p}),\nonumber \\
&=&  \int_{DC} \frac{d^2 p}{(2 \pi)^2} {\bar \Psi}(\mathbf{p})[m_+
\sigma_0 \gamma^0 + m_- \sigma_0 \gamma^3] \Psi(\mathbf{p}).
\end{eqnarray}
As was already mentioned after Eq.~(\ref{H_1-def}), the term $m_+
\sigma_0 \gamma^0$ changes the total carrier density of both
$\mathrm{A}$ and $\mathrm{B}$ sublattices and can thus be absorbed
in the term with the chemical potential $\mu \sigma_0 \gamma^0$.
However, the term $m_- \sigma_0 \gamma^3$ breaks $\mathcal{P}$
defined by Eq.~(\ref{inversion-QED-spinor}) and is an example of one
of the Dirac masses introduced above. The operator
\begin{eqnarray}
\label{excitonic} \int \! d^2x   \bar \Psi\sigma_0 \gamma^3 \Psi  =
\sum_\sigma \! \! \! \! && \int_{DC} \!\!\frac{d^2 p}{(2 \pi)^2}
\left[ a_{\sigma}^\dagger (\mathbf{K}_{+} +\mathbf{p}) a_{\sigma}
(\mathbf{K}_{+} + \mathbf{p}) + a_{\sigma}^\dagger (\mathbf{K}_{-} +
\mathbf{p})
a_{\sigma} (\mathbf{K}_{-} + \mathbf{p}) \right. \nonumber \\
&&\left. - b_{\sigma}^\dagger (\mathbf{K}_{+} + \mathbf{p})
b_{\sigma} (\mathbf{K}_{+} + \mathbf{p})-b_{\sigma}^\dagger
(\mathbf{K}_{-} + \mathbf{p}) b_{\sigma} (\mathbf{K}_{-} +
\mathbf{p})\right]
\end{eqnarray}
determines the magnitude of the order parameter $\langle\bar
\Psi\sigma_{0}\gamma^{3}\Psi\rangle$ (and hence the fermion gap)
proportional to the electron density imbalance between the
$\mathrm{A}$ and $\mathrm{B}$ sublattices and corresponds to the
formation of a site-centered charge density wave (CDW) in the
excitonic insulating ground state\cite{Khveshchenko2001PRL} which
lifts sublattice degeneracy. As we have already mentioned in
Sec.~\ref{sec:inversion-lattice}, it has been
suggested\cite{Giovannetti2007} that this kind of Dirac mass can be
introduced in graphene by placing it on top an appropriate substrate
which breaks the graphene sublattice symmetry between $\mathrm{A}$
and $\mathrm{B}$ and so generates an intrinsic Dirac mass for the
fermions.

The Dirac mass can also be generated dynamically. Historically the
phenomenon of the electron-hole (fermion-antifermion) pairing in a
magnetic field called {\em magnetic catalysis\/} was revealed in
field theory.\cite{Gusynin1995PRD} Later, the experiments on graphite and
their interpretation in terms of the Dirac fermions\cite{Kopelevich}
inspired a theoretical condensed matter consideration of this kind
of Dirac
mass.\cite{Khveshchenko2001PRL,Gorbar2002PRB,Gorbar2003PLA,Khveshchenko2004nb}
Note that in the notations of these papers, it corresponds to $\bar
\Psi \Psi$. In the present conventions, the term $\bar \Psi \Psi$
yields a chiral mixing between $\mathbf{K}_{\pm}$ points:
\begin{eqnarray}
\label{Kekule}  \int \! d^2x  \bar \Psi\sigma_0  \Psi  = \sum_\sigma
\! \! \! \! && \int_{DC} \! \frac{d^2 p}{(2 \pi)^2} \left[
a_{\sigma}^\dagger (\mathbf{K}_{+} +\mathbf{p}) b_{\sigma}
(\mathbf{K}_{-} + \mathbf{p}) + b_{\sigma}^\dagger (\mathbf{K}_{+} +
\mathbf{p})
a_{\sigma} (\mathbf{K}_{-} + \mathbf{p}) \right. \nonumber \\
&&\left. + b_{\sigma}^\dagger (\mathbf{K}_{-} + \mathbf{p})
a_{\sigma} (\mathbf{K}_{+} + \mathbf{p})+ a_{\sigma}^\dagger
(\mathbf{K}_{-} + \mathbf{p}) b_{\sigma} (\mathbf{K}_{+} +
\mathbf{p})\right]
\end{eqnarray}
This term preserves $\mathcal{P}$, $\mathcal{T}$ and $\mathcal{C}$
symmetries.

Magnetic catalysis is one of the candidates (see
Ref.~\refcite{Yang2007} for an overview of all scenarios) for an
explanation of the new QHE states observed in graphene in high
magnetic fields.\cite{Zhang2006PRL,Abanin2007,Kim2007APS} In
particular,
Refs.~\refcite{Gusynin2006catalysis,Herbut2006PRL,Fuchs2007PRB,Ezawa2006}
are devoted to the further development of this scenario. The latest
experimental results\cite{Abanin2007,Kim2007APS,Jiang2007} pose new
questions that have to be addressed in the future theoretical work
on  spontaneous symmetry breaking and other competing models that
attempt to explain the origin of $\nu=0, \pm1$ IQHE states. Here we
will not go into the details of magnetic catalysis but instead
consider all theoretically possible Dirac masses. Their physical
meaning becomes clear when we write them in terms of the
tight-binding model of graphene. Also we investigate their discrete
symmetry properties.

Similarly to the mass in Eq.~(\ref{Kekule}), the mass term with
$\Gamma=\gamma^5$ also mixes $\mathbf{K}_{\pm}$ points
\begin{eqnarray}
\label{gamma5}  \int \! d^2x    \bar \Psi\sigma_0 i \gamma^5\Psi  =
-i \sum_\sigma \!\!\!\! && \int \! \!\frac{d^2 p}{(2 \pi)^2}
\left[a_{\sigma}^\dagger (\mathbf{K}_{+} +\mathbf{p}) b_{\sigma}
(\mathbf{K}_{-} + \mathbf{p}) + b_{\sigma}^\dagger (\mathbf{K}_{+} +
\mathbf{p})
a_{\sigma} (\mathbf{K}_{-} + \mathbf{p}) \right. \nonumber \\
&&\left. - b_{\sigma}^\dagger (\mathbf{K}_{-} + \mathbf{p})
a_{\sigma} (\mathbf{K}_{+} + \mathbf{p})- a_{\sigma}^\dagger
(\mathbf{K}_{-} + \mathbf{p}) b_{\sigma} (\mathbf{K}_{+} +
\mathbf{p})\right]
\end{eqnarray}
and breaks $\mathcal{P}$ and $\mathcal{C}$ symmetries. The order
parameters in Eqs.~(\ref{Kekule}) and (\ref{gamma5}) are related to a
Kekul\'{e} distortion\cite{Hou2007PRL} which is $\mbox{Re}\Delta
\bar\Psi\Psi - \mbox{Im}\Delta \bar\Psi i\gamma^5\Psi$ and breaks in
general $\mathcal{P}$ symmetry.

Another interesting example is the mass term with $\Gamma=\gamma^3
\gamma^5$ for which  one has
\begin{eqnarray}
\label{mass-T-breaking} \int \! d^2x   \bar \Psi \sigma_0 \gamma^3
\gamma^5 \Psi
  =    \sum_\sigma \! \! \! \!&& \int \! \!\frac{d^2 p}{(2 \pi)^2} \left[
a_{\sigma}^\dagger (\mathbf{K}_{+} +\mathbf{p}) a_{\sigma}
(\mathbf{K}_{+} + \mathbf{p}) - a_{\sigma}^\dagger (\mathbf{K}_{-} +
\mathbf{p})
a_{\sigma} (\mathbf{K}_{-} + \mathbf{p}) \right. \nonumber \\
&&- \left. b_{\sigma}^\dagger (\mathbf{K}_{+} + \mathbf{p})
b_{\sigma} (\mathbf{K}_{+} + \mathbf{p})+b_{\sigma}^\dagger
(\mathbf{K}_{-} + \mathbf{p}) b_{\sigma} (\mathbf{K}_{-} +
\mathbf{p})\right].
\end{eqnarray}
In contrast to the gap (mass) in Eq.~\ref{excitonic}), the gap in
Eq.~(\ref{mass-T-breaking}) corresponds to a gap with the opposite
sign at $\mathbf{K}_{-}$ point [cf. Eqs.~(\ref{mass-T-breaking}) and
(\ref{excitonic})]. As pointed out in the second paper of
Ref.~\refcite{Kane2005PRL} it is related to a model introduced by
Haldane\cite{Haldane1988PRL} as a realization of the parity anomaly
in $2+1$ dimensional field theory (see also
Refs.~\refcite{Fradkin1986PRL,Schakel1991PRD,Tchernyshyov2000PRB}).
Notice another definition of the spinor
\begin{equation}
\label{Psi-prime} \Psi^{\prime\dagger}_\sigma = (a_{\sigma}^\dagger
(\mathbf{K}_{+} +\mathbf{p}), b_{\sigma}^\dagger (\mathbf{K}_{+}
+\mathbf{p}), a_{\sigma}^\dagger (\mathbf{K}_{-} +\mathbf{p}),
b_{\sigma}^\dagger (\mathbf{K}_{-} +\mathbf{p}))
\end{equation}
in Ref.~\refcite{Kane2005PRL}, so that the gap in Eq.~(\ref{mass-T-breaking})
term becomes $\Psi^{\prime \dagger} \sigma_0
\otimes {\tilde \tau}_3 \otimes \tau_3 \Psi^{\prime}$ ($\tilde \tau$
and $\tau$ matrices act in valley and sublattice spaces,
respectively). In our conventions (\ref{4-spinor}) this corresponds
to $\Psi^\dagger \sigma_0 \otimes {\tilde \tau}_0 \otimes \tau_3
\Psi = {\bar \Psi} \sigma_0 \gamma^3 \gamma^5 \Psi$ gap. Comparing
the masses in Eqs.~(\ref{excitonic}) and (\ref{mass-T-breaking}), one finds
that the second mass preserves $\mathcal{P}$. On the other hand,
considering that $\mathbf{K}_{\pm}$ points are related by
time-reversal $\mathcal{T}$ (see Eq.~(\ref{time-QED-illustration})),
one can easily see that while the mass term $\Delta \sigma_0
\gamma^3$ preserves $\mathcal{T}$, the mass $\Delta_{\mathcal{T}}
\sigma_0 \gamma^3 \gamma^5$ breaks it. This is also illustrated by
the the corresponding energies of the LLL given by Eqs.~(\ref{LL})
and (\ref{LL-anom}), respectively. In contrast, the mass $\sim
\sigma_3 \gamma^3 \gamma^5$ which is related to the spin-orbit
interaction\cite{Kane2005PRL} does not break $\mathcal{T}$. In fact,
the gaps $\sigma_3 \gamma^3\gamma^5$ and $\sigma_0\otimes I_4$ are
the only gaps that respect all the discrete symmetries.

Yet another option for the Dirac mass which involves spin degrees of
freedom was recently considered in Ref.~\refcite{Herbut2006PRL}
\begin{eqnarray}
\label{excitonic-Herbut}  \int \! d^2x   \bar \Psi\sigma_3 \gamma^3
\Psi  = \sum_\sigma \! \! \! \!&& \int \! \! \frac{d^2 p}{(2 \pi)^2}
\sigma \left[ a_{\sigma}^\dagger (\mathbf{K}_{+} +\mathbf{p})
a_{\sigma} (\mathbf{K}_{+} + \mathbf{p}) + a_{\sigma}^\dagger
(\mathbf{K}_{-} + \mathbf{p})
a_{\sigma} (\mathbf{K}_{-} + \mathbf{p}) \right. \nonumber \\
&&- \left. b_{\sigma}^\dagger (\mathbf{K}_{+} + \mathbf{p})
b_{\sigma} (\mathbf{K}_{+} + \mathbf{p})-b_{\sigma}^\dagger
(\mathbf{K}_{-} + \mathbf{p}) b_{\sigma} (\mathbf{K}_{-} +
\mathbf{p})\right].
\end{eqnarray}
Since $\mathcal{P}$ transformation does not affect the spin
variable, this mass  breaks this symmetry as does the mass
(\ref{excitonic}). However, in contrast to this latest case, the
mass (\ref{excitonic-Herbut}) is also $\mathcal{T}$ breaking.

All possible Dirac mass terms $\bar \Psi M_D \Psi$ and their
transformation properties under $\mathcal{P}$, $\mathcal{T}$,
$\mathcal{C}$ symmetries are summarized in Table~\ref{tab:1}. For
completeness we also included matrices $M_\mu$, $M_Z$ and
$\mathcal{M}_{\mu 1}, \mathcal{M}_{\mu 2}$ that correspond to the
chemical potential, Zeeman term and generalized chemical potentials,
respectively. For convenience of comparison with
Ref.~\refcite{Kane2005PRL} we provide equivalents $M^\prime$ of
these terms for $\Psi^\prime$ and $\Psi^{\prime \dagger}$ spinors
(\ref{Psi-prime}).

\begin{table}[ht]
\ttbl{25pc}{Transformation properties of the various Dirac mass
terms $\bar \Psi M_D \Psi = \Psi^\dagger M \Psi = \Psi^{\prime
\dagger} M^\prime \Psi^\prime$ with $M_D = \sigma_{0,3} \otimes
\Gamma$ under $\mathcal{P}$, $\mathcal{T}$ and $\mathcal{C}$
transformations. The matrices $M_\mu$, $M_Z$  and $\mathcal{M}_{\mu
1}, \mathcal{M}_{\mu 2}$ correspond to
the chemical potential, Zeeman term and generalized chemical potentials, respectively.}\\
\vspace{2mm}
\begin{center}
{\begin{tabular}{|c|c|c|c|c|c|c|}
  \hline
  $M_D$ & $M$ & $M^\prime$ & $\mathcal{P}$ & $\mathcal{T}$ & $\mathcal{C}$
   \\ \hline \hline
  $\sigma_0 \otimes I_4$ & $\sigma_0 \otimes {\tilde \tau}_1 \otimes \tau_0$ &
  $\sigma_0 \otimes {\tilde \tau}_1 \otimes \tau_1$ & $1$ & $1$ & $1$ \\[0.8ex]
  $\sigma_0 \otimes \gamma^3$ & $\sigma_0 \otimes {\tilde \tau}_3 \otimes \tau_3$
  & $\sigma_0 \otimes {\tilde \tau}_0 \otimes \tau_3$ & $-1$ & $1$ & $1$  \\[0.8ex]
  $\sigma_0 \otimes i \gamma^5$ & $\sigma_0 \otimes {\tilde \tau}_2 \otimes \tau_0$ &
  $\sigma_0 \otimes {\tilde \tau}_2 \otimes \tau_1$ & $-1$ & $1$ & $-1$ \\[0.8ex]
  $\sigma_0 \otimes \gamma^3 \gamma^5$ & $\sigma_0 \otimes {\tilde \tau}_0 \otimes \tau_3$
  & $\sigma_0 \otimes {\tilde \tau}_3 \otimes \tau_3$ & $1$ & $-1$ & $1$  \\[0.8ex]
  $\sigma_3 \otimes I_4$ & $\sigma_3 \otimes {\tilde \tau}_1 \otimes \tau_0$ &
  $\sigma_3 \otimes {\tilde \tau}_1 \otimes \tau_1$ & $1$ & $-1$ & $1$  \\[0.8ex]
  $\sigma_3 \otimes \gamma^3$ & $\sigma_3 \otimes {\tilde \tau}_3 \otimes \tau_3$ &
  $\sigma_3 \otimes {\tilde \tau}_0 \otimes \tau_3$ & $-1$ & $-1$ & $1$  \\[0.8ex]
  $\sigma_3 \otimes i \gamma^5$ & $\sigma_3 \otimes {\tilde \tau}_2 \otimes \tau_0$ &
  $\sigma_3 \otimes {\tilde \tau}_2 \otimes \tau_1$  & $-1$ & $-1$ & $-1$  \\[0.8ex]
  $\sigma_3 \otimes \gamma^3 \gamma^5$ & $\sigma_3 \otimes {\tilde \tau}_0 \otimes \tau_3$ &
  $\sigma_3 \otimes {\tilde \tau}_3 \otimes \tau_3$ & $1$ & $1$ & $1$  \\[0.8ex]
  \hline \hline
  $M_\mu =\sigma_0\otimes\gamma^0$ & $\sigma_0 \otimes {\tilde \tau}_0 \otimes \tau_0$ &
  $\sigma_0 \otimes {\tilde \tau}_0 \otimes \tau_0$
  & $1$ & $1$ & $-1$ \\[0.8ex]
  $M_{Z} =\sigma_3\otimes\gamma^0$ &  $\sigma_3 \otimes {\tilde \tau}_0 \otimes \tau_0$ &
  $\sigma_3 \otimes {\tilde \tau}_0 \otimes \tau_0$
  & $1$ & $-1$ & $-1$ \\[0.8ex]
  $\mathcal{M}_{\mu1} =\sigma_0\otimes\gamma^0\gamma^5 $ & $\sigma_0\otimes -i{\tilde
  \tau}_2\otimes\tau_0$ & $\sigma_0\otimes -i{\tilde
  \tau_2}\otimes\tau_1$ & $-1$ & $-1$ & $-1$ \\[0.8ex]
  $\mathcal{M}_{\mu2} =\sigma_3\otimes\gamma^0\gamma^5 $ & $\sigma_3\otimes -i{\tilde
  \tau}_2\otimes\tau_0$ & $\sigma_3\otimes -i{\tilde
  \tau_2}\otimes\tau_1$ & $-1$ & $1$ & $-1$ \\[0.8ex]
  \hline
\end{tabular}}
\end{center}
\label{tab:1}
\end{table}
Concerning the continuous $U(4)$ symmetry group, all Dirac gaps
except $\sigma_{0}\otimes \gamma^{3}\gamma^{5}$ break it down to
$U(2)\otimes U(2)$ subgroup though the pattern of breaking depends
on a concrete  gap. For example, in case of the gap $\Delta \bar
\Psi \gamma^{3}\Psi$ we have breakdown of $U(4)$ to the
$U_{a}(2)\otimes U_{b}(2)$ with the generators
\begin{equation}
\label{pattern-SB1} \frac{\sigma^{a}}{2}\otimes I_{4}, \quad
\frac{\sigma^{a}}{2}\otimes T_{2}.
\end{equation} Subgroups $U_{a,b}(2)$ act
in the state spaces $(\mathbf{K}_{+}, \uparrow)$, $(\mathbf{K}_{+},
\downarrow)$ and $(\mathbf{K}_{-}, \uparrow)$, $(\mathbf{K}_{-},
\downarrow)$, not mixing these two spaces.

An interesting example of  $U(4)$ symmetry breaking is provided by
the gap $\bar \Psi\sigma_3 \gamma^{3}\Psi$. In this case, we have $8$
unbroken generators
\begin{equation}
\frac{\sigma^{1}}{2}\otimes T_{1},\quad \frac{\sigma^{2}}{2}\otimes
T_{1},\quad \frac{\sigma^{1}}{2}\otimes T_{3},\quad
\frac{\sigma^{2}}{2}\otimes T_{3},\quad \frac{\sigma^{0}}{2}\otimes
T_{2}, \quad \frac{\sigma^{3}}{2}\otimes T_{2},\quad
\frac{\sigma^{0}}{2}\otimes I_{4}, \quad \frac{\sigma^{3}}{2}\otimes
I_{4},
\end{equation}
which form another $U_{a^{\prime}}(2)\otimes U_{b^{\prime}}(2)$
group acting in the state spaces $(\mathbf{K}_{+}, \uparrow)$,
$(\mathbf{K}_{-}, \downarrow)$ and $(\mathbf{K}_{+}, \downarrow),
(\mathbf{K}_{-}, \uparrow) $, respectively. While this gap is
$\mathcal{P}$-odd, $\mathcal{T}$-odd it is invariant under combined
$\mathcal{P} \mathcal{T}$-inversion.

Finishing our classification of the Dirac masses, we mention that
the QED$_{2+1}$ description of $d$-wave superconductivity also
involves various Dirac
masses\cite{Vafek2001PRB,Khveshchenko2001aPRL,Herbut2002PRB,Gusynin2004EPJB,Sharapov2006PRB}
when the possible opening of a secondary gap is considered (for
example, a charge or spin density wave or a second superconducting
order parameter with different symmetry).

\subsubsection{Dirac Landau levels}

\begin{figure}[bt]
\centerline{\psfig{file=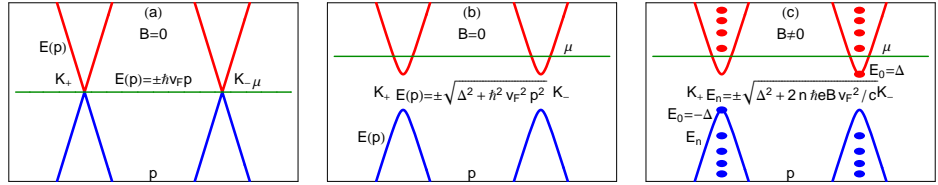,width=5.1in}} \vspace*{8pt}
\caption{(Colour online) Band structure of graphene. Electrons are
shown in red and holes in blue. (a) The low-energy linear-dispersion
$E(\mathbf{k})$ near the Dirac $\mathbf{K}_+$ and $\mathbf{K}_-$
points for $B=0$. (b) A possible modification of the quasiparticle
spectrum by the finite gap (Dirac mass) $\Delta$. The chemical
potential (indicated by horizontal line) $\mu$ is shifted from zero
by the gate voltage. (c) Landau levels $E_n$ in the Dirac theory of
graphene. Spin degree of freedom is ignored. For a given direction
of the magnetic field $\mathbf{B}$ applied perpendicular to
graphene's plane, the lowest $(n=0)$ Landau level has the energy
$E_0=-\Delta$ at $\mathbf{K}_+$ and $E_0=\Delta$ at $\mathbf{K}_-$.
} \label{fig:3}
\end{figure}

In an external magnetic field $B$ Dirac Landau levels are formed. In
the presence of a finite  gap amplitude $\Delta \neq 0$ and ignoring
the Zeeman splitting, for the matrices $\Gamma = I_4,
\gamma^3,\gamma^5$ the energies of Landau levels are
\begin{equation}
\label{LL} E_{n } =
 \left\{
\begin{array}{lc}
\pm \Delta \mbox{sgn}(eB),  & n=0, \\
\mbox{sgn}(n)\sqrt{\Delta^2 + 2 |n|\hbar v_F^2 |eB|/c}, &   \qquad n=\pm 1,\pm2,\ldots.\\
\end{array}
\right.
\end{equation}
To understand how the spectrum (\ref{LL}) emerges, it is instructive
to write the Dirac equation for one of these masses, considering
separately each $\mathbf{K}$ point. For example, for the $\Delta
\sigma_0 \gamma^3$ mass, one obtains
\begin{eqnarray}
\label{Dirac-excitonic1} && \left[i\tau_0\hbar\partial_t + i \hbar
v_F \tau_1 D_x + i \hbar v_F \tau_2 D_y - \tau_3 \Delta
\right]\psi_{\mathbf{K}_{+}}(t,{\mathbf
r})=0,  \\
&& \label{Dirac-excitonic2} \left[i\tau_0\hbar\partial_t - i \hbar
v_F \tau_1 D_x - i \hbar v_F \tau_2 D_y + \tau_3 \Delta
\right]\psi_{\mathbf{K}_{-}}(t,{\mathbf r})=0.
\end{eqnarray}
Multiplying the left side of Eq.~(\ref{Dirac-excitonic1}) by
$\tau_3$ and the left side of Eq.~(\ref{Dirac-excitonic2}) by $-\tau_3$,
one obtains
\begin{eqnarray}
\label{Dirac-excitonic-gamma1} &&
\left[i\hat{\gamma}^0\hbar\partial_t + i \hbar v_F \hat{\gamma}^1
D_x + i \hbar v_F \hat{\gamma}^2 D_y - \Delta
\right]\psi_{\mathbf{K}_{+}}(t,{\mathbf
r})=0,  \\
\label{Dirac-excitonic-gamma2} && \left[i \tilde{\gamma}^0
\hbar\partial_t + i\hbar  v_F \tilde{\gamma}^1 D_x + i \hbar v_F
\tilde{\gamma}^2 D_y -  \Delta
\right]\psi_{\mathbf{K}_{-}}(t,{\mathbf r})=0,
\end{eqnarray}
where in Eq.~(\ref{Dirac-excitonic-gamma1}), $2\times 2$ gamma
matrices are given by Eq.~(\ref{gamma-old}) and in
Eq.~(\ref{Dirac-excitonic-gamma2}) $\tilde{\gamma}^0 =-
\hat{\gamma}^0$, $\tilde{\gamma}^{1,2} =\hat{\gamma}^{1,2}$ and have
the opposite signature $\eta$. For reference, the solution of
Eqs.~(\ref{Dirac-excitonic-gamma1}) and
(\ref{Dirac-excitonic-gamma2}) in an external magnetic field is
written down in Appendix~A. The energies of the Landau levels with $n$
and $-n$ for $|n| \geq 1$ are symmetric (see Eqs.~(\ref{LL}) and
(\ref{LL-anom})). The LLL is, however, asymmetric and anomalous,
because for Eq.~(\ref{Dirac-excitonic-gamma1}) its energy is
$-\Delta \, \mbox{sgn}(eB)$, while for
Eq.~(\ref{Dirac-excitonic-gamma2}), its energy is $\Delta \,
\mbox{sgn}(eB)$ (see Fig.~\ref{fig:3}~(c) where the case $eB >0$ is
shown).
It turns out that the sign of the LLL energy is defined by the
relative sign of the signature $\eta$ of the $2\times 2$ Dirac
matrices (\ref{signature}) and the Dirac mass
$\Delta$.\cite{Hosotani1993PLB} This implies that because the sign
before the mass $\Delta$ in Eqs.~(\ref{Dirac-excitonic-gamma1}) and
(\ref{Dirac-excitonic-gamma2}) is the same, the sign of the LLL
energy depends on the sign of the signature of $\hat{\gamma}^\nu$ and
$\tilde{\gamma}^\nu$ matrices. Thus altogether the combined energy
spectrum of Eqs.~(\ref{Dirac-excitonic1}) and
(\ref{Dirac-excitonic2}) is symmetric and given by Eq.~(\ref{LL}).
This reflects the fact that the Dirac mass $\Delta\sigma_0 \gamma^3$
does not break the time-reversal symmetry.

In the limit $\Delta \to 0$, the Landau levels $\pm \Delta
\mbox{sgn}(eB)$ corresponding to $n=0$ Landau levels merge together
to form a single level. However, for $\mu =0$ and ignoring Zeeman
splitting, it remains half-filled. This property of the LLL equally
shared by particles and antiparticles is at the heart of the
unconventional QHE in graphene.\cite{Gusynin2005PRL}

For $\Gamma = \gamma^3\gamma^5$, the spectrum is asymmetric
\begin{equation}
\label{LL-anom} E_{n} = \left\{
\begin{array}{lc}
-\Delta \mbox{sgn}(eB),  & n=0, \\
\mbox{sgn}(n) \sqrt{\Delta^2 + 2 |n|\hbar v_F^2 |eB|/c},
&  \quad n=\pm 1,\pm 2,\ldots \\
\end{array}
\right.
\end{equation}
Indeed, the $\mathcal{T}$ breaking $\Delta_{\mathcal{T}} \sigma_0
\gamma^3 \gamma^5$ mass  results in two equations
\begin{eqnarray}
\label{Dirac-mass-T-breaking} && \left[i\tau_0\hbar\partial_t + i
\hbar v_F \tau_1 D_x + i \hbar v_F \tau_2 D_y - \tau_3
\Delta_{\mathcal{T}} \right]\psi_{\mathbf{K}_{+}}(t,{\mathbf
r})=0, \nonumber \\
&& \left[i\tau_0\hbar\partial_t - i \hbar v_F \tau_1 D_x - i \hbar
v_F \tau_2 D_y - \tau_3 \Delta_{\mathcal{T}}
\right]\psi_{\mathbf{K}_{-}}(t,{\mathbf r})=0
\end{eqnarray}
which have the same sign before $\Delta_{\mathcal{T}}$. Accordingly,
because the corresponding $2\times 2$ gamma matrices are the same as
for Eqs.~(\ref{Dirac-excitonic-gamma1}) and
(\ref{Dirac-excitonic-gamma2}), one can readily see that both
equations in (\ref{Dirac-mass-T-breaking}) lead to the same sign of
the LLL energy, viz. $-\Delta_{\mathcal{T}} \, \mbox{sgn}(eB)$.
Accordingly, the LLL remains doubly degenerate. Therefore, the
combined energy spectrum (\ref{LL-anom}) is obviously asymmetric.
Thus, the time-reversal breaking character of the Dirac mass
$\Delta_{\mathcal{T}} \sigma_0 \gamma^3 \gamma^5$ is revealed by an
external magnetic field. At temperatures well below
$\Delta_{\mathcal{T}}$ this leads to the QHE even in zero magnetic
field.\cite{Haldane1988PRL} Indeed, since the transverse Hall
conductivity $\sigma_{xy}$ is odd under time reversal, a nonzero
$\sigma_{xy}$ can occur if time-reversal invariance is broken.

In the limit $\Delta_{\mathcal{T}}\to 0$, the $n=0$ Landau level $-
\Delta \mbox{sgn}(eB)$ in Eq.~(\ref{LL-anom}) is either
electron-like or hole-like depending on the relative sign of
$\Delta$ and $eB$. Thus, the generic zero gap limit is indefinite in
terms of a clear particle or hole character of quasiparticles
occupying the LLL. Nevertheless, from a physical point of view, the
property of the LLL to be equally shared by particles and
antiparticles can be argued by the presence of the Zeeman term.

The degeneracy of Landau levels with $n=\pm 1,\pm 2,\ldots$ in
Eqs.~(\ref{LL}) and (\ref{LL-anom}) is $|eB|/(\pi\hbar)$ (per unit
area and per spin), while the degeneracy of the lowest Landau levels
(LLL) $n=0$ for Eq.~(\ref{LL}) is $|eB|/(2\pi\hbar)$ and  for
Eq.~(\ref{LL-anom}) is $|eB|/(\pi\hbar)$.

It is worth emphasizing that it is the spectacular ``relativistic''
energy Landau scale $L(B)= \sqrt{|eB| \hbar v_F^2/c}$ which defines
the distance between Landau levels
\begin{equation}
\label{L-scale} \Delta E= E_1 (\Delta=0)- E_0 (\Delta=0) = \sqrt{2
L^2(B)} \approx 424 \sqrt{B [\mbox{T}]} \mbox{K}.
\end{equation}
For instance, it corresponds to $\Delta E \approx 2800\, \mbox{K}$
at $B=45\, \mbox{T}$ which makes it possible for the QHE in graphene
to be observed at room temperature.\cite{GeimKim2007Science}

\begin{figure}[bt]
\centerline{\psfig{file=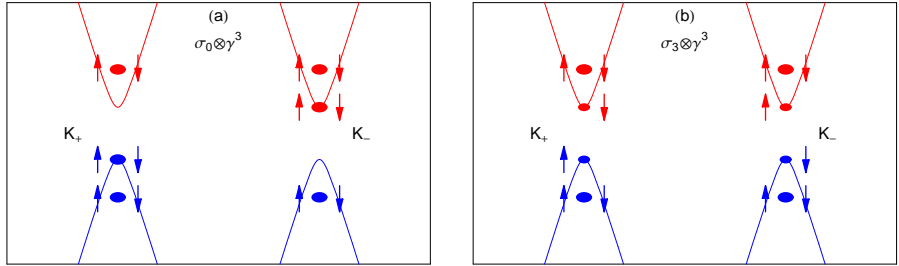,width=4.9in}} \vspace*{8pt}
\caption{(Colour online) Landau levels including spin degree of
freedom, but taking zero value of Zeeman splitting. Electrons are
shown in red and holes in blue. (a) for $\sigma_0 \otimes \gamma^3$
gap. (b) for $\sigma_3 \otimes \gamma^3$ gap. } \label{fig:4}
\end{figure}

Including the real spin degree of freedom, we have the following
situation for Landau levels in a magnetic field with the
$\mathcal{P}$-odd, $\mathcal{T}$-even gap $\Delta\bar
\Psi\gamma^{3}\Psi$.\footnote{We are grateful to D.V.~Khveshchenko
for a discussion on Dirac masses considered in this section.} For
levels $n\geq1$ all Landau levels are $4$-fold degenerate (states
$|\mathbf{K}_{+},\uparrow\rangle$,
$|\mathbf{K}_{+},\downarrow\rangle$,
$|\mathbf{K}_{-},\uparrow\rangle$,
$|\mathbf{K}_{-},\downarrow\rangle$). On the other hand, for the
$n=0$ Landau levels, as shown in Fig.~\ref{fig:4}~(a), we have
$2$-fold degeneracy: states $|\mathbf{K}_{+},\uparrow\rangle,
|\mathbf{K}_{+},\downarrow\rangle$ and
$|\mathbf{K}_{-},\uparrow\rangle, |\mathbf{K}_{-},\downarrow\rangle$
with energies $\mp\Delta$, respectively.
For $\mathcal{P}$-odd, $\mathcal{T}$-odd gap $\Delta\bar
\Psi\sigma_3 \gamma^{3}\Psi$ (see Fig.~\ref{fig:4}~(b)) the
situation with levels $n\geq1$ is the same, but two Landau levels
with $n=0$ with energies $\mp\Delta$ now have different $2$-fold
degeneracies: $|\mathbf{K}_{+},\uparrow\rangle$,
$|\mathbf{K}_{-},\downarrow\rangle$ and
$|\mathbf{K}_{-},\uparrow\rangle$,
$|\mathbf{K}_{+},\downarrow\rangle$ corresponding to a symmetry
breaking pattern described by Eq.~(\ref{pattern-SB1}). In both
cases, switching on the Zeeman interaction leads to further
splitting of the level $n=0$ resulting in four different levels,
thus completely lifting the $4$-fold degeneracy of this level in
noninteracting theory. For both gaps considered we expect QHE with
the filling factors $\nu=0,\pm1,\pm2k$ where $k$ is a positive
integer.

Let us now consider  the $\mathcal{P}$-even, $\mathcal{T}$-odd gap
$\Delta\bar \Psi\gamma^{3}\gamma^{5}\Psi$ which does not break the
$U(4)$ symmetry. All Landau levels have $4$-fold degeneracy. More
precisely, for the levels with $n=0$, the states of either spin from
$\mathbf{K}_{+}$ - valley residing on the sublattice $\mathrm{B}$
are degenerate with the states of both spins from $\mathbf{K}_{+}$ -
valley residing on the sublattice $\mathrm{A}$. Applying the Zeeman
term, one now obtains only $2$ different (doubly degenerate) levels.

Finally, for $\mathcal{P}$-even, $\mathcal{T}$-even gap $\Delta\bar
\Psi\sigma_3 \gamma^{3}\gamma^{5}\Psi$ one has two pairs of states
for the $n=0$ Landau levels: $|\mathbf{K}_{+},\uparrow\rangle,
|\mathbf{K}_{-},\uparrow\rangle$ and
$|\mathbf{K}_{+},\downarrow\rangle,
|\mathbf{K}_{-},\downarrow\rangle$, which remain degenerate even
when the Zeeman interaction is turned on. Hence, for the last two gaps we
expect QHE with the fillings factors $\nu=2k$ with $k=0,\pm1,\pm2,
\dots$.

\section{Electromagnetic response}
\label{sec:conductivity}

\subsection{Electrical conductivity}

The frequency-dependent electrical conductivity tensor
$\sigma_{\alpha \beta}(\Omega)$ is calculated using the Kubo formula
\begin{equation}\label{Kubo}
\sigma_{\alpha \beta}(\Omega)= \frac{K_{\alpha
\beta}(\Omega+i0)}{-i(\Omega+i0)}, \qquad K_{\alpha
\beta}(\Omega+i0) \equiv \frac{\langle\tau_{\alpha
\beta}\rangle}{V}+ \frac{\Pi_{\alpha \beta}^R(\Omega+i0)}{\hbar V},
\end{equation}
where the retarded correlation function for currents is given by
\begin{equation}
\Pi_{\alpha \beta}^R(\Omega)=\int_{-\infty}^\infty dt\,e^{i\Omega
t}\Pi_{\alpha \beta}^R(t),\quad \Pi_{\alpha
\beta}^R(t)=-i\theta(t){\rm
Tr}\left(\hat{\rho}[J_\alpha(t),J_\beta(0)]\right),
\end{equation}
$V$ is the volume (area) of the system, $\hat{\rho}=\exp(-\beta
H_0)/Z$ is the density matrix of the grand canonical ensemble,
$\beta=1/T$ is the inverse temperature, $Z={\rm Tr}\exp(-\beta H_0)$
is the partition function, and $J_\alpha$ are the total paramagnetic
current operators with
\begin{equation}
\label{current_t=0}
J_\alpha(t)=e^{iHt/\hbar}J_\alpha(0)e^{-iHt/\hbar}, \qquad
J_\alpha(t)=\sum_{\mathbf{n}} j_\alpha^P(t,\mathbf{n}),
\end{equation}
expressed via the paramagnetic current density
(\ref{current-param}). The diamagnetic or stress tensor $\langle
\tau_{\alpha \beta}\rangle$ in the Kubo formula (\ref{Kubo}) is a
thermal average of the diamagnetic part (\ref{diamagnetic-term})
\begin{equation}
\label{diamagnetic-term-average} \langle \tau_{\alpha \beta}\rangle
= \langle \sum_{\mathbf{n}} \tau_{\alpha \beta} (\mathbf{n})
\rangle.
\end{equation}
In many cases, the effective low-energy QED$_{2+1}$ description
provides a very good starting point for the investigation of the
various transport properties of graphene making it possible to use
the powerful field theoretical methods that allow us to obtain
rather simple analytical expressions. In this case, the conductivity
tensor in Eq.~(\ref{Kubo}) is calculated for the model Lagrangian
(\ref{Lagrangian-M}) rather than the full tight-binding model
in Eq.~(\ref{Hamilton-lattice}).

\subsection{Zero field AC conductivity in the continuum Dirac model}
\label{sec:AC-conductivity}

One of the interesting results for the zero magnetic field case is
that in the high-frequency limit, the interband contribution to
longitudinal conductivity is
constant\cite{Ando2002JPSJ,Gusynin2006micro,Gusynin2007JPCM,Falkovsky2006,Ryu2006}
\begin{equation}
\label{diagonal-high-frequency} \mbox{Re} \sigma_{xx}(\Omega) \simeq
\frac{\pi e^2}{2 h}, \qquad \Omega \gg |\mu|, T.
\end{equation}
It is remarkable that the conductivity is universal, independent of
the band structure parameters $t$ and $v_F$. Moreover,
Eq.~(\ref{diagonal-high-frequency}) is even valid for a finite
wave-vector $\mathbf{k}$, $k v_F \ll T \ll
\Omega$.\cite{Falkovsky2006} The flatness of the conductivity
results from the interband transitions between electron and hole
bands at $\mathbf{K}_{\pm}$ points. Of course, this result needs
modification if the photon energy becomes comparable to the band
edge energy, when the assumption of linear momentum dispersion
relation ceases to be valid. We will return to this important
question in Sec.~\ref{sec:sum}.

A more general expression\cite{Gusynin2006micro} for the
conductivity $\sigma_{xx}(\Omega,T)$ valid in the limit of small
impurity scattering rate $\Gamma(\omega)$ and neglecting the real part
of the impurity self-energy reads
\begin{eqnarray}
\label{B=0.cond-intra-inter}
\sigma_{xx}(\Omega,T)&=&\frac{e^2}{\pi^2
\hbar}\int\limits_{-\infty}^\infty
d\omega\frac{[n_F(\omega)-n_F(\omega^\prime)]}{\Omega}\frac{\pi}{4\omega\omega^\prime} \nonumber \\
&\times&\left[\frac{2\Gamma(\omega)}{\Omega^2+4\Gamma^2(\omega)}-\frac{2\Gamma(\omega)}
{(\omega+\omega^\prime)^2+4\Gamma^2(\omega)}\right]\nonumber \\
&\times& (|\omega|+|\omega^\prime|)(\omega^2+{\omega^\prime}^2),
\qquad \omega^\prime = \omega + \Omega.
\end{eqnarray}
The first term in square brackets of
Eq.~(\ref{B=0.cond-intra-inter}) describes the intraband transitions
and the second term, which reduces to
Eq.~(\ref{diagonal-high-frequency}),describes the interband transitions. An essential feature
of Eq.~(\ref{B=0.cond-intra-inter}) is that we kept the energy
dependence of $\Gamma(\omega)$. In deriving this equation, we have
assumed the small $\Omega$ limit, so that $\Gamma(\omega^\prime)
\simeq\Gamma(\omega)$ and for $\Omega \ll T$, the difference
$[n_F(\omega)-n_F(\omega^\prime)]/\Omega$ can be replaced by the
derivative $-\partial n_F(\omega)/\partial \omega$. The intraband
term of Eq.~(\ref{B=0.cond-intra-inter}) in this case results in the
following expression for Drude conductivity
\begin{equation}
\label{B=0.cond-intra}
\sigma_{xx}(\Omega,T)=\sigma_{00}\int\limits_{-\infty}^\infty
d\omega \left(-\frac{\partial\, n_F(\omega)}{\partial\omega}\right)
\frac{2\pi|\omega|\Gamma(\omega)}{\Omega^2+4\Gamma^2(\omega)},
\end{equation}
with $\sigma_{00}=e^2/(\pi^2 \hbar)$. As discussed in
Ref.~\refcite{Gusynin2006micro}, the frequency dependence of
$\Gamma(\omega)$ could be probed by measuring $\sigma_{xx}(\Omega)$
at the different values of the gate voltage $V_g$.

In fact, if it is assumed that $\Gamma(\omega)$ is linear in $\omega$,
as can be expected in the Born approximation for weak scattering when
the chemical potential $\mu=0$, i.e. $\Gamma(\omega)= \gamma_{00}+
\alpha |\omega|$ with a small value of $\gamma_{00}$, then one can
show
\begin{equation}
\label{B=0.cond-intra-mu=0} \sigma_{xx}(\Omega,T) \simeq \frac{\pi
\sigma_{00}}{2 \alpha}\left[1- \frac{\pi}{8
\alpha}\frac{\Omega}{T}\right], \qquad \gamma_{00}< \Omega \ll T.
\end{equation}
Note the linear dependence on microwave frequency $\Omega$ with
slope inversely proportional to the temperature. A similar formula
has been derived for the microwave conductivity in a $d$-wave
superconductor,\cite{Kim2004PRB} a system which, as mentioned in
Sec.~\ref{sec:Lagrangian}, can also be described by QED$_{2+1}$. It
was used to explain the cusp like behavior seen in pure samples of
ortho II YBCO$_{6.5}$ in which every second chain is complete and
the others entirely missing. Eq.~(\ref{B=0.cond-intra}) offers the
possibility of observing directly through microwave experiments the
transport quasiparticle scattering rates in graphene and
establishing whether the scattering is weak and the Born approximation
applies, or it is strong (unitary limit) and a better model for
$\Gamma(\omega)$ is constant $\Gamma_0$ within the energy range of
importance in microwave experiments.  In this case,
Eq.~(\ref{B=0.cond-intra}) reduces to a Drude profile even for
$\mu=0$.

Another prediction of theory is that for $|\mu|>T$,
Eq.~(\ref{B=0.cond-intra-mu=0}) is replaced by
\begin{equation}
\label{B=0.cond-intra-mu} \sigma_{xx}(\Omega,T) = \sigma_{00} 2 \pi
|\mu| \frac{(\gamma_{00} + \alpha |\mu|)}{\Omega^2 + 4 (\gamma_{00}
+ \alpha |\mu|)^2}
\end{equation}
which has the Drude form and shows the remarkable property that its
width can be increased continuously by increasing the gate voltage,
although the impurity content is left unchanged. This arises due to
the relation $\mu \varpropto \mbox{sgn}\,V_g \sqrt{|V_g|}$ (see
Sec.~\ref{sec:diagonal-sum} below), because of the increase in
residual scattering which is proportional to the final state density
of states  which varies as $|\omega|$.

In the context of massive Dirac quasiparticles, we note that for
$T=0$ and in the limit of zero scattering $\Gamma \to 0$, one can
obtain a remarkably simple form for the conductivity, viz.
\begin{eqnarray} \label{B=0.cond-Delta-T=0}
\sigma_{xx}(\Omega)&=&\frac{2\pi
e^2}{h}\delta(\Omega)\frac{(\mu^2-\Delta^2)\theta(\mu^2 -
\Delta^2)}{|\mu|} \\&+&\frac{\pi
e^2}{2h}\frac{\Omega^2+4\Delta^2}{\Omega^2}\theta\left(\frac{|\Omega|}{2}
-{\rm max}(|\mu|,\Delta)\right). \nonumber
\end{eqnarray}
Here, the theta function $\theta\left(|\Omega|/2 -{\rm
max}(|\mu|,\Delta)\right)$ cuts off the low $\Omega$ part of the
interband contribution at $2|\mu|$ or $2\Delta$ whichever is the
largest. The expression in Eq.~(\ref{B=0.cond-Delta-T=0}) was obtained
considering as an example the excitonic gap in Eq.~(\ref{excitonic}). The
factor $(\Omega^2 + 4 \Delta^2)/\Omega^2$  in the second term of
Eq.~(\ref{B=0.cond-Delta-T=0}) modifies the interband contribution
from its constant value of the massless Dirac case. At large
$\Omega$, this modulating factor goes to $1$, but for $\Omega = 2
\Delta$ it is equal to $2$. This shows that the existence of a
finite $\Delta$ does have a clear signature in the AC conductivity.
So far only ARPES measurements show some evidence for a finite value
of the gap $\Delta$ at $B=0$ in graphene epitaxially grown on
SiC.\cite{Lanzara.private} We will consider the situation for finite magnetic fields
in the next Section.

\subsection{Magneto-optical conductivity}
\label{sec:magnetoconductivity}

Motivated by recent experimental advances in infrared spectroscopy
of one layer graphene,\cite{Jiang2007} a few layer epitaxial
graphite\cite{Sadowski2006PRL} and in highly oriented pyrolytic
graphite,\cite{Li2006PRB} we discuss some results obtained in
Refs.~\refcite{Gusynin2007PRL,Gusynin2007JPCM} for magneto-optical
conductivity of graphene here. Starting from the Lagrangian
(\ref{Lagrangian-M}), one can obtain a simple representation for the
complex diagonal conductivity
\begin{eqnarray}
\label{sigma_xx-complex-corrected}
\sigma_{xx}(\Omega)&&=-\frac{e^2v_F^2|eB|}{2\pi ci}\\
&\times& \sum_{n=0}^\infty
\left\{\left(1-\frac{\Delta^2}{M_nM_{n+1}}\right)\left([n_{F}(M_n) -
n_F(M_{n+1})] + [n_F(-M_{n+1}) - n_F(-M_{n})]\right)
\right. \nonumber \\
&&\times \frac{1}{M_{n+1}-M_n}\nonumber \\
&\times &\left.\left(\frac{1}{M_n - M_{n+1}+\Omega +i (\Gamma_n +
\Gamma_{n+1})}-\frac{1}{M_n - M_{n+1}-\Omega -i (\Gamma_n +
\Gamma_{n+1})}\right)\right. \nonumber\\
&+&
\left.\left(1+\frac{\Delta^2}{M_nM_{n+1}}\right)\left([n_{F}(-M_n) -
n_F(M_{n+1})] + [n_F(-M_{n+1}) - n_F(M_{n})]\right) \right.
\nonumber\\
&& \times
\frac{1}{M_{n+1}+M_n} \nonumber \\
& \times&\left.\left(\frac{1}{M_n +M_{n+1} +\Omega+ i(\Gamma_n +
\Gamma_{n+1})} -\frac{1}{M_n +M_{n+1} -\Omega- i(\Gamma_n +
\Gamma_{n+1})}\right)\right\}, \nonumber
\end{eqnarray}
where $M_n = \sqrt{\Delta^2 +  2 n v_F^2|eB|/c}$ are the absolute
values of the energies of the Landau levels.\footnote{The levels
with the energies $E_n = - M_n$ are explicitly included in
Eq.~(\ref{sigma_xx-complex-corrected}) and we set $\hbar =1$.} The
Zeeman splitting was not taken into account, because it can only
shift the positions of the lines if the value of the spitting
depends on Landau level index $n$. The Lorentzian representation
(\ref{sigma_xx-complex-corrected}) is derived in the Appendix of
Ref.~\refcite{Gusynin2007JPCM} using the assumption that the Landau
level half-width $\Gamma_n$ depends only on the Landau level index
$n$ and is independent of the energy $\omega$. In general, the
scattering rate $\Gamma_n(\omega)$ is expressed via the retarded
fermion self-energy, $\Gamma_n(\omega) = -
\mbox{Im}\Sigma^R_n(\omega)$, which depends on the energy,
temperature, field and the Landau levels index $n$. This
self-energy, which in general has a real part also, has to be
determined self-consistently from the Schwinger-Dyson equation. This
equation can be solved
analytically\cite{Khveshchenko2001aPRL,Dora2007} and numerically, as
was done in Refs.~\refcite{Zheng2002PRB,Peres2006PRB}. When the
frequency dependence of $\mbox{Im}\Sigma^R_n(\omega)$ is retained,
the $\mbox{Re}\Sigma^R_n(\omega)$ is nonzero, which results in the
shift of Landau level positions. An analytical expression for the
optical conductivity can also be obtained for this
case.\cite{Gusynin2005PRL,Peres2006PRB} Similarly to
Eq.~(\ref{B=0.cond-Delta-T=0}), the expression
(\ref{sigma_xx-complex-corrected}) was obtained by including the
excitonic gap (\ref{excitonic}).

Jiang {\em et al.}\cite{Jiang2007} have recently performed optical
transmission experiment in an external field on a single layer
graphene. Earlier measurements were on several layer systems, namely
ultra thin epitaxial graphite.\cite{Sadowski2006PRL} Another recent
work reported a magneto-reflectance study\cite{Li2006PRB} of highly
oriented pyrolytic graphite in a magnetic field up to $18\,
\mbox{T}$. In this experiment the Landau level energies are found to
be linear in $B$, while in graphene as well as in the epitaxial
graphite samples they were found to go like $\sqrt{B}$ instead, which
is  consistent with Dirac quasiparticles. While this is expected in
a single sheet graphene, the situation is not as clear for the
multilayer case. Consequently the data in this case were analyzed in
some detail in Ref.~\refcite{Gusynin2007PRL} and the discrepancies
with our Eq.~(\ref{sigma_xx-complex-corrected}) were noted. In
particular, it is hard to understand the observed ratio of the
intensity of the first to the second interband line. We do not
repeat details here, rather we comment on the newest data of Jiang
{\em et al.}\cite{Jiang2007}. In particular, these authors find that
the distance between the Landau levels with the energies $-M_1$ and
$M_2$ is somewhat bigger than predicted on the basis of the
independent particle model such as
Eq.~(\ref{sigma_xx-complex-corrected}). They examine the possible
effect of Coulomb interaction as described by A.~Iyengar {\em at
al.}\cite{Iyengar2007PRB} on Landau levels and tentatively conclude
that this could possibly explain the data. On the other hand, they
observe that the model with a gap for the $\nu=0$ plateau included
as in Eq.~(\ref{sigma_xx-complex-corrected}) is not compatible with
their observations, at least in the present form. It predicts a
reduction rather than the observed small increase in spacing between
first and second interband optical line. At this early stage,
however, it is hard to be definitive about correlation effects.
\begin{figure}[bt]
\centerline{\psfig{file=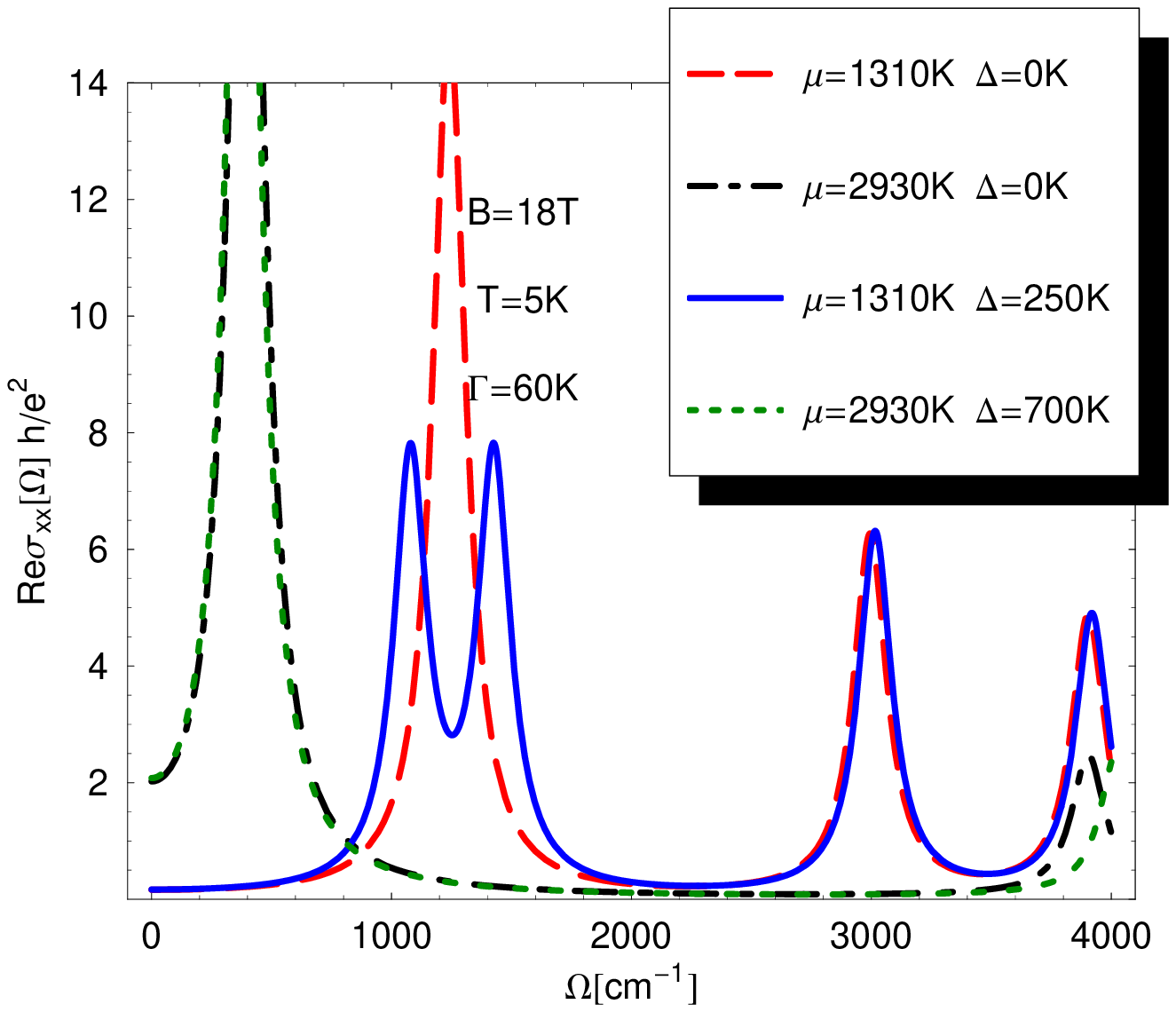,width=2.4in}\psfig{file=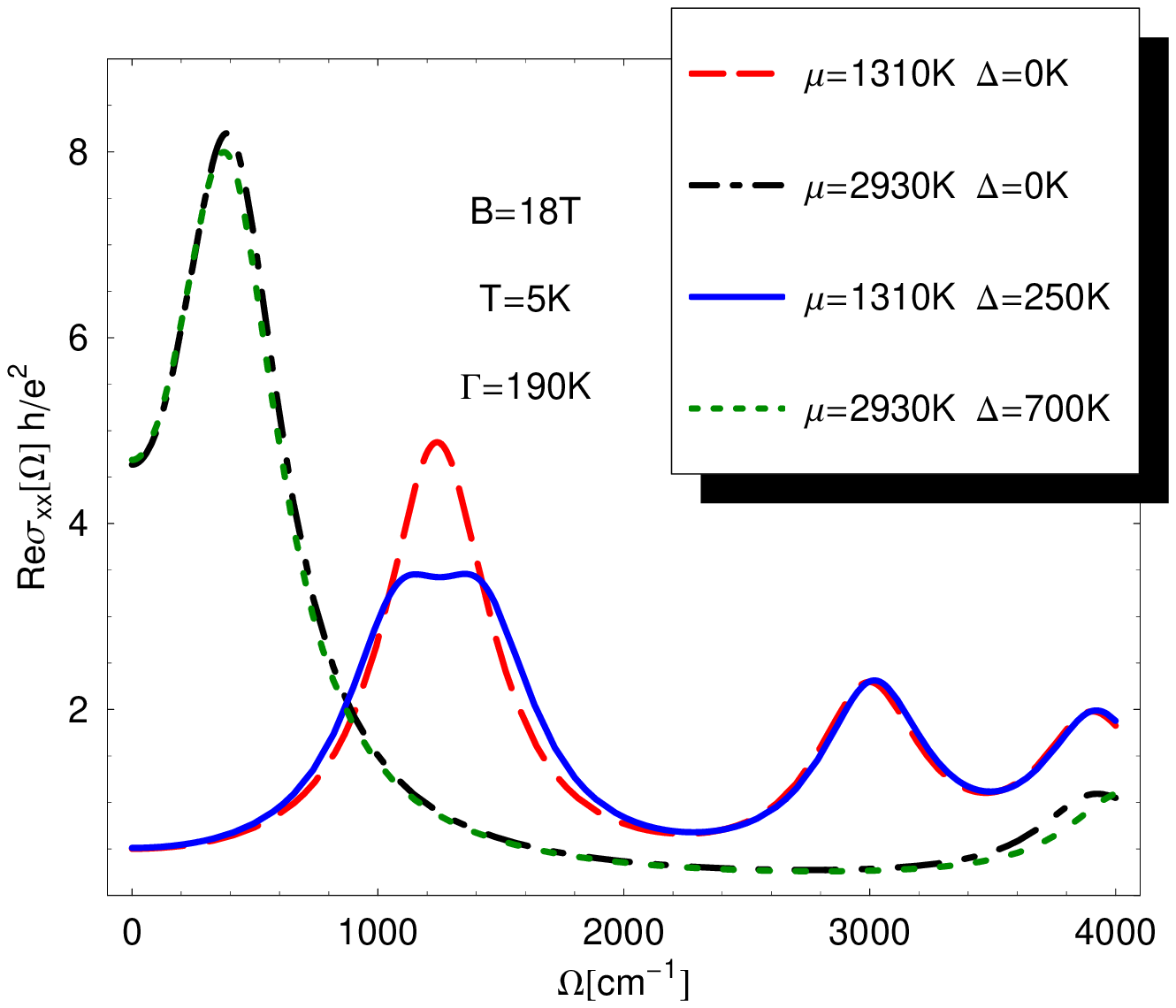,width=2.4in}}
\vspace*{8pt} \caption{(Colour online) Real part of the longitudinal
conductivity, $\mbox{Re} \, \sigma_{xx}(\Omega)$ in units of $e^2/h$
vs frequency $\Omega$ in $\mbox{cm}^{-1}$ for field $B=18\,
\mbox{T}$, temperature $T=5 \mbox{K}$. Long dashed, the chemical
potential $\mu = 1310 \mbox{K}$ and the gap $\Delta = 0\, \mbox{K}$,
dash-dotted $\mu = 2930 \mbox{K}$ and $\Delta = 0\, \mbox{K}$, solid
$\mu = 1310 \mbox{K}$ and $\Delta = 250\, \mbox{K}$, short dashed
$\mu = 2930 \mbox{K}$ and $\Delta = 700\, \mbox{K}$. For the left
frame the scattering rate $\Gamma =60 \mbox{K}$ and for the right
frame scattering rate $\Gamma =190 \mbox{K}$. } \label{fig:5}
\end{figure}
In Fig.~\ref{fig:5} we show results for the real part of the
longitudinal conductivity $\mbox{Re} \, \sigma_{xx}(\Omega)$ for
$B=18\, \mbox{T}$, $T=5 \mbox{K}$ and two possible Landau level
half-width $\Gamma=60 \, \mbox{K}$ (roughly 4 times larger than that
observed in dc measurements for $\nu=4$ plateaux in
Ref.~\refcite{Zhang2006PRL}) and $\Gamma=190 \, \mbox{K}$, which
corresponds to the observed (in Ref.~\refcite{Jiang2007}) $20\,
\mbox{fs}$ half-widths of the transmission peaks. We show results
for two values of chemical potential $\mu=1310 \, \mbox{K}$ and
$\mu=2930 \, \mbox{K}$. The first corresponds to the line $T_1$ in
Ref.~\refcite{Jiang2007} measured at the filling factor $\nu=2$ with
the carrier imbalance $\rho =9.4\times 10^{11} \, \mbox{cm}^{-2}$,
and falls  between $M_0$ and $M_1$ Landau level energies in the
notation of Eq.~(\ref{sigma_xx-complex-corrected}). The second
corresponds to the line $T_2$ in Ref.~\refcite{Jiang2007}, $\nu=10$
with the carrier imbalance $\rho =4.7\times 10^{12} \,
\mbox{cm}^{-2}$ and falls between $M_2$ and $M_3$ Landau level
energies.\footnote{The actual measurements in
Ref.~\refcite{Jiang2007} were made for holes with $\nu = -2$ and
$-10$, but from theoretical consideration the particle-hole symmetry
is obeyed because the conductivity
(\ref{sigma_xx-complex-corrected}) is an even function of $\mu$.}
The long dashed (red) curve includes no gap $\Delta$. It displays an
interband absorption peak at $\Omega = 1248 \, \mbox{cm}^{-1}$
corresponding to the transition from $n=0$ to $n=1$ levels, with a
second at $\Omega = 3013 \, \mbox{cm}^{-1}$ (from $-M_1$ to $M_2$
and $-M_2$ to $M_1$), a third at $\Omega = 3927 \, \mbox{cm}^{-1}$
(from $-M_2$ to $M_3$ and $-M_3$ to $M_2$) etc.  The solid (blue)
line is for the same case, but now a gap $\Delta = 250 \, \mbox{K}$
is introduced to illustrate its effect on the absorption line. We
see no change in the higher energy peaks but the peak at $\Omega =
1248 \, \mbox{cm}^{-1}$ is split into two, one shifted to lower
energies and the other to higher energies. Each one is broadened by
the peak half-width $2 \Gamma = 120\, \mbox{K}$. While our
parameters were chosen with the experiments of Jiang {\em et
al.}\cite{Jiang2007} in mind, no splitting of the first interband
line is observed, which is consistent with the fact that their
$\Gamma$ is considerably larger. If we increase $\Gamma$ instead to
their value $190\, \mbox{K}$, the results are shown in the right
hand frame, which differs from the left hand counterpart only
through the value of $\Gamma$. We see broader peaks associated with
the various optically induced allowed transitions between Landau
levels. Also, the broadening of the first interband line in the
solid (blue) curve is now sufficient, so that a single peak is seen.
However, it does not have the single Lorentzian lineshape of the
dashed (red) line. There is a slight depression at the position of
the original peak, a flat top and slight overall broadening of the
line. There is no visible increase in the distance between the first
and second absorption peak.

We have included in Fig.~\ref{fig:5}, for comparison, results for
another value of chemical potential $\mu=2930 \, \mbox{K}$. This
corresponds to the case when $\mu$ lies between the second and third
Landau levels. As described in Ref.~\refcite{Gusynin2007PRL}, in
this instance the first interband line has faded into the background,
as has the second, and the third has halved its original intensity
(i.e. compared with the long dashed (red) curve). This is seen in
the dash-dotted (black) curve which has $\Delta=0$. Note that a new
line appears at lower energies $\Omega = 397\, \mbox{cm}^{-1}$ which
corresponds to an intraband transition between the second and third
Landau levels. To observe even a small change in this curve in the
region above $\sim 3500\, \mbox{cm}^{-1}$, we need to include a large
gap $\Delta = 700 \, \mbox{K}$ as seen in the dotted (green) curve.

One of the important predictions of
Eq.~(\ref{sigma_xx-complex-corrected}) which has not been
addressed yet in any of the reports on optical experiments done so far,
is the behavior of the absorption lines as the chemical potential is
changed. As we saw in Fig.~\ref{fig:5} for $\mu$ falling in the
energy range between $n=0$ and $n=1$ Landau levels, there is a line
at $\Omega = M_1$. This line is anomalous in the sense that it
disappears when $\mu$ falls beyond this range. If $\mu$ falls
between $n=1$ and $n=2$, there will still be an interband line at
$M_1 + M_2$, but it will have only half the intensity it had in the
previous case. The other lines remain. If we further increase $\mu$
to fall between $M_2$ and $M_3$, this would correspond to the
dash-dotted (black) curve in Fig.~\ref{fig:5} discussed above, i.e.
the first and second interband lines have disappeared and the third
halved its intensity.

It is hoped that future experiments will verify this predicted
pattern of change with increase of $\mu$ and in particular verify
that the first interband line is anomalous.\cite{Gusynin2007PRL} It
appears at full intensity or not at all, while all other lines first
fall to half intensity before disappearing entirely. Its optical
spectral weight is transferred to an intraband line at $M_{n+1}
-M_n$.

We note that the present consideration is based on a perfect
symmetry between particles and antiparticles, which is built into the
Dirac formalism. Accordingly, as we saw above, two different
transitions can contribute equally  to the same optical line. A most
recent experiment\cite{Deacon2007} shows, however, clear evidence
for the breaking of particle-antiparticle symmetry in the graphene
system at the level of $± 2.5 \%$, approximately five times larger
than expected.

Coming back to the role of correlation effects, we mention another
recent paper by  Jiang {\em et al.}\cite{Kim2007APS} which also
points to possible limitations of an independent particle theory.
Based on magnetoresistance measurements in fields up to $45 \,
\mbox{T}$ it is concluded that the observed activation gap is much
smaller than expected on the basis of the known spacing between
Landau levels, which is set by the value of the Fermi velocity $v_F$.
These results point to a need to include in the theory, additional
effects which go beyond the independent particle model. In this work,
we have classified and discussed several possibilities related to
different Dirac masses and,  as an example, considered the effect of
an excitonic gap on the optical conductivity. We hope that this can
help when pursuing future developments.

\subsection{Sum rules for the optical and Hall conductivity}
\label{sec:sum}

It is instructive to consider the case when QED$_{2+1}$
approximation is not sufficient and one should consider a full
tight-binding model (\ref{Hamilton-lattice}). This example is given
by the optical conductivity sum rules.\cite{Gusynin2007PRB} The real
part of the frequency dependent optical conductivity
$\sigma_{xx}(\Omega)$ is its absorptive part and its spectral weight
distribution as a function of energy ($\hbar \Omega$) is encoded
with information on the nature of the possible electronic
transitions resulting from the absorption of a photon. Even though
the relationship of the conductivity to the electronic structure and
transport lifetimes is not straightforward, much valuable
information can be obtained from such data. In particular, the
$f$-sum rule on the real part of $\sigma_{xx}(\Omega)$ states that
(see e.g.
Refs.~\refcite{Millis:book,Marel:book,Carbotte:review,Benfatto:review}
for a review)
\begin{equation}
\label{sum-rule-sigma} \frac{1}{\pi}\int\limits_{-\infty}^\infty
d\Omega \mathrm{ Re} \sigma_{xx}(\Omega)=
\frac{\langle\tau_{xx}\rangle}{V},
\end{equation}
where $\langle\tau_{xx}\rangle$ is the thermal average of the
diamagnetic term (\ref{diamagnetic-term-average}). The optical
conductivity sum rule is a consequence of gauge invariance and
causality. Gauge invariance dictates the way that the vector
potential enters Eq.~(\ref{Hamilton-lattice}) and determines the
diamagnetic and paramagnetic terms in the expansion
(\ref{Hamitonian-expand}) as well as the form of Kubo formula
(\ref{Kubo}). Causality implies that the conductivity,
Eq.~(\ref{Kubo}) satisfies the Kramers-Kr\"onig  relation. Note that
defining the plasma frequency $\omega_P$ via
$\omega_P^2/(4\pi)\delta_{\alpha \beta} \equiv \langle\tau_{\alpha
\beta}\rangle/V$, we can rewrite the RHS of the sum rule
(\ref{sum-rule-sigma}) in terms of $\omega_P$.

For a finite single tight-binding band, one can explicitly calculate
the thermal average (\ref{diamagnetic-term-average}) and obtain the
following representation of the optical sum
rule\cite{Millis:book,Marel:book,Carbotte:review,Benfatto:review}
\begin{equation} \label{sum-rule-1band} \frac{2}{\pi}\int_{0}^{\Omega_M}
d\Omega \mbox{Re} \, \sigma_{xx}(\Omega) =  \frac{e^2}{\hbar^2V}
\sum_{\mathbf{k},\sigma} n_{\mathbf{k},\sigma} \frac{\partial^2
\epsilon_\mathbf{k}}{\partial k_x^2},
\end{equation}
with $\Omega_M$ being a cutoff energy on the band of interest and only the
contribution to $\mbox{Re} \, \sigma_{xx}(\Omega)$ of this
particular band is to be included in the integral. Here
$\epsilon_\mathbf{k}$ is the electronic dispersion, $\mathbf{k}$ is
the wave vector in the Brillouin zone, and $n_{\mathbf{k},\sigma}$
is the probability of occupation of the state $|\mathbf{k}, \sigma
\rangle$. For tight-binding dispersion with nearest neighbor hopping
on a square lattice, it is easy to show that the right-hand side
(RHS) of Eq.~(\ref{sum-rule-1band}) reduces to $e^2/\hbar^2$
multiplied by minus one-half of the kinetic energy,
$W_{\mathrm{K.E.}}$ per atom.

When a constant external magnetic field  is applied to a metallic
system in the $z$-direction, the optical conductivity acquires a
transverse component $\sigma_{xy}(\Omega)$ in addition to the
longitudinal component $\sigma_{xx}(\Omega)$. This quantity gives
additional information on the electronic properties modified by the
magnetic field. In this case,\cite{Drew1997PRL} there is a new sum
rule on the optical Hall angle $\theta_H(\Omega)$. If we define
\begin{equation}\label{t_H}
t_H(\Omega) \equiv \tan \theta_H(\Omega) =
\frac{\sigma_{xy}(\Omega)}{\sigma_{xx}(\Omega)},
\end{equation}
then
\begin{equation}
\label{sum-rule-Hall} \frac{2}{\pi} \int_0^\infty d \Omega
\mathrm{Re} t_H(\Omega) = \omega_H,
\end{equation}
where $\omega_H$ is the Hall frequency which corresponds to the
cyclotron frequency $\omega_c = eB/m c$ for free electrons.

\subsubsection{Limitations of the Dirac approximation}
\label{sec:Dirac-limitations}

The fact mentioned above that the effective low-energy Dirac
description of graphene is insufficient for the derivation of the
sum rules can be easily understood from two examples. As we have
seen the RHS of Eq.~(\ref{sum-rule-sigma}) is normally equal
to\cite{Millis:book,Marel:book,Benfatto:review,Carbotte:review}) the
thermal average of the diamagnetic term
(\ref{diamagnetic-term-average}) that is defined as the second
derivative of the Hamiltonian [see Eq.~(\ref{diamagnetic-term})
above] with respect to the vector potential. This term is zero if
one tries to obtain it from the  approximated Dirac Lagrangian
(\ref{Lagrangian}). On the other hand, it follows from the same
Dirac approximation that, in the high-frequency limit, the interband
contribution  to conductivity is
constant\cite{Ando2002JPSJ,Gusynin2006micro,Gusynin2007JPCM,Falkovsky2006}
given by Eq.~(\ref{diagonal-high-frequency}). The frequency $\Omega$
is unbounded in the Dirac approximation, although physically
$\Omega$ should be well below the band edge. This example indicates
that when considering sum rules, one should go beyond the Dirac
approximation.

The second example is that the cyclotron frequency $\omega_c$ for
the Dirac quasiparticles\cite{Zheng2002PRB} is defined in a
different way, $\omega_c = eB v^2_F/( c |\mu|)$, where $v_F$ is the
Fermi velocity. This definition follows from the fact that a
fictitious ``relativistic'' mass
\cite{Geim2005Nature,Kim2005Nature,Gusynin2005PRL} $m_c =
|\mu|/v_F^2$ plays the role of the cyclotron mass in the temperature
factor of the Lifshits-Kosevich formula for
graphene.\cite{Sharapov2004PRB,Gusynin2005PRB} This cyclotron
frequency diverges as $\mu \to 0$, also posing the question as what
one should use as a Hall frequency on the RHS of
Eq.~(\ref{sum-rule-Hall}).

\subsubsection{Diagonal conductivity sum rule}
\label{sec:diagonal-sum}

These problems were resolved in Ref.~\refcite{Gusynin2007PRB}, where
considering a tight-binding model (\ref{Hamilton-lattice}) we
obtained the RHS of the sum rule (\ref{sum-rule-sigma}) to be
equal to
\begin{equation} \label{sum-rule-phase}
\frac{\langle\tau_{\alpha
\alpha}\rangle}{V}=\frac{2e^2}{\hbar^2}\int_{BZ}\frac{d^2{\bf
k}}{(2\pi)^2}\left[n_F\left( \epsilon(\mathbf{k})
\right)-n_F(-\epsilon(\mathbf{k}))\right]\left[\frac{\partial^2}{\partial
k_\alpha^2} -\left(\frac{\partial\varphi(\mathbf{k})}{\partial
k_\alpha}\right)^2\right]\epsilon(\mathbf{k}).
\end{equation}
Here $n_{F}(\omega) =1/[\exp((\omega-\mu)/T)+1]$ is the Fermi
distribution, the phase $\varphi(\mathbf{k})$ is defined below
Eq.~(\ref{H_0}) and the energy $\epsilon(\mathbf{k})$ is given by
Eq.~(\ref{dispersion}). The momentum integration in
Eq.~(\ref{sum-rule-phase}) is over the entire BZ and the thermal
factors $n_F(\epsilon(\mathbf{k}))$ and $n_F(-\epsilon(\mathbf{k}))$
refer to the upper and lower Dirac cones (see Fig.~\ref{fig:2}),
respectively. We stress that a simple generalization of
Eq.~(\ref{sum-rule-1band}) for a two band case would miss the term
with the derivative of the phase,
$\left(\partial\varphi(\mathbf{k})/\partial k_\alpha\right)^2$. This
term occurred due to the fact that the Peierls substitution was made
in the initial Hamiltonians (\ref{Hamilton-lattice}) and (\ref{H_0})
rather than after the diagonalization of Eq.~(\ref{H_0}). The second
comment on Eq.~(\ref{sum-rule-phase}) is that $\langle\tau_{\alpha
\alpha}\rangle$ vanishes if $\epsilon(\mathbf{k})$ is taken in the
linear approximation. This reflects the absence of the diamagnetic
term in the Dirac approximation. The correct way is to firstly  take
the derivatives in Eq.~(\ref{sum-rule-phase}). This leads to the
final result\cite{Gusynin2007PRB}
\begin{equation} \label{sum-rule-main}
\frac{\langle\tau_{\alpha \alpha}\rangle}{V}
=-\frac{e^2a^2}{3\hbar^2}\int_{BZ}\frac{d^2{\bf
k}}{(2\pi)^2}\left[n_F(\epsilon(\mathbf{k})) -
n_F(-\epsilon(\mathbf{k}))\right] \epsilon(\mathbf{k}).
\end{equation}
Eq.~(\ref{sum-rule-main}) is equivalent to
Eq.~(\ref{sum-rule-phase}). Note that $\langle\tau_{\alpha
\alpha}\rangle$ is always positive and does not depend on the
arbitrary choice of the sign before $t$ in
Eq.~(\ref{Hamilton-lattice}). Now Eq.~(\ref{sum-rule-main}) is
$e^2/\hbar^2$ times $-2/(3\sqrt{3})(\sim - 0.39)$ of the kinetic
energy per atom instead of $-1/2$ for the usual square lattice.

It is useful to separate explicitly the contribution
$\langle\tau_{xx}(\mu=T=0)\rangle$ of the Dirac sea
 from Eq.~(\ref{sum-rule-main}):
\begin{equation}
\langle\tau_{xx}\rangle = \langle\tau_{xx}(\mu=T=0)\rangle +
\langle\tau_{xx}^{eh}(\mu,T)\rangle,
\end{equation}
where
\begin{equation}
\label{Dirac-sea} \frac{\langle\tau_{xx}(\mu=T=0)\rangle}{V}
=-\frac{e^2a^2}{3\hbar^2}\int_{BZ}\frac{d^2{\bf k}}{(2\pi)^2} (-
\epsilon(\mathbf{k}))
\end{equation}
is the contribution of the Dirac sea (the energy of the filled
valence band) and
\begin{equation}
\label{e-h}
\frac{\langle\tau_{xx}^{eh}(\mu,T)\rangle}{V}=-\frac{e^2a^2}{3\hbar^2}\int_{BZ}\frac{d^2{\bf
k}}{(2\pi)^2}\left[n_F(\epsilon(\mathbf{k})) + 1-
n_F(-\epsilon(\mathbf{k}))\right] \epsilon(\mathbf{k})
\end{equation}
is the electron-hole contribution. The numerical calculation of the
Dirac sea contribution (\ref{Dirac-sea}) with the full dispersion
(\ref{dispersion}) gives
\begin{equation}
\label{Dirac-sea-num} \frac{\langle\tau_{xx}(\mu=T=0)\rangle}{V} =
\alpha \frac{e^2 t}{\hbar^2}, \qquad \alpha \approx 0.61.
\end{equation}
The same answer also follows from the linearized Dirac approximation
with the trigonal density of states if the band width $W$ is given
by Eq.~(\ref{cutoff-energy}). The electron-hole contribution
(\ref{e-h}) can be estimated analytically in the linear
approximation for the dispersion law. In particular, for $\mu=0$ the
temperature dependence of the diagonal conductivity sum rule is
$\sim T^3$, in contrast to the often found $T^2$
dependence\cite{Millis:book,Marel:book,Benfatto:review,Carbotte:review}
in tight-binding model of conventional quasiparticles.

On the other hand, for $|\mu| \gg T$ one obtains
\begin{equation}
\label{e-h-approx} \frac{\langle\tau_{xx}^{eh}(\mu,T)\rangle}{V}=-
\frac{e^2a^2}{9\pi\hbar^4v_F^2} \left[|\mu|^3+\pi^2|\mu|T^2 \right].
\end{equation}
We note that the $|\mu|^3$ behavior can be observed in an
experimental configuration which has very recently been used in
Ref.~\refcite{Jiang2007} by incorporating the specimen into a
field-effect device. In this case the chemical potential $\mu$ is
easily changed by varying the gate voltage $V_g$. Unless $\mu$ is
chosen to be large, the change in the sum rule from its $\mu=0$
value is small [$\sim (|\mu|/t)^3$]. Using Eq.~(\ref{rho}) for
carrier imbalance $\rho$ which is proportional to $V_g$, we obtain
that $\mu \varpropto \mbox{sgn}\,V_g \sqrt{|V_g|}$ and, therefore,
$\langle\tau_{xx}^{eh}(V_g)\rangle/V \sim -|V_g|^{3/2}$.

\subsubsection{Hall conductivity sum rule}

The RHS of the Hall-angle sum rule (\ref{sum-rule-Hall}) for
graphene\cite{Gusynin2007PRB} is equal to
\begin{equation}
\label{omega_H-final} \omega_H = - \frac{1}{4\alpha} \frac{eB}{c}
\frac{t a^2}{ \hbar^2}\rho a^2.
\end{equation}
Here $\rho$ is the carrier imbalance ($\rho = n_e - n_h$), where
$n_e$ and $n_h$ are the densities of electrons and holes, $\alpha$
is the numerical constant from Eq.~(\ref{Dirac-sea-num}). The
carrier imbalance for $B=T=0$ and in the absence of impurities is
\begin{equation} \label{rho} \rho = \frac{\mu^2 \mbox{sgn}
\mu}{\pi \hbar^2 v_F^2}.
\end{equation}
Since $t a^2/\hbar^2$ has the dimensionality of the inverse mass and
$\rho a^2$ is dimensionless, Eq.~(\ref{omega_H-final}) has the
correct dimensionality of the Hall frequency. Substituting
Eq.~(\ref{rho}), expressing $v_F$ via $t$ and employing the Landau
scale $L(B)$ used in Eq.~(\ref{L-scale}), one can rewrite
\begin{equation}
\label{omega_H-final-K} \omega_H = - \frac{4\,\mbox{sgn}\,(eB)}{9
\pi\alpha} L^2(B) \frac{\mu^2 \mbox{sgn} \mu}{\hbar t^3 }.
\end{equation}

As mentioned in Sec.~\ref{sec:Dirac-limitations}, in the recent
interpretation of Shubnikov de Haas measurements,  a gate
voltage-dependent cyclotron mass was introduced
\cite{Geim2005Nature,Kim2005Nature} through the relationship $|\mu|
= m_c v_F^2$. If this is used in Eq.~(\ref{omega_H-final-K}), we get
\begin{equation}
\label{omega_H-filled} \omega_H = -\frac{eB}{c m_c}
\left(\frac{\mu}{1.62 t}\right)^3.
\end{equation}
Since a full upper Dirac band corresponds to a value $\mu =
W=\sqrt{\sqrt{3} \pi} t$ (see Eq.~(\ref{cutoff-energy})), in this
case formula (\ref{omega_H-filled}) resembles the formula $\omega_H
= \omega_c = eB/mc$ from Ref.~\refcite{Drew1997PRL} for a
two-dimensional electron gas with $m_c$ replacing the free electron
mass. In graphene, however, $m_c$ varies as the square root of the
carrier imbalance $|\rho|$ and the two cases look the same only
formally.

When the spectrum becomes gapped with $E = \pm \sqrt{\hbar^2
v_F^2\mathbf{p}^2 + \Delta^2}$, where $\Delta$ is the magnitude of
one of the Dirac masses (\ref{masses-Dirac}), the carrier imbalance
is
\begin{equation}
\label{rho-gap} \rho = \frac{1}{\pi \hbar^2 v_F^2}(\mu^2 -
\Delta^2)\theta(\mu^2 - \Delta^2)\mbox{sgn} \mu.
\end{equation}
This implies that the gap $\Delta$ can be extracted from the change
in $\omega_H$ obtained from  magneto-optical measurements. This kind
of measurement  which reveals gapped behavior has  already been
done\cite{Rigal2004PRL} on the underdoped high-temperature
superconductor YBa$_2$Cu$_3$O$_{6 + x}$.

\section{Conclusion}
\label{sec:concl}

The number of articles devoted to graphene grows very fast and the
seminal experimental papers where the IQHE was
reported\cite{Geim2005Nature,Kim2005Nature} have already been cited
almost 300 times in arXiv, while the total list of references of our
review has about 100 papers. This is partly caused by the
restriction on the size of the manuscript and by the amount of time
we decided to invest in this project. We intentionally omitted such
interesting and important topics as bilayer and multilayer graphene,
the role of impurities,  defects and their gauge field description,
minimal conductivity at the Dirac point, edge states, weak
(anti)localization, fractional quantum Hall effect, graphene
nanoribbons, etc. Each of these topics may,  in the future, become a
subject of a review on its own. Instead we decided to restrict
ourselves to the specific topics covered in the review and to
discuss them in detail. We hope that this choice is indeed
complementary to already existing textbooks and
reviews.\cite{review1,review2,Saito:book,Ando2005JPSJ}

During our work on the review, we found that in the literature on
graphene,  the word {\it ``chirality''} is becoming as fashionable as
{\it ``Berry's phase''}, but depending on researcher's background, it
is used to denote different concepts, so that it is getting hard to
grasp its meaning. Here we followed the  commonly accepted
definition in  field theory\cite{Itzykson.book} and demonstrated
that the chirality quantum number corresponds to the valley index in
graphene. In Sec.~\ref{sec:chirality} we also introduced the {\em
pseudohelicity\/} which characterizes the projection of a
quasiparticle pseudospin on the direction of its momentum and
explained it specificifically in $2+1$ dimension  and the relationship
between pseudohelicity and chirality. We hope that this solid state
reincarnation of chirality will lead, in  future, to a deeper
understanding of chirality in field theory.

It is also very instructive to see how the spinors emerge from the
two atom per unit cell description of graphene. Because these
spinors are not related to a real spin, the corresponding degree of
freedom is called pseudospin. This solid state realization of
pseudospin may help us understand the origin of a real spin better.

Also we  mentioned a link between the QED$_{2+1}$ description of
graphene and $d$-wave superconductivity. In both cases, the
low-energy quasiparticle excitations are described by massless
QED$_{2+1}$. Massless excitations in graphene are often compared
with  massless neutrinos. However, as we pointed out in
Sec.~\ref{sec:discrete-difference}, this analogy in not complete,
because their discrete symmetries are different. Massless
QED$_{2+1}$ has  $U(4)$ symmetry and a rather common way in the
field theory  to break this symmetry is a dynamical generation of
the Dirac mass. It is unclear at this stage whether this rather
generic mechanism works in graphene or the observed new $\pm 1$ QHE
states can be explained by other mechanisms. From our point of view,
this is one of the most interesting open questions which demands
further experimental and theoretical work. Accordingly, in our
review we simply tried to facilitate the future theoretical work by
providing  a classification for possible Dirac masses.

\section*{Acknowledgements}

We thank D.~Basov, L.~Benfatto, L.~Brey, A.K.~Geim, M.O.~Goerbig,
E.V.~Gor\-bar, E.~Henriksen, I.F.~Her\-but, P.I.~Holod, A.~Iyengar,
P.~Kim, D.V.~Khve\-shchen\-ko, Z.~Li, V.M.~Loktev, A.H.~MacDonald,
V.A.~Miransky, I.A.~Shovkovy and M.A.H.~Vozmediano for illuminating
discussions. The work of V.P.G. was supported by the SCOPES-project
IB 7320-110848 of the Swiss NSF, the grant 10/07-N "Nanostructure
systems, nanomaterials, nanotechnologies", and by the Program of
Fundamental Research of the Physics and Astronomy Division of the
National Academy of Sciences of Ukraine and by Ukrainian State
Foundation for Fundamental Research. J.P.C. and S.G.Sh. were
supported by the Natural Science and Engineering Research Council of
Canada (NSERC) and by the Canadian Institute for Advanced Research
(CIFAR). S.G.Sh. is also grateful to the Kavli Institute for
Theoretical Physics at the University of California Santa Barbara
(the National Science Foundation under Grant No. PHY05-51164) for
hospitality during its graphene workshop.

\appendix{Solution of the Dirac equation in the Landau gauge}
\label{sec:Appendix}

The Dirac equation in symmetric gauge and in the representation
(\ref{gamma-old}) is solved, for example, in Appendix~D of the
second paper in Ref.~\refcite{Gusynin2005PRL} (see also
Ref.~\refcite{Melrose} for a solution of 3D Dirac equation in the
Landau and symmetric gauges). For $e B , \Delta >0$ in the Landau
gauge $\mathbf{A} = (0,Bx)$ the positive
($\psi^{(+)}_{\mathbf{K}_{+}}$) and negative
($\psi^{(-)}_{\mathbf{K}_{+}}$) energy solutions of
Eq.~(\ref{Dirac-excitonic-gamma1}) at $\mathbf{K}_{+}$ point are
\begin{eqnarray}
&&\psi^{(+)}_{\mathbf{K}_{+}}(x,y;n,p)=\frac{1}{\sqrt{2\pi
l}}e^{-i|E_{n}|t+ipy}\frac{1}{\sqrt{2|E_{n}|}}
\left(\begin{array}{c}\sqrt{|E_{n}|+\Delta}\,w_{n-1}(\xi) \nonumber \\
i\sqrt{|E_{n}|-\Delta}\,w_{n}(\xi)\end{array}\right),\quad n\geq1,\\
&&\psi^{(-)}_{\mathbf{K}_{+}}(x,y;n,p)=\frac{1}{\sqrt{2\pi
l}}e^{i|E_{n}|t+ipy}\frac{1}{\sqrt{2|E_{n}|}}
\left(\begin{array}{c}\sqrt{|E_{n}|-\Delta}\,w_{n-1}(\xi)\nonumber \\
-i\sqrt{|E_{n}|+\Delta}\,w_{n}(\xi)\end{array}\right),\quad n\geq1,\\
&&\psi^{(-)}_{\mathbf{K}_{+}}(x,y;n=0,p)=\frac{1}{\sqrt{2\pi
l}}e^{i|E_{0}|t+ipy}
\left(\begin{array}{c}0\\
-i w_{0}(\xi)\end{array}\right),\quad n=0.
\end{eqnarray}
The solutions of Eq.~(\ref{Dirac-excitonic-gamma2}) at
$\mathbf{K}_{-}$ point are
\begin{eqnarray}
&&\psi^{(+)}_{\mathbf{K}_{-}}(x,y;n=0,p)=\frac{1}{\sqrt{2\pi
l}}e^{-i|E_{0}|t+ipy}
\left(\begin{array}{c}0\\
iw_{0}(\xi)\end{array}\right),\quad n=0, \\
&&\psi^{(+)}_{\mathbf{K}_{-}}(x,y;n,p)=\frac{1}{\sqrt{2\pi
l}}e^{-i|E_{n}|t+ipy}\frac{1}{\sqrt{2|E_{n}|}}
\left(\begin{array}{c}\sqrt{|E_{n}|-\Delta}w_{n-1}(\xi)\\
i\sqrt{|E_{n}|+\Delta}w_{n}(\xi)\end{array}\right),\quad n\geq1,\nonumber \\
&&\psi^{(-)}_{\mathbf{K}_{-}}(x,y;n,p)=\frac{1}{\sqrt{2\pi
l}}e^{i|E_{n}|t+ipy}\frac{1}{\sqrt{2|E_{n}|}}
\left(\begin{array}{c}-\sqrt{|E_{n}|+\Delta}w_{n-1}(\xi)\\
i\sqrt{|E_{n}|-\Delta}w_{n}(\xi)\end{array}\right),\quad n\geq1.
\nonumber
\end{eqnarray}
Here $\xi = x/l + p l$, where $l\equiv(\hbar c/|eB|)^{1/2}$ is the
magnetic length and $p$ is the momentum along the $y$ axis,
\begin{eqnarray} && E_{n}=\pm\sqrt{\Delta^{2}+\hbar
v_{F}^{2}2n|eB|/c}, \\
&&w_{n}(\xi)=(\pi^{1/2}2^{n}n!)^{-1/2} e^{-\xi^2/2}H_n(\xi), \quad
\int\limits_{-\infty}^{\infty} d\xi w_{n}(\xi)w_{m}(\xi)=\delta_{nm}
\nonumber
\end{eqnarray}
with $H_n(\xi)$ being the Hermite polynomial and we defined
$w_{-1}(\xi)\equiv0$. The lowest Landau level $n=0$ is special:
while at $n\geq1$, there are solutions corresponding to both
electron ($E_{n}>0$) and hole ($E_{n}<0$) states at both
$\mathbf{K}_{+}$ and $\mathbf{K}_{-}$ points, the solution with
$n=0$ describes holes at $\mathbf{K}_{+}$ and electrons at
$\mathbf{K}_{-}$ point. The corresponding structure of Landau levels
is shown in Fig.~\ref{fig:3}~(c). See also Fig.~2  of
Ref.~\refcite{Bernevig2006IJMPB} for a graphical representation of
the dependence of the LLL position on the relative signs of $eB$ and
$\Delta$. One can also say that for $eB>0$ the solution of
Eq.~(\ref{Dirac-excitonic-gamma1}) for LLL at $\mathbf{K}_{+}$
describes holes that live on the $\mathrm{B}$ sublattice, while the
solution of Eq.~(\ref{Dirac-excitonic-gamma2}) for LLL at
$\mathbf{K}_{-}$ corresponds to electrons living on the $\mathrm{A}$
sublattice (we remind that in the $\mathbf{K}_{-}$ valley, the spinor
components are inverted in comparison to the spinor in the
$\mathbf{K}_{+}$ valley). Finally, we notice that the relation
in Eq.~(\ref{tau_2}) is violated in the finite field. Indeed
\begin{equation}
\label{tau_2-gen} \tau_2\mathcal{H}^\ast_0(k,eB)\tau_2 = -
\mathcal{H}_0(k,-eB), \quad \mathcal{H}_0(k,eB)= \tau_1k_x +
\tau_2(k_y +eBx)+\Delta\tau_3.
\end{equation}
This indicates that the wave function $\tau_2 | \psi\rangle^\ast$ is
the solution of the equation $\mathcal{H}_0(k,-eB)\tau_2 |\psi
\rangle^\ast = -E \tau_2| \psi \rangle^\ast$, i.e. the
generalization (\ref{tau_2-gen}) of the relation (\ref{tau_2}) for a
finite magnetic field does not describe the symmetry of the spectrum
about $E=0$, but rather relates the solutions of
Eqs.~(\ref{Dirac-excitonic-gamma1}) and
(\ref{Dirac-excitonic-gamma2}) obtained for $eB>0$ with the
solutions of the same equations for $eB<0$.


\section*{References}

\begin{thebibliography}{00}


\bibitem{Semenoff1984PRL} G.W.~Semenoff,
{\it Phys. Rev. Lett.} {\bf 53}, 2449 (1984).

\bibitem{Novoselov2004Science} K.S.~Novoselov, A.K.~Geim, S.V.~Morozov,
D.~Jiang, Y.~Zhang, S.V.~Dubonos, I.V.~Grigorieva, and A.A.~Firsov,
{\it Science} {\bf 306}, 666 (2004); K.S.~Novoselov, D.~Jiang,
T.~Booth, V.V.~Khotkevich, S.M.~Morozov, and A.K.~Geim, {\it
Proc.~Nat.~Acad.~Sc.} {\bf 102}, 10451 (2005).

\bibitem{Fradkin1986PRL} E.~Fradkin, E.~Dagotto, and D.~Boyanovsky,
{\it Phys. Rev. Lett.} {\bf 57}, 2967 (1986), {\it ibid} {\bf 58},
961(E) (1987); D.~Boyanovsky, E.~Dagotto, and E.~Fradkin, {\it Nucl.
Phys. B} {\bf 285}[FS19], 340 (1987).

\bibitem{Haldane1988PRL} F.D.M.~Haldane, {\it Phys. Rev. Lett.} {\bf 61}, 2015 (1988).

\bibitem{Schakel1991PRD} A.M.J.~Schakel, {\it Phys.~Rev.~D} {\bf 43}, 1428
(1991).


\bibitem{Gonzales1993NP} J.~Gonz{\`a}lez, F.~Guinea, and M.A.H.~Vozmediano,
{\it Nucl.~Phys. B} {\bf 406}, 771 (1993); {\em ibid.} {\bf 424},
595 (1994).

\bibitem{Gonzales1996PRL} J.~Gonz{\`a}lez, F.~Guinea, and M.A.H.~Vozmediano,
{\it Phys. Rev. Lett.} {\bf 77}, 3589 (1996); {\it Phys. Rev. B}
{\bf 59}, R2474 (1999); {\it Phys. Rev. B} {\bf 63}, 134421 (2001).


\bibitem{Khveshchenko2001PRL} D.V.~Khveshchenko,
{\it Phys. Rev. Lett.} {\bf 87}, 206401 (2001); {\em ibid.} {\bf
87}, 246802 (2001).

\bibitem{Gorbar2002PRB} E.V.~Gorbar, V.P.~Gusynin, V.A.~Miransky,
and I.A.~Shovkovy, {\it Phys. Rev. B} {\bf 66}, 045108 (2002).

\bibitem{Ludwig1994PRB} A.W.W.~Ludwig, M.P.A.~Fisher, R.~Shankar and
G.~Grinstein, {\it Phys. Rev. B} {\bf 50}, 7526 (1994).


\bibitem{Hou2007PRL} C.-Y.~Hou, C.~Chamon, C.~Mudry,
{\it Phys. Rev. Lett.} {\bf 98}, 186809  (2007).

\bibitem{Jackiw2007} R.~Jackiw and S.-Y.~Pi, Phys. Rev. Lett. {\bf 98}, 266402 (2007).

\bibitem{Herbut2007frac} I.F.~Herbut, arXiv:0704.2234.

\bibitem{Geim2005Nature}
K.S.~Novoselov, A.K.~Geim, S.V.~Morozov, D.~Jiang, M.I.~Katsnelson,
I.V.~Grigorieva, S.V.~Dubonos, and A.A.~Firsov, {\it Nature} {\bf 438},
197 (2005).

\bibitem{Kim2005Nature} Y.~Zhang, Y.-W.~Tan, H.L.~Stormer, and P.~Kim, {\it Nature} {\bf 438}, 201
(2005).


\bibitem{Zheng2002PRB} Y.~Zheng and T.~Ando, {\it Phys. Rev. B} {\bf 65}, 245420 (2002).

\bibitem{Gusynin2005PRL} V.P.~Gusynin and S.G.~Sharapov,
{\it Phys. Rev. Lett.} {\bf 95}, 146801 (2005);
{\it Phys. Rev. B} {\bf 73}, 245411 (2006).

\bibitem{Peres2006PRB} N.M.R. Peres, F. Guinea, and A.H. Castro Neto,
{\it Phys. Rev. B} {\bf 73}, 125411 (2006).

\bibitem{review1} A.K.~Geim and K.S.~Novoselov, {\it Nature Materials} {\bf 6}, 183
(2007).

\bibitem{GeimKim2007Science} K.S.~Novoselov, Z.~Jiang, Y.~Zhang, S.V.~Morozov, H.L.~Stormer,
U.~Zeitler, J.C.~Maan, G.S.~Boebinger, P.~Kim, and A.K.~Geim, {\it
Science} {\bf 315}, 1379 (2007).


\bibitem{Katsnelson2006NatPhys}
M.I.~Katsnelson, K. S. Novoselov and A.K.~Geim, {\it Nature Phys.}
{\bf 2}, 620 (2006).

\bibitem{Calogeracos2006NatPhys} A.~Calogeracos, {\it Nature Phys.}
{\bf 2}, 579 (2006).

\bibitem{review2} M.I.~Katsnelson and K.S.~Novoselov,
{\it Solid State Comm.} {\bf 143}, 3 (2007).


\bibitem{Cserti2006PRB} J.~Cserti and G.~D\'{a}vid, {\it Phys. Rev. B} {\bf 74}, 172305
(2006).


\bibitem{Wu2007PRL} X.~Wu, X.~Li, Z.~Song, C.~Berger, and
W.A.~de~Heer, {\it Phys. Rev. Lett.} {\bf 98}, 136801 (2007).


\bibitem{Zhang2006PRL} Y.~Zhang, Z.~Jiang, J.P.~Small, M.S.~Purewal,
Y.-W.~Tan, M.~Fazlollahi, J.D.~Chudow, J.A.~Jaszczak, H.L.~Stormer, and
P. Kim, {\it Phys. Rev. Lett.} {\bf 96}, 136806 (2006).

\bibitem{Abanin2007} D.A.~Abanin, K.S.~Novoselov, U.~Zeitler, P.A.~Lee, A.K.~Geim,
L.S.~Levitov, {\it Phys. Rev. Lett.} {\bf 98}, 196806 (2007).

\bibitem{Kim2007APS} Z.~Jiang, Y.~Zhang, Y.~Tan, H. Stormer, and P.~Kim,
Phys. Rev. Lett. {\bf 99}, 106802 (2007).

\bibitem{Jiang2007} Z.~Jiang, E.A.~Henriksen, L.C.~Tung, Y.-J.~Wang,
M.E.~Schwartz, M.Y.~Han, P.~Kim, and  H.L.~Stormer, {\it Phys. Rev.
Lett.} {\bf 98}, 197403 (2007).

\bibitem{Kohn1961PRev} W.~Kohn, {\it Phys. Rev.} {\bf 123}, 1241
(1961).

\bibitem{Yang2007} K.~Yang, {\it Solid State
Comm.} {\bf 143}, 27 (2007).


\bibitem{Wilson2006PT} M.~Wilson, {\it Phys. Today}, January 2006, 21.

\bibitem{Neto2006PW} A.H.~Castro Neto, F.~Guinea, and N.M.~Peres, {\it Physics World} {\bf
19}, 33 November (2006).

\bibitem{Chakraborty2006PC} T.~Chakraborty, {\it Physics in Canada} {\bf 62}, 351 (2006).

\bibitem{Katsnelson2007MT} M.I.~Katsnelson, {\it Mater. Today} {\bf 10}, Issue 1\&2, 20 (2007).


\bibitem{Saito:book} R.~Saito, G.~Dresselhaus and M.S.~Dresselhaus,
{\em Physical Properties of Carbon Nanotubes}, Imperial College
Press, London, 1998.

\bibitem{Ando2005JPSJ} T.~Ando, {\it J.~Phys.~Soc.~Jpn.} {\bf 74}, 777
(2005).




\bibitem{Lomer1955PRS} W.H.~Lomer, {\it Proc.~Roy.~Soc.} {\bf A227}, 330
(1955).

\bibitem{Slonczewski1958PRev} J.C.~Slonczewski and P.R.~Weiss, {\it Phys.
Rev.} {\bf 109}, 272 (1958).

\bibitem{Millis:book} A. J. Millis, in {\it Strong Interactions in Low Dimensions}, edited by D.
Baeriswyl and L. De Giorgi (Kluver Academic, Berlin, 2003).

\bibitem{Meyer2007Nature} J.C.~Meyer, A.K.~Geim, M.I.~Katsnelson, K.S.~Novoselov,
T.J.~Booth, and S.~Roth, {\it Nature} {\bf 446}, 60 (2007).

\bibitem{Fukuyama2007JPSJ} H.~Fukuyama, {\it J.~Phys.~Soc.~Jpn.} {\bf
76} 043711 (2007).

\bibitem{Wallace1947PRev} P.R.~Wallace, {\it Phys.~Rev.} {\bf 77}, 622
(1947).

\bibitem{Deacon2007} R.S.~Deacon, K-C.~Chuang, R.J.~Nicholas, K.S.~Novoselov, and A.K.
Geim, Phys. Rev. B {\bf 76}, 081406(R) (2007).

\bibitem{Manes2007PRB} J.L.~Ma\~{n}es, F.~Guinea, and M.A.H.~Vozmediano,
{\it Phys. Rev. B} {\bf 75}, 155424 (2007).

\bibitem{Hatsugai2007} Y.~Hatsugai, T.~Fukui, and H.~Aoki,
cond-mat/0701431.

\bibitem{Giovannetti2007} G.~Giovannetti, P.A.~Khomyakov, G.~Brocks, P.J.~Kelly, and J.~van~den~Brink,
Phys. Rev B {\bf 76}, 073103 (2007).

\bibitem{Davydov.book} A.S.~Davydov, {\it Quantum Mechanics},
2nd Edition, Pergamon, New York, 1976.


\bibitem{Ryu2002PRL} S.~Ryu and Y.~Hatsugai,
{\it Phys. Rev. Lett.} {\bf 89}, 077002 (2002).

\bibitem{Ziegler2007} K.~Ziegler, cond-mat/0703628.

\bibitem{Volovik1}G.E.~Volovik, in {\it Quantum Analogues:
From Phase Transitions to Black Holes and Cosmology}, Eds.
W.G.~Unruh and R.~Schutzhold, Springer Lecture Notes in Physics
718/2007, pp. 31-73; cond-mat/0601372.

\bibitem{Gaididei2006FNT}
Yu.B.~Gaididei and V.M.~Loktev, {\it Fiz. Nizk. Temp.} {\bf 32}, 923
(2006) [{\it Low Temp. Phys.} {\bf 32}, 703 (2006).]


\bibitem{DiVincenzo1984PRB} D.P.~DiVincenzo and E.J.~Mele, {\it Phys. Rev. B} {\bf 29},
1685 (1984).


\bibitem{Khveshchenko2004nb} H. Leal and D.V.~Khveshchenko,
{\it Nucl. Phys. B} {\bf 687}, 323 (2004);
D.V.~Khveshchenko and W.F.~Shively, {\it Phys. Rev. B} {\bf 73},
115104 (2006).

\bibitem{Gusynin2006catalysis} V.P.~Gusynin, V.A.~Miransky, S.G.~Sharapov, and
I.A.~Shovkovy, {\it Phys. Rev. B} {\bf 74}, 195429 (2006);
cond-mat/0612488.


\bibitem{Herbut2006PRL} I.F. Herbut,
{\it Phys. Rev. Lett.} {\bf 97}, 146401 (2006);
{\it Phys. Rev. B} {\bf 75}, 165411 (2007).


\bibitem{McCann2004JPCM} E.~McCann and V.I.~Fal'ko, {\it J. Phys.: Condens. Matter.}
{\bf 16}, 2371 (2004).

\bibitem{Kane2005PRL} C.L.~Kane and E.J.~Mele,
{\it Phys. Rev. Lett.} {\bf 95}, 146802 (2005); {\it ibid.} {\bf
95}, 226801 (2005).

\bibitem{Cheianov2006PRL} V.V.~Cheianov and V.I.~Fal'ko,
{\it Phys. Rev. Lett.} {\bf 97},  226801  (2006); {\it Phys. Rev. B}
{\bf 74}, 041403(R) (2006).

\bibitem{McCann2006PRL} E.~McCann, K.~Kechedzhi, V.I.~Fal'ko, H.~Suzuura, T.~Ando, and
B.L.~Altshuler, {\it Phys. Rev. Lett.} {\bf 97}, 146805 (2006).

\bibitem{Ostrovsky2006PRB} P.M.~Ostrovsky, I.V.~Gornyi, and
A.D.~Mirlin, {\it Phys. Rev. B} {\bf 74}, 235443 (2006).

\bibitem{Khveshchenko2007PRB} D.V.~Khveshchenko,
{\it Phys. Rev. B} {\bf 75}, 153405 (2007).

\bibitem{Volovik2001PR} G.E.~Volovik, {\it Phys.~Rep.} {\bf 351}, 195 (2001).

\bibitem{Vafek2001PRB}
O.~Vafek, A.~Melikyan, M.~Franz, and Z.~Te\v{s}anovi\'{c}, {\it
Phys. Rev. B} {\bf 63}, 134509 (2001); M.~Franz,
Z.~Te\v{s}anovi\'{c}, and O.~Vafek, {\it Phys. Rev. B} {\bf 66},
054535 (2002).

\bibitem{Khveshchenko2001aPRL} D.V.~Khveshchenko and J.~Paaske, {\it Phys.~Rev.~Lett.} {\bf 86}, 4672
(2001).

\bibitem{Herbut2002PRB} I.F.~Herbut, {\it Phys.~Rev.~B} {\bf 66}, 094504 (2002).

\bibitem{Gusynin2004EPJB} V.P.~Gusynin and V.A.~Miransky,
{\it Eur.~Phys.~J.~B} {\bf 37}, 363 (2004).

\bibitem{Sharapov2006PRB} S.G.~Sharapov and J.P.~Carbotte, {\it Phys. Rev. B} {\bf 73}, 094519 (2006).


\bibitem{Alicea2006PRB} J.~Alicea and M.P.A.~Fisher, {\it Phys. Rev. B} {\bf 74}, 075422 (2006).

\bibitem{Khveshchenko2006PRB} D.V.~Khveshchenko, {\it Phys. Rev. B} {\bf 74}, 161402R
(2006).

\bibitem{Mishchenko2006} E.G.~Mishchenko, {\it Phys.~Rev.~Lett.} {\bf 98}, 216801
(2007).

\bibitem{Barlas2007} Y.~Barlas, T.~Pereg-Barnea, M.~Polini, R.~Asgari, and
A.H.~MacDonald, {\it Phys. Rev. Lett.} {\bf 98}, 236601 (2007).


\bibitem{Appelquist1986PRD}
T.~Appelquist, M.~Bowick, D.~Karabali and L.C.R.~Wijewardhana, {\it
Phys. Rev. D} {\bf 33}, 3704 (1986).

\bibitem{Ando1998JPSJ} T.~Ando, T.~Nakanishi, and R.~Saito, {\it J.~Phys.~Soc.~Jpn.} {\bf 67}, 2857
(1998); H.~Suzuura and T.~Ando, {\it Phys. Rev. Lett.} {\bf 89},
266603 (2002).


\bibitem{Itzykson.book} C.~Itzykson and J.-B.~Zuber, {\it Quantum Field Theory},
McGraw-Hill Inc., 1980.


\bibitem{Binegar1981JMP} B.~Binegar, {\it J.~Math.~Phys.} {\bf 23},
1511 (1981).

\bibitem{Boyanovsky1986NPB} D.~Boyanovsky, R.~Blankenbecler and
R.~Yahalon, {\it Nucl. Phys B.} {\bf 270}, 483 (1986).

\bibitem{McEuen1999PRL} P.L.~McEuen, M.~Bockrath, D.H.~Cobden, Y.-G.~Yoon, and
S.G.~Louie, {\it Phys. Rev. Lett.} {\bf 83}, 5098 (1999).

\bibitem{Jackiw1981PRD}
R.~Jackiw and S.~Templeton, {\it Phys. Rev. D} {\bf 23}, 2291
(1981).

\bibitem{Hosotani1993PLB} Y.~Hosotani, {\it Phys. Lett. B} {\bf
319}, 332 (1993);
{\it Phys. Rev. D} {\bf 51}, 2022 (1995).

\bibitem{Altland2002PR} A.~Altland, B.D.~Simons and M.R.~Zirnbauer,
{\it Phys.~Rep.} {\bf 359}, 283 (2002).

\bibitem{Garcia2006PRB} A.M.~Garcia-Garcia and E.~Cuevas, {\it Phys. Rev. B} {\bf 74},
113101 (2006).


\bibitem{Tchernyshyov2000PRB} O.~Tchernyshyov, Phys. Rev. B {\bf 62}, 16751 (2000).


\bibitem{Heersche2007Nature} H.B.~Heersche, P.~Jarillo-Herrero, J.B.~Oostinga,
L.M.K.~Vandersypen  and  A.F.~Morpurgo, Nature {\bf 446}, 56 (2007).

\bibitem{Titov2006PRB} M.~Titov and C.W.J.~Beenakker, {\it Phys.~Rev.~B} {\bf 74},
041401(R) (2006).


\bibitem{Abanin2006PRL} D.A.~Abanin, P.A.~Lee, and L.S.~Levitov, {\it Phys. Rev. Lett.} {\bf
96}, 176803 (2006), {\em ibid.} {\bf 98}, 156801 (2007); {\it Solid
State Comm.} {\bf 143}, 77 (2007).

\bibitem{Nomura2006PRL} K. Nomura and A. H. MacDonald, {\it Phys. Rev.
Lett.} {\bf 96}, 256602 (2006).

\bibitem{Yang2006PRB} K.~Yang, S.~Das~Sarma, and A. H. MacDonald, {\it Phys. Rev. B} {\bf 74}, 075423 (2006).

\bibitem{Goerbig2006PRB} M.O.~Goerbig, R.~Moessner, and B.~Dou\c{c}ot, {\it Phys. Rev. B} {\bf 74}, 161407(R)
(2006).

\bibitem{Fuchs2007PRB} J.-N. Fuchs and P. Lederer, {\it Phys. Rev. Lett.} {\bf 98},
016803 (2007).

\bibitem{Ezawa2006} M.~Ezawa, cond-mat/0606084;
cond-mat/0609612.

\bibitem{Herbut2007} I.F.~Herbut, Phys. Rev B {\bf 76}, 085432 (2007).

\bibitem{Sheng2007} L.~Sheng, D.N.~Sheng, F.D.M.~Haldane,
L.~Balents,  arXiv:0706.0371.



\bibitem{Gusynin1995PRD} V.P.~Gusynin, V.A.~Miransky, and I.A.~Shovkovy,
{\it Phys. Rev. Lett.} {\bf 73}, 3499 (1994);
{\it Phys. Rev. D} {\bf 52}, 4718 (1995);
{\it Nucl.\ Phys.\ B} {\bf 462}, 249 (1996).

\bibitem{Kopelevich}
Y.~Kopelevich, V.V. Lemanov, S. Moehlecke, and J. H. S. Torres, {\it
Fiz. Tverd. Tela} {\bf 41}, 2135 (1999) [{\it Phys. Solid State}
{\bf 41}, 1959 (1999)]; H. Kempa, Y.~Kopelevich, F. Mrowka, A.
Setzer, J. H. S. Torres, R. H\"{o}hne, and P. Esquinazi, {\it Solid
State Commun.} {\bf 115}, 539 (2000); M. S. Sercheli, Y.~Kopelevich,
R. R. da Silva, J. H. S. Torres, and C. Rettori, {\it Solid State
Commun.} {\bf 121}, 579 (2002). For the latest results of this
group, see Y.~Kopelevich, J. C. Medina Pantoja, R. R. da Silva, F.
Mrowka, and P. Esquinazi, {\it Phys. Lett. A} {\bf 355}, 233 (2006).

\bibitem{Gorbar2003PLA} E.V.~Gorbar, V.P.~Gusynin, V.A.~Miransky,
and I.A.~Shovkovy, {\it Phys. Lett. A} {\bf 313}, 472 (2003).



\bibitem{Ando2002JPSJ} T.~Ando, Y.~Zheng, and H.~Suzuura {\it J.~Phys.~Soc.~Jpn.} {\bf 71},
1318 (2002).

\bibitem{Gusynin2006micro} V.P.~Gusynin, S.G.~Sharapov and J.P.~Carbotte,
{\it Phys.~Rev.~Lett.} {\bf 96}, 256802 (2006).

\bibitem{Gusynin2007JPCM} V.P.~Gusynin, S.G.~Sharapov and
J.P.~Carbotte, {\it J. Phys.: Condens. Matter.} {\bf 19}, 026222
(2007).

\bibitem{Falkovsky2006} L.A.~Falkovsky and A.A.~Varlamov,
{\it Eur. Phys. J. B} {\bf 56}, 281 (2007).

\bibitem{Ryu2006} S.~Ryu, C.~Mudry, A.~Furusaki, and A.W.W.~Ludwig,
{\it Phys. Rev. B} {\bf 75}, 205344 (2007).


\bibitem{Kim2004PRB} W.~Kim, F.~Marsiglio, and J.~P.~Carbotte, Phys.~Rev.~B {\bf 70},
060505(R) (2004).

\bibitem{Lanzara.private} A.~Lanzara, private communication and
S.Y. Zhou, G.-H. Gweon, A.V. Fedorov, P.N. First, W.A. de Heer, D.-H. Lee, F. Guinea,
A.H. Castro Neto, and  A. Lanzara, {\it Nature Materials} {\bf 6}, 770 (2007).

\bibitem{Sadowski2006PRL} M.L.~Sadowski, G.~Martinez, M.~Potemski, C.~Berger, and
W.A.~de~Heer, {\it Phys. Rev. Lett.} {\bf 97}, 266405 (2006); {\it
Solid State Comm.} {\bf 143}, 123 (2007).

\bibitem{Li2006PRB} Z.Q.~Li, S.-W.~Tsai, W.J.~Padilla, S.V.~Dordevic,
K.S.~Burch, Y.J.~Wang, and D.N. Basov, {\it Phys. Rev. B} {\bf 74},
195404 (2006).

\bibitem{Gusynin2007PRL} V.P.~Gusynin, S.G.~Sharapov and J.P.~Carbotte,
{\it Phys.~Rev.~Lett.} {\bf 98}, 157402 (2007).

\bibitem{Dora2007} B.~D\'{o}ra and P.~Thalmeier,
Phys. Rev. B {\bf 76}, 035402 (2007).

\bibitem{Iyengar2007PRB}
A.~Iyengar, J.~Wang, H.A.~Fertig, and L.~Brey, {\it Phys. Rev. B}
{\bf 75}, 125430 (2007).





\bibitem{Gusynin2007PRB} V.P.~Gusynin, S.G.~Sharapov and J.P.~Carbotte,
{\it Phys.~Rev. B} {\bf 75}, 165407 (2007).

\bibitem{Marel:book} D. van der Marel,
in {\em Strong Interactions in Low Dimensions, Series: Physics and
Chemistry of Materials with Low-Dimensional Structures}, Vol. 25,
edited by D.~Baeriswyl and L. De Giorgi, (Kluver Academic, Berlin,
2005); arXiv:cond-mat/0301506.

\bibitem{Benfatto:review} L.~Benfatto and S.~Sharapov, {\it Fiz. Nizk. Temp.} {\bf 32}, 700 (2006)
[{\it Low Temp. Phys.} {\bf 32}, 533 (2006).]

\bibitem{Carbotte:review} J.P.~Carbotte and E.~Schachinger, {\it J. Low Temp.
Phys.} {\bf 144}, 61 (2006).

\bibitem{Drew1997PRL} H.D.~Drew and P.~Coleman, {\it Phys. Rev. Lett.} {\bf 78}, 1572 (1997).


\bibitem{Sharapov2004PRB} S.G.~Sharapov, V.P.~Gusynin, and H.~Beck,
{\it Phys. Rev. B} {\bf 69}, 075104 (2004).

\bibitem{Gusynin2005PRB} V.P.~Gusynin and S.G.~Sharapov,
{\it Phys. Rev. B} {\bf 71}, 125124 (2005).

\bibitem{Rigal2004PRL} L.B.~Rigal, D.C.~Schmadel, H.D.~Drew, B.~Maiorov, E.~Osquigil,
J.S.~Preston, R.~Hughes, and G.D.~Gu, {\it Phys. Rev. Lett.} {\bf
93}, 137002 (2004).

\bibitem{Melrose} D.B.~Melrose and A.J.~Parle, Aust.~J.~Phys. {\bf 36}, 755 (1983).

\bibitem{Bernevig2006IJMPB} B.A.~Bernevig, T.L.~Hughes, H.-D.~Chen, C.~Wu,
S.-C.~Zhang, {\it Int.~J. Mod.~Phys. B} {\bf 20}, 3257 (2006).

\end{thebibliography}
\end{document}